\documentclass[journal,onecolumn,10pt]{IEEEtran}
\usepackage{amsmath,amssymb}

\usepackage{amsthm}
\usepackage{bm}
\usepackage{iftex}
\ifPDFTeX\usepackage{graphicx}\else\ifLuaTeX\usepackage{graphicx}\else\ifXeTeX\usepackage{graphicx}\else
  \usepackage[dvipdfmx]{graphicx}% (u)pLaTeX + dvipdfmx
\fi\fi\fi
\usepackage{booktabs}
\usepackage{tikz}
\usetikzlibrary{decorations.pathreplacing,arrows.meta}
\usepackage{cite}

\theoremstyle{plain}
\newtheorem{theorem}{Theorem}
\newtheorem{lemma}{Lemma}
\newtheorem{proposition}{Proposition}
\newtheorem{corollary}{Corollary}
\theoremstyle{definition}
\newtheorem{definition}{Definition}
\theoremstyle{definition}% bold-face heads with upright text for Remarks and Examples (as in Parts I and III)
\newtheorem{remark}{Remark}

\newtheorem{remarkA}{Remark}% Remark A: separate counter, so that the numbering of the other Remarks is unchanged

\newcommand{\bX}{\mathbf{X}}
\newcommand{\bY}{\mathbf{Y}}
\newcommand{\bx}{\mathbf{x}}
\newcommand{\by}{\mathbf{y}}
\newcommand{\bz}{\mathbf{z}}
\newcommand{\ba}{\mathbf{a}}
\newcommand{\bb}{\mathbf{b}}
\newcommand{\bu}{\mathbf{u}}
\newcommand{\bv}{\mathbf{v}}
\newcommand{\bone}{\mathbf{1}}
\newcommand{\R}{\mathbb{R}}
\newcommand{\E}{\mathbb{E}}
\newcommand{\Cl}{\mathrm{Cl}}
\newcommand{\cX}{\mathcal{X}}
\newcommand{\cY}{\mathcal{Y}}
\newcommand{\cZ}{\mathcal{Z}}
\newcommand{\cV}{\mathcal{V}}
\newcommand{\cU}{\mathcal{U}}
\newcommand{\cA}{\mathcal{A}}
\newcommand{\cB}{\mathcal{B}}
\newcommand{\cC}{\mathcal{C}}
\newcommand{\cS}{\mathcal{S}}
\newcommand{\cT}{\mathcal{T}}
\newcommand{\cK}{\mathcal{K}}
\newcommand{\cI}{\mathcal{I}}
\newcommand{\cO}{\mathcal{O}}
\newcommand{\cP}{\mathcal{P}}
\newcommand{\essinf}{w\text{-}\mathrm{ess.inf}\,}
\newcommand{\tth}{F}% realized component (random variable with law w)
\newcommand{\cL}{\mathcal{L}}
\newcommand{\cD}{\mathcal{D}}
\newcommand{\cG}{\mathcal{G}}
\newcommand{\cE}{\mathcal{E}}
\newcommand{\bj}{\mathbf{j}}
\newcommand{\bGamma}{\boldsymbol{\Gamma}}

\begin{document}

\title{Information-Spectrum Methods for Multiple-Access\\ Channels---Part~II: $\varepsilon$-Capacity Regions\\ in the Mixed Finite-Alphabet Case}
\author{Te Sun Han, \textit{Life Fellow, IEEE} and Hideki Yagi, \textit{Member, IEEE}%
\thanks{T. S. Han is with The University of Electro-Communications, Chofu, Tokyo 182-8585, Japan (e-mail: han@is.uec.ac.jp).}%
\thanks{H. Yagi is with The University of Electro-Communications, Chofu, Tokyo 182-8585, Japan (e-mail: h.yagi@uec.ac.jp).}}

\maketitle

\begin{abstract}
In a nonergodic channel such as a slowly fading link, the channel law is drawn once and held fixed for the whole codeword, and
the relevant performance measure is the $\varepsilon$-capacity. We study the $\varepsilon$-capacity regions of two-user mixed
multiple-access channels (MACs) with finite alphabets, additive cost constraints and an arbitrary mixing measure. The direct parts
are derived from a single information spectrum formula; the converse parts use, in addition, Fano's inequality and Ahlswede's strong
converse applied to each component. Without knowledge of the realized component, we prove a single-letter inner bound for every
$0\le\varepsilon<1$, in which one input serves all components; for $\varepsilon=0$ it is tight, and the capacity region is given by
the essential infima of the component mutual informations. We prove a single-letter outer bound, the outage region, in which each
component may use its own input; it holds with or without channel state information (CSI) at either terminal, and it is the
$\varepsilon$-capacity region when the encoders know the component and the parameter space is countable. Both bounds coincide, in all
four CSI scenarios, for well-ordered mixed MACs, which include degradedly ordered families and mixtures with a common optimal input,
such as mixed additive MACs. Without CSI the outer bound can be strictly loose, and the no-CSI $\varepsilon$-capacity region for
$0<\varepsilon<1$ remains open in general. Finally, we extend Ahlswede's strong converse to $K$ users with cost constraints and show
that most of these results extend to $K$-user mixed MACs.
\end{abstract}

\begin{IEEEkeywords}
Information spectrum, mixed channel, multiple-access channel, $\varepsilon$-capacity region, outage capacity,
strong converse, channel state information.
\end{IEEEkeywords}

%% ====================================================================
\section{Introduction}
\subsection{Motivation}

In many wireless systems the channel varies slowly compared with the duration of a codeword. The fading
coefficient, or more generally the channel law, is then drawn once at the start of transmission and kept fixed until
the end. Such a channel is \emph{nonergodic}: coding over a long block does not average over channel states, and for many
fading distributions no positive rate can be supported with vanishing error probability. The engineering response
is the notion of \emph{outage} \cite{OSW94}, \cite{BPS98}: one fixes a target rate, accepts that the channel will occasionally be too weak to
support it, and asks for the largest rate whose outage probability does not exceed a prescribed $\varepsilon$.

The information-theoretic model of this situation is the \emph{mixed channel}, also called the averaged \cite{Ahl68} or composite
\cite{EGL10} channel. It is a weighted mixture of component channels $W_\theta$, the weight being a probability measure $w$ on
a parameter space $\Theta$; the component in force is drawn according to $w$ and held fixed for the whole block. The
operational quantity that corresponds to outage is the \emph{$\varepsilon$-capacity}, the supremum of the rates achievable with error
probability at most $\varepsilon$. For a mixed channel it genuinely depends on $\varepsilon$. For a multiple-access channel (MAC) the
corresponding object is the $\varepsilon$-capacity region.

The information spectrum method \cite{VH94}, \cite{Han98}, \cite{Han03} gives the $\varepsilon$-capacity region of a completely general MAC, without
any assumption of stationarity, ergodicity or memorylessness, in terms of the limiting distribution of information
densities. This general formula is the starting point of the present series (see \cite[Fig.~1]{HY-I} for an overview). In Part I \cite{HY-I} it is extended to cost constraints
and applied to Gaussian and quasi-static fading MACs and to mixed MACs with general components and CSI; in Part III \cite{HY-III} the random-coding and converse lemmas from which it is derived are used to prove the strong converse for the
$K$-user Gaussian MAC. The price of the generality is that the formula is not directly computable. For engineering
use one wants to know when, and in what form, it reduces to a single-letter expression of the outage type.

\begin{table}[t]
\centering
\caption{Summary of results for finite-alphabet mixed MACs with an arbitrary mixing measure.}
\label{tab:summary}
\small
\begin{tabular}{llll}
\toprule
Channel & CSI & $\varepsilon$ & Result\\
\midrule
Mixed additive & any & $0\le\varepsilon<1$ & exact, single-letter (Thm.~\ref{thm:mixadd}, Cor.~\ref{cor:verifiable}(a))\\
Mixed memoryless & none & $0\le\varepsilon<1$ & inner bound $\cI_{\Gamma}(\varepsilon)$ (Thm.~\ref{thm:inner})\\
Mixed memoryless & none & $\varepsilon=0$ & exact, essential infima (Thm.~\ref{thm:cap})\\
Mixed memoryless & any & $0\le\varepsilon<1$ & outer bound $\cO_{\Gamma}(\varepsilon)$ (Thm.~\ref{thm:outer})\\
Mixed memoryless, $\Theta$ countable & CSIT, CSIRT & $0\le\varepsilon<1$ & exact, $=\cO_{\Gamma}(\varepsilon)$ (Thm.~\ref{thm:csit})\\
Well-ordered & any & $0\le\varepsilon<1$ & exact, $\cI_{\Gamma}(\varepsilon)=\cO_{\Gamma}(\varepsilon)$ (Thm.~\ref{thm:wellordered})\\
Mixed memoryless & none & $0<\varepsilon<1$ & open; $\cO_{\Gamma}(\varepsilon)$ not tight in general (Prop.~\ref{prop:gap})\\
\bottomrule
\end{tabular}
\end{table}

This paper answers that question for finite-alphabet mixed memoryless MACs with additive cost constraints and
an arbitrary mixing measure, and determines what knowledge of the realized component at either terminal can and
cannot buy in this setting. Mixed MACs with completely general components are treated in Part I \cite[Sec.~VIII]{HY-I}.

\subsection{One input for all components, or one input per component}\label{sec:oneinput}

The central object is a pair of single-letter regions, stated precisely in Sections~\ref{sec:inner} and \ref{sec:outer}. For an input triple
$(X,Y,T)$, with time-sharing variable $T$, and a component $\theta$, let $\Lambda_\theta(X,Y,T)$ be the Ahlswede--Liao pentagon
\[
R_1\le I(X;Z_\theta\mid YT),\qquad R_2\le I(Y;Z_\theta\mid XT),\qquad R_1+R_2\le I(XY;Z_\theta\mid T),
\]
where $Z_\theta$ is the output of $W_\theta$, and let $C_{\Gamma}(W_\theta)$ be the capacity region of $W_\theta$ under the cost constraint $\Gamma$. Consider
\begin{align*}
\cI_{\Gamma}(\varepsilon)&=\Cl\Bigl(\bigcup_{(X,Y,T)}\bigl\{R\in\R^2_+:w\{\theta:R\notin\Lambda_\theta(X,Y,T)\}\le\varepsilon\bigr\}\Bigr),\\
\cO_{\Gamma}(\varepsilon)&=\Bigl\{R\in\R^2_+:w\Bigl\{\theta:R\notin\bigcup_{(X,Y,T)}\Lambda_\theta(X,Y,T)\Bigr\}\le\varepsilon\Bigr\}=\bigl\{R\in\R^2_+:w\{\theta:R\notin C_{\Gamma}(W_\theta)\}\le\varepsilon\bigr\},
\end{align*}
where $\Cl$ denotes closure (see \eqref{eq:inner-region} and \eqref{eq:outage-region}). The two regions differ only in where the union over inputs is taken.
In $\cI_{\Gamma}(\varepsilon)$ it is outside the probability: a single input must serve every component, as it must when the encoders do
not know $\theta$. In $\cO_{\Gamma}(\varepsilon)$ it is inside: each component may use its own input, so $\cO_{\Gamma}(\varepsilon)$ is the familiar outage region
of the engineering literature. Always $\cI_{\Gamma}(\varepsilon)\subseteq\cO_{\Gamma}(\varepsilon)$.

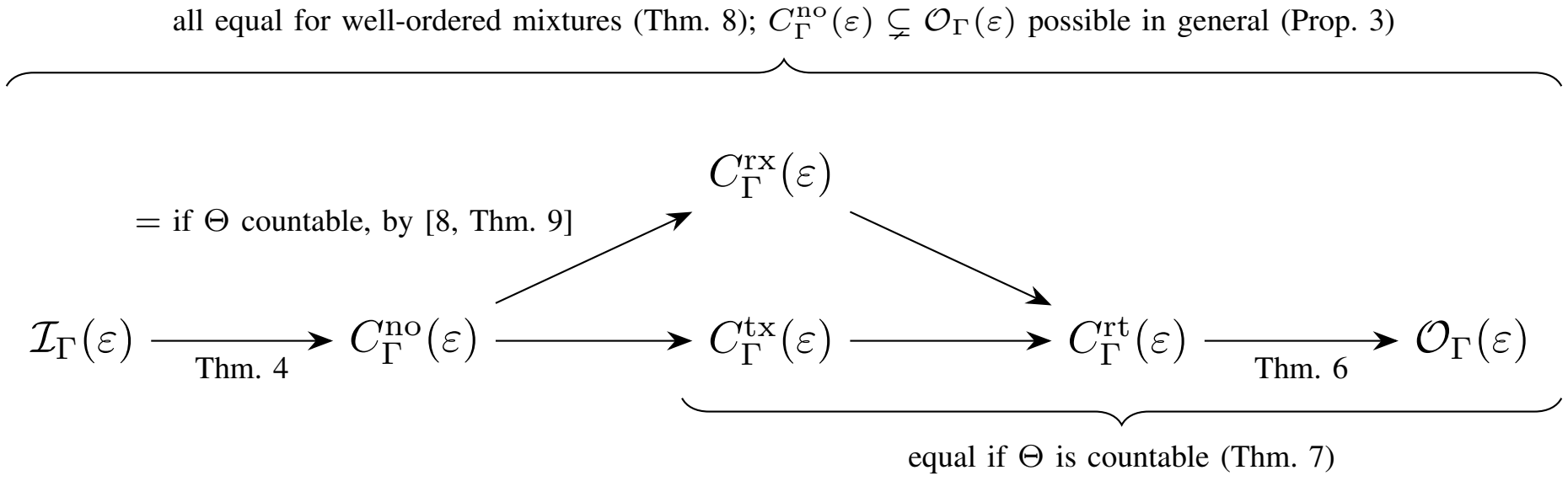
\begin{figure}[t]
\centering
\begin{tikzpicture}[x=1cm,y=1cm,>={Stealth[length=2mm]}]
\node (I) at (0,0) {$\cI_{\Gamma}(\varepsilon)$};
\node (no) at (2.6,0) {$C^{\rm no}_{\Gamma}(\varepsilon)$};
\node (tx) at (5.4,0) {$C^{\rm tx}_{\Gamma}(\varepsilon)$};
\node (rt) at (8.2,0) {$C^{\rm rt}_{\Gamma}(\varepsilon)$};
\node (O) at (10.9,0) {$\cO_{\Gamma}(\varepsilon)$};
\node (rx) at (5.4,1.3) {$C^{\rm rx}_{\Gamma}(\varepsilon)$};
\draw[->] (I) -- node[below,font=\scriptsize]{Thm.~\ref{thm:inner}} (no);
\draw[->] (no) -- (tx);
\draw[->] (tx) -- (rt);
\draw[->] (rt) -- node[below,font=\scriptsize]{Thm.~\ref{thm:outer}} (O);
\draw[->] (no) -- node[above left=1pt,font=\scriptsize]{$=$ if $\Theta$ countable, by \cite[Thm.~9]{HY-I}} (rx);
\draw[->] (rx) -- (rt);
\draw[decorate,decoration={brace,amplitude=6pt,mirror}] (4.7,-0.45) -- (11.6,-0.45) node[midway,below=7pt,font=\scriptsize]{equal if $\Theta$ is countable (Thm.~\ref{thm:csit})};
\draw[decorate,decoration={brace,amplitude=6pt}] (-0.6,2.0) -- (11.6,2.0) node[midway,above=7pt,font=\scriptsize]{all equal for well-ordered mixtures (Thm.~\ref{thm:wellordered}); $C^{\rm no}_{\Gamma}(\varepsilon)\subsetneq\cO_{\Gamma}(\varepsilon)$ possible in general (Prop.~\ref{prop:gap})};
\end{tikzpicture}
\caption{The four $\varepsilon$-capacity regions and the single-letter bounds $\cI_{\Gamma}(\varepsilon)$ and $\cO_{\Gamma}(\varepsilon)$. An arrow $A\to B$ means $A\subseteq B$. If $\Theta$ is
countable, $C^{\rm no}_{\Gamma}(\varepsilon)=C^{\rm rx}_{\Gamma}(\varepsilon)$ \cite[Theorem~9]{HY-I} and $C^{\rm tx}_{\Gamma}(\varepsilon)=C^{\rm rt}_{\Gamma}(\varepsilon)=\cO_{\Gamma}(\varepsilon)$ (Theorem~\ref{thm:csit}).}
\label{fig:chain}
\end{figure}

Our main results locate the $\varepsilon$-capacity regions between these two sets. We consider four CSI scenarios: no CSI,
CSI at the receiver (CSIR), at the transmitters (CSIT), or at both (CSIRT); the corresponding regions are denoted
$C^{\rm no}_{\Gamma}$, $C^{\rm rx}_{\Gamma}$, $C^{\rm tx}_{\Gamma}$ and $C^{\rm rt}_{\Gamma}$. Fig.~\ref{fig:chain} shows the resulting chain of inclusions, and Table~\ref{tab:summary} lists the results.
\begin{itemize}
\item \emph{Inner bound} (Theorem~\ref{thm:inner}). $\cI_{\Gamma}(\varepsilon)\subseteq C^{\rm no}_{\Gamma}(\varepsilon)$ for every $0\le\varepsilon<1$ and every mixing measure. The parameter
space need not be finite or countable.
\item \emph{Capacity region} (Theorem~\ref{thm:cap}). For $\varepsilon=0$ the inner bound is tight. The capacity region is described by the
essential infima, with respect to $w$, of the three component mutual informations. It reduces to the Ahlswede--Liao
region \cite{Ahl71}, \cite{Liao72} for a single component and to the formulas of Ahlswede \cite{Ahl68} and Han \cite[Theorem 3.6.5]{Han03}
for a single user.
\item \emph{Outer bound} (Theorem~\ref{thm:outer}). $C^s_{\Gamma}(\varepsilon)\subseteq\cO_{\Gamma}(\varepsilon)$ for all four scenarios $s$. The proof applies Ahlswede's strong
converse \cite{Ahl82} to each component separately.
\item \emph{CSI at the transmitters} (Theorem~\ref{thm:csit}). If $\Theta$ is countable, $C^{\rm tx}_{\Gamma}(\varepsilon)=C^{\rm rt}_{\Gamma}(\varepsilon)=\cO_{\Gamma}(\varepsilon)$. Knowledge of the component
at the encoders is exactly what is needed to realize the outage region, and knowledge at the receiver adds
nothing further. In combination with Part I \cite[Theorem~9]{HY-I}, which gives these regions for mixed MACs with
general components as a multi-letter information spectrum formula, Theorem~\ref{thm:csit} single-letterizes that formula.
\item \emph{Well-ordered mixtures} (Theorem~\ref{thm:wellordered}, Corollaries~\ref{cor:verifiable} and \ref{cor:womc}). If, for every target rate pair, one input serves all components
that can support it, then $\cI_{\Gamma}(\varepsilon)=\cO_{\Gamma}(\varepsilon)$ and all four $\varepsilon$-capacity regions coincide. This holds for degradedly
ordered families, where the $\varepsilon$-capacity region is the capacity region of the component at the $\varepsilon$-quantile of the
noise level, and for mixtures with a common optimal input. Mixed additive MACs (Theorem~\ref{thm:mixadd}) are an instance
of the latter.
\item \emph{A gap} (Proposition~\ref{prop:gap}). Without CSI the outer bound is not tight in general.
\end{itemize}

\paragraph*{Relation to earlier work}
For a single user, first- and second-order $\varepsilon$-capacities of mixed memoryless channels with a general mixture were obtained by Yagi,
Han and Nomura \cite{YHN16}, and the essential-infimum formula for $\varepsilon=0$ goes back to Ahlswede \cite{Ahl68} and \cite[Theorem 3.6.5]{Han03}.
Theorems~\ref{thm:inner} and \ref{thm:cap} extend the direct part and the $\varepsilon=0$ formula to the MAC with an arbitrary mixing measure. What is
new, to our knowledge, beyond this extension is the treatment of CSI for finite-alphabet mixed MACs: the outer bound in all four CSI scenarios
(Theorem~\ref{thm:outer}), the exact CSIT and CSIRT regions for a countable parameter space (Theorem~\ref{thm:csit}), the classes of mixed MACs for which the
no-CSI inner bound is tight (Theorem~\ref{thm:wellordered}), and an example showing that the outer bound is not tight without CSI
(Proposition~\ref{prop:gap}). The capacity versus outage of Effros, Goldsmith and Liang \cite{EGL10} assumes CSI at most at the receiver, and Part I
\cite[Sec.~VIII]{HY-I} gives the regions with CSI only as multi-letter information spectrum formulas.
For $0<\varepsilon<1$ and general mixed MACs without CSI, the gap between $\cI_{\Gamma}(\varepsilon)$ and $\cO_{\Gamma}(\varepsilon)$ remains open. Section~\ref{sec:discussion}
explains why a per-component strong converse cannot close it.

\subsection{Related work}

The general formula for single-user channels is due to Verd\'u and Han \cite{VH94}, and its MAC version to Han \cite{Han98}. For
single-user mixed memoryless channels with a general mixture, first- and second-order $\varepsilon$-capacities were obtained
by Yagi, Han and Nomura \cite{YHN16}, who also introduced a well-ordered condition for single-user mixed channels;
Definition~\ref{def:wellordered} is a MAC counterpart of a simplified form of it (see Proposition~\ref{prop:singleuser}). Capacity versus outage for
composite channels is studied by Effros, Goldsmith and Liang \cite{EGL10}, with CSI available at most at the receiver; CSI
at the transmitter, the setting of Theorem~\ref{thm:csit}, is not considered there. For compound channels, where no probability
is assigned to the states and a single code must be reliable for every state, Loyka and Charalambous \cite{LC16} derived
general formulas for the capacity and the $\varepsilon$-capacity with CSI at the receiver by the information spectrum approach.
Their formula extends \cite[Theorem 3.3.5]{Han03}, which treats finite-state compound channels through mixed channels, to
arbitrary state sets. They also consider the worst-case capacity $\inf_s\sup_pI(X;Y\mid s)$ \cite[(29)]{LC16}, which
for discrete memoryless compound channels is the capacity with full CSI at the transmitter \cite{Wol78}, and conditions under
which it equals the compound capacity \cite[Sec.~IV-B]{LC16}. Our setting differs in that a mixing measure $w$ is given and the
error probability is averaged over it, as in \eqref{eq:csi-avgerr}. For $\varepsilon=0$, the capacity region of Theorem~\ref{thm:cap} has the same sup--inf
form as the capacity of a uniform compound channel \cite[Theorem 12]{LC16}, with the infimum replaced by an essential
infimum with respect to $w$. Part I \cite[Sec.~VIII]{HY-I} extends to the MAC with CSI the information spectrum treatment of mixed
channels with general components \cite[Lemma 7]{YHN16}, \cite[Lemma 4]{Han98}. Its finite-alphabet MAC version, with CSI at the transmitters, is new here (see the end of Section~\ref{sec:oneinput}). Section~\ref{sec:outer}
is the finite-alphabet counterpart of the outage analysis of the quasi-static fading Gaussian MAC in Part I \cite{HY-I}.

\subsection{Method}

Specializing the general formula to a concrete channel amounts to computing tail probabilities of information
densities, a step sometimes called information spectrum calculus. Besides the general formula, we need only three
elementary tools, collected in Section~\ref{sec:tools}: a change-of-measure bound (Lemma~\ref{lem:com}); a type-counting bound that
compares a mixture of memoryless channels with almost all of its components (Lemma~\ref{lem:mixcomp}); and the uniform continuity
of mutual information (Lemma~\ref{lem:cont}). The idea is simple. By Lemma~\ref{lem:mixcomp}, the information densities of the mixture are,
for almost every realized component, close to those of that component, and the latter obey the law of large numbers.
In engineering terms, the decoder of the mixture loses nothing in rate by not knowing which component is in force,
so that each component either supports the rate pair or is in outage, exactly as in the Gaussian case of Part I.
The main line of every direct part is therefore the same three steps as in Part I \cite[Sec.~I-B]{HY-I}: the general formula,
a single-letter input lemma (Lemma~\ref{lem:single}), and the reduction from the mixture to the components (Lemma~\ref{lem:reduction}).
Section~\ref{sec:outer} additionally uses Ahlswede's non-asymptotic strong converse
for the discrete memoryless MAC \cite{Ahl82}, extended to cost constraints, and, for the direct part with CSI at the transmitters, the
CSIT coding lemmas of Part I \cite[Lemmas~7 and 8]{HY-I}. Additive MACs (Section~\ref{sec:additive}) serve as a warm-up
in which every step of the calculus can be carried out explicitly.

\subsection{Organization}

Section~\ref{sec:prelim} introduces the coding problem, recalls the general formula, and states the three tools. Section~\ref{sec:additive} treats
additive MACs. Section~\ref{sec:inner} proves the single-letter inner bound, and Section~\ref{sec:cap} the capacity region. Section~\ref{sec:outer} introduces the CSI
scenarios, derives the single-letter outer bound, treats CSI at the transmitters and well-ordered mixtures, and discusses the open case. Section~\ref{sec:kuser} extends the results to $K$
users. Section~\ref{sec:conc} concludes the
paper. Technical proofs are collected in the appendices.

%% ====================================================================
\section{General MAC and Basic Tools}\label{sec:prelim}
\subsection{Notation}

Logarithms are natural. For a set $A\subseteq\R^2$ we write
\begin{equation}\label{eq:clplus}
\Cl_+(A)\equiv\Cl(A)\cap\R^2_+,
\end{equation}
where $\Cl$ denotes the closure in $\R^2$; that is, rate pairs with negative components are admitted before the closure
is taken, and the result is then restricted to $\R^2_+$. This convention, used in \cite[Theorem~1 and Remark~1]{HY-I}, is
needed in the general formula (Theorem~\ref{thm:general}) and in Theorem~\ref{thm:additive}, which are stated for completely general channels; Remark~\ref{rem:clplus} explains why it cannot be dropped there. The
single-letter results (Theorems~\ref{thm:mixadd}--\ref{thm:cap} and Section~\ref{sec:outer}) are stated with rate pairs in $\R^2_+$ and the ordinary closure
$\Cl$; equivalent forms with $\Cl_+$ are noted where useful (Theorem~\ref{thm:mixadd} and \eqref{eq:cap-clplus}). For a pair $R=(R_1,R_2)$
and $\gamma\in\R$ we write $R-\gamma\equiv(R_1-\gamma,R_2-\gamma)$. For pmfs $P,Q$ on a finite set $\cU$, $d(P,Q)\equiv\frac12\sum_u|P(u)-Q(u)|$ is the variational
distance, and $h(\cdot)$ denotes the binary entropy function. For a measurable function $f$ on a probability space $(\Theta,w)$,
$\essinf f\equiv\sup\{a:w\{\theta:f(\theta)<a\}=0\}$.

\subsection{Coding system}

Let $\cX$, $\cY$ and $\cZ$ be the input alphabets of users 1 and 2 and the output alphabet, respectively. A general two-user
MAC is a sequence $W=\{W^n\}_{n=1}^\infty$ of transition probabilities $W^n:\cX^n\times\cY^n\to\cZ^n$, which may be nonstationary
and nonergodic. An $(n,M_n^{(1)},M_n^{(2)},\varepsilon_n)$ code consists of codebooks $\{\bu_j\}_{j=1}^{M_n^{(1)}}\subseteq\cX^n$ and $\{\bv_k\}_{k=1}^{M_n^{(2)}}\subseteq\cY^n$ and
mutually disjoint decoding regions $D_{jk}\subseteq\cZ^n$, with average error probability
\begin{equation}\label{eq:avgerr}
\varepsilon_n\equiv\frac{1}{M_n^{(1)}M_n^{(2)}}\sum_{j=1}^{M_n^{(1)}}\sum_{k=1}^{M_n^{(2)}}W^n(D^c_{jk}\mid\bu_j,\bv_k).
\end{equation}

\subsection{$\varepsilon$-capacity region with cost constraint}

Let $c_1:\cX^n\to[0,\infty)$ and $c_2:\cY^n\to[0,\infty)$ be cost functions and $\Gamma=(\Gamma_1,\Gamma_2)$ a cost constraint. A code
satisfies the cost constraint if $\frac1nc_1(\bu_j)\le\Gamma_1$ and $\frac1nc_2(\bv_k)\le\Gamma_2$ for all $j,k$.

\begin{definition}\label{def:achievable}
Let $0\le\varepsilon<1$. A pair $(R_1,R_2)\in\R^2_+$ is \emph{$\varepsilon$-achievable with cost constraint} $\Gamma$ if there exists a sequence
of $(n,M_n^{(1)},M_n^{(2)},\varepsilon_n)$ codes satisfying the cost constraint with
\[
\limsup_{n\to\infty}\varepsilon_n\le\varepsilon,\qquad\liminf_{n\to\infty}\frac1n\log M_n^{(\alpha)}\ge R_\alpha\quad(\alpha=1,2).
\]
The closure of the set of $\varepsilon$-achievable pairs is the $\varepsilon$-capacity region $C_{\Gamma}(\varepsilon\mid W)$, and $C_{\Gamma}(0\mid W)$ is the capacity
region.
\end{definition}

The cost functions and the cost constraint $\Gamma$ are fixed throughout the paper. As in Parts I and III, $\Gamma$ is nevertheless kept in
the notation of the $\varepsilon$-capacity regions and of the set of general inputs $\cG_{\Gamma}$. In the present paper it is kept also in that of the
set of single-letter inputs $\cP_{\Gamma}$ and of every quantity defined below that depends on $\Gamma$ through $\cP_{\Gamma}$: the inner bound $\cI_{\Gamma}(\varepsilon)$, the component capacity
regions $C_{\Gamma}(W_\theta)$ and $\widehat C_{\Gamma}(W_\theta)$, the outage function $o_{\Gamma}$ and the outage region $\cO_{\Gamma}(\varepsilon)$ ($\bGamma$ in place of $\Gamma$ for $K$ users).
The pentagons $\Lambda_\theta(X,Y,T)$ and $\Lambda^*(X,Y,T)$ depend only on the input and carry no subscript; neither does $C_{\rm ad}$ of Section~\ref{sec:additive},
where there is no cost constraint.

A general input is a pair $(\bX,\bY)$ of sequences $\bX=\{X^n\}$ and $\bY=\{Y^n\}$ with $X^n$ and $Y^n$ independent for
each $n$. Let $\cG_{\Gamma}$ denote the set of general inputs with $\Pr\{\frac1nc_1(X^n)\le\Gamma_1\}=\Pr\{\frac1nc_2(Y^n)\le\Gamma_2\}=1$ for all $n$. Let
$Z^n$ be the output of $W^n$ due to $(X^n,Y^n)$. Define the information densities
\begin{equation}\label{eq:densities}
\iota_1^n\equiv\frac1n\log\frac{W^n(Z^n\mid X^n,Y^n)}{P_{Z^n|Y^n}(Z^n\mid Y^n)},\qquad\iota_2^n\equiv\frac1n\log\frac{W^n(Z^n\mid X^n,Y^n)}{P_{Z^n|X^n}(Z^n\mid X^n)},\qquad
\iota_3^n\equiv\frac1n\log\frac{W^n(Z^n\mid X^n,Y^n)}{P_{Z^n}(Z^n)}.
\end{equation}
For a threshold vector $a=(a_1,a_2,a_3)\in\R^3$ let
\begin{equation}\label{eq:En}
E_n(a\mid\bX,\bY)\equiv\bigcup_{k=1}^3\{\iota_k^n\le a_k\},
\end{equation}
and for $R=(R_1,R_2)\in\R^2$ define
\begin{equation}\label{eq:J}
J_W(R\mid\bX,\bY)\equiv\limsup_{n\to\infty}\Pr\bigl\{E_n(R_1,R_2,R_1+R_2\mid\bX,\bY)\bigr\}.
\end{equation}
Since $E_n(a\mid\bX,\bY)$ grows with each $a_k$, $J_W(R\mid\bX,\bY)$ is nondecreasing in each of $R_1$ and $R_2$. Let
\begin{equation}\label{eq:Aeps}
A_\varepsilon(\bX,\bY)\equiv\bigl\{R\in\R^2:J_W(R\mid\bX,\bY)\le\varepsilon\bigr\}.
\end{equation}
The following general formula is the starting point of this paper. It is \cite[Theorem~1]{HY-I}, which extends \cite[Theorem 5]{Han98}
and \cite[Theorem 7.11.1]{Han03} to cost constraints; we use it without proof.

\begin{theorem}[General formula]\label{thm:general}
For a general MAC $W$, every cost constraint $\Gamma$ and every $0\le\varepsilon<1$,\footnote{Equivalently, the operation $\Cl_+$ may be placed
in front of each $A_\varepsilon(\bX,\bY)$: $C_{\Gamma}(\varepsilon\mid W)=\bigcup_{(\bX,\bY)\in\cG_{\Gamma}}\Cl_+\bigl(A_\varepsilon(\bX,\bY)\bigr)$, which is the form of the general
formula stated in the footnote to Theorem~1 of the preliminary version of Part I \cite{HY-I}. The equivalence is proved in Remark~\ref{rem:clplus-swap} below (cf.\ \cite[Remark~2]{HY-I}).}
\begin{equation}\label{eq:general}
C_{\Gamma}(\varepsilon\mid W)=\Cl_+\Bigl(\bigcup_{(\bX,\bY)\in\cG_{\Gamma}}A_\varepsilon(\bX,\bY)\Bigr).
\end{equation}
\end{theorem}

The cost constraint enters the proof only through the set $\cG_{\Gamma}$. In the direct part of \cite[Theorem 5]{Han98}, codewords
are drawn independently from an input in $\cG_{\Gamma}$ and therefore satisfy the constraint. In the converse part, the input
distributed uniformly on the codewords of a code that satisfies the constraint belongs to $\cG_{\Gamma}$. With these two
observations the proof of \cite[Theorem 5]{Han98} applies without change; the single-user analogue is \cite{VH94}.

\begin{remarkA}[Equivalence of the two forms of the general formula]\label{rem:clplus-swap}
In \eqref{eq:general}, the operation $\Cl_+$ may equivalently be placed in front of each $A_\varepsilon(\bX,\bY)$:
\[
\bigcup_{(\bX,\bY)\in\cG_{\Gamma}}\Cl_+\bigl(A_\varepsilon(\bX,\bY)\bigr)=\Cl_+\Bigl(\bigcup_{(\bX,\bY)\in\cG_{\Gamma}}A_\varepsilon(\bX,\bY)\Bigr).
\]
This is the form in which the general formula was stated in the preliminary version of Part I (see \cite[Remark~2]{HY-I}), so that results stated in
either form are the same. The inclusion ``$\subseteq$'' is obvious. For ``$\supseteq$'', since intersecting with $\R^2_+$ commutes with the union, it
suffices to show that $\bigcup_{(\bX,\bY)\in\cG_{\Gamma}}\Cl(A_\varepsilon(\bX,\bY))$ is closed. Let $R^{(k)}=(R_1^{(k)},R_2^{(k)})\in\Cl(A_\varepsilon(\bX_k,\bY_k))$, $k=1,2,\ldots$,
converge to $R=(R_1,R_2)$, where $(\bX_k,\bY_k)=\{(X_k^n,Y_k^n)\}_{n=1}^\infty\in\cG_{\Gamma}$. Since $R^{(k)}$ is a limit point of $A_\varepsilon(\bX_k,\bY_k)$ and
$J_W$ is nondecreasing in each of $R_1$ and $R_2$, we have $J_W(R^{(k)}-\frac1k\mid\bX_k,\bY_k)\le\varepsilon$, and hence there is an integer $n_k$ such that
\[
\Pr\bigl\{E_n\bigl(R^{(k)}_1-\tfrac1k,R^{(k)}_2-\tfrac1k,R^{(k)}_1+R^{(k)}_2-\tfrac2k\bigm|\bX_k,\bY_k\bigr)\bigr\}\le\varepsilon+\frac1k\quad\mbox{for all }n\ge n_k.
\]
We may assume $n_1<n_2<\cdots$. Define the diagonal input $(\bX,\bY)$ by $(X^n,Y^n)\equiv(X_k^n,Y_k^n)$ for $n_k\le n<n_{k+1}$ (and $(X^n,Y^n)\equiv(X_1^n,Y_1^n)$
for $n<n_1$). Then $(\bX,\bY)\in\cG_{\Gamma}$, because the cost constraint and the independence of $X^n$ and $Y^n$ are required for each $n$
separately, and, since the densities \eqref{eq:densities} at blocklength $n$ depend only on $P_{X^nY^n}$ and $W^n$, the event $E_n$ for $(\bX,\bY)$ at
$n\in[n_k,n_{k+1})$ is that for $(\bX_k,\bY_k)$. Fix $\gamma>0$. Since $R^{(k)}\to R$, there is $k_0$ such that $R-\gamma\le R^{(k)}-\frac1k$ componentwise for all $k\ge k_0$. By the bound above
and the monotonicity of $E_n$, $\Pr\{E_n(R_1-\gamma,R_2-\gamma,R_1+R_2-2\gamma\mid\bX,\bY)\}\le\varepsilon+\frac1k$ for $n_k\le n<n_{k+1}$ and $k\ge k_0$, so that
$J_W(R-\gamma\mid\bX,\bY)\le\varepsilon$, i.e., $R-\gamma\in A_\varepsilon(\bX,\bY)$. Letting $\gamma\downarrow0$ yields $R\in\Cl(A_\varepsilon(\bX,\bY))$, which proves the closedness.
\end{remarkA}

Theorem~\ref{thm:general} has a simple interpretation. The quantity $J_W(R\mid\bX,\bY)$ is the asymptotic probability that at least
one of the three information densities falls below the corresponding rate, that is, that the realized channel behavior
cannot support the rate pair $R$ with the input $(\bX,\bY)$. The $\varepsilon$-capacity region consists of the rate pairs for which
this ``spectral outage probability'' can be made at most $\varepsilon$ by some input. For a mixed channel the densities do not
concentrate at a single value but spread over the component mutual informations; the task of the following sections
is to identify this spread in single-letter terms.

\begin{remark}[Why $\Cl_+$ is needed in \eqref{eq:general}]\label{rem:clplus}
In \eqref{eq:general} the sets $A_\varepsilon(\bX,\bY)$ must be allowed to contain rate pairs with negative
components; restricting them to $\R^2_+$ before taking the closure would make the formula false. Consider a MAC in
which the input of user 2 does not affect the output, $W^n(\bz\mid\bx,\by)=V^n(\bz\mid\bx)$. For every independent input,
$P_{Z^n|X^n}=W^n$, so $\iota_2^n\equiv0$ and $J_W(R\mid\bX,\bY)\ge\Pr\{\iota_2^n\le R_2\}=1$ whenever $R_2\ge0$. Thus $A_\varepsilon(\bX,\bY)\cap\R^2_+=\emptyset$
for every input and every $\varepsilon<1$, and $\Cl(\bigcup A_\varepsilon(\bX,\bY)\cap\R^2_+)=\emptyset$. Yet every pair $(R_1,0)$ with $R_1$ below the capacity
of $V$ is achievable with $M_n^{(2)}=1$; these pairs are recovered by $\Cl_+$ as limits of pairs in $A_\varepsilon(\bX,\bY)$ with $R_2<0$.
The single-letter regions below do not have this defect, because their constraints are of the form $R_\alpha\le I$
with $I\ge0$ nondecreasing in the set of users (see Lemma~\ref{lem:proj}).
\end{remark}

\subsection{Three elementary tools}\label{sec:tools}

The proofs of the three lemmas below are elementary and are given in Appendix~\ref{app:tools}.

The first tool is a change-of-measure bound \cite[Lemma 3.2.1]{Han03}.

\begin{lemma}[Change of measure]\label{lem:com}
Let $(S,U)$ be random variables with $U$ taking values in a countable set, and let
$Q(\cdot\mid s)$ be a pmf for every $s$. Then, for every $a>0$,
\begin{equation}\label{eq:com}
\Pr\bigl\{Q(U\mid S)>e^aP_{U|S}(U\mid S)\bigr\}\le e^{-a}.
\end{equation}
The same holds when $U$ takes values in a general space and $P_{U|S}(\cdot\mid s)$ and $Q(\cdot\mid s)$ are densities with respect to a common
$\sigma$-finite measure \cite[Lemma~1]{HY-I}.
\end{lemma}

The second tool compares a mixture of memoryless channels with its components. Let $\cA,\cB$ be finite sets,
let $(\Theta,w)$ be a probability space, and let $\{Q_\theta:\cA\to\cB\}_{\theta\in\Theta}$ be stochastic matrices such that $\theta\mapsto Q_\theta(b\mid a)$ is
measurable for all $(a,b)$. Write $Q^n_\theta(\bb\mid\ba)\equiv\prod_{i=1}^nQ_\theta(b_i\mid a_i)$ and $Q^n(\bb\mid\ba)\equiv\int_\Theta Q^n_\theta(\bb\mid\ba)\,dw(\theta)$. A
finite-alphabet source is the special case $|\cA|=1$.

\begin{lemma}[Mixture versus components]\label{lem:mixcomp}
Let
\begin{equation}\label{eq:Thetastar}
\Theta^*_n\equiv\Bigl\{\theta\in\Theta:Q^n_\theta(\bb\mid\ba)\le e^{\sqrt n}\,Q^n(\bb\mid\ba)\mbox{ for all }(\ba,\bb)\in\cA^n\times\cB^n\Bigr\}.
\end{equation}
Then $w(\Theta\setminus\Theta^*_n)\le(n+1)^{|\cA||\cB|}e^{-\sqrt n}\to0$ as $n\to\infty$.
\end{lemma}

In words, for all components outside a set of vanishing $w$-measure, the $n$-letter component channel exceeds the
mixture by at most a factor $e^{\sqrt n}$, which is negligible on the exponential scale $e^{n\gamma}$ relevant to information densities.
The bound holds simultaneously for all sequences, because $Q^n_\theta(\bb\mid\ba)$ depends on $(\ba,\bb)$ only through the joint
type, and there are only polynomially many types.

Lemma~\ref{lem:mixcomp} is a variant of \cite[proof of Lemma 3.3.3]{Han03}; see also \cite{YHN16}.

The third tool is the uniform continuity of mutual information.

For a positive integer $N$ and $0\le\delta\le1/2$ let
\begin{equation}\label{eq:tau}
\tau_N(\delta)\equiv3\bigl(\delta\log N+h(\delta)\bigr),
\end{equation}
which is nondecreasing in $\delta$ and tends to $0$ as $\delta\downarrow0$ for each fixed $N$. In the applications below, $N$ is the product
of the alphabet sizes of the two random variables whose mutual information is compared, e.g., $N=|\cX||\cY||\cZ|$.

\begin{lemma}[Continuity of mutual information]\label{lem:cont}
Let $\cU,\cV$ be finite sets, and let $P,\tilde P$ be pmfs on $\cU\times\cV$ with
$d(P,\tilde P)\le\delta\le1/2$. Then
\begin{equation}\label{eq:cont}
\bigl|I_P(U;V)-I_{\tilde P}(U;V)\bigr|\le\tau_{|\cU||\cV|}(\delta).
\end{equation}
\end{lemma}

%% ====================================================================
\section{$\varepsilon$-Capacity Region of Additive MACs}\label{sec:additive}

Before turning to general mixed MACs, we treat additive MACs, for which the information spectrum calculus
can be carried out completely. Uniform inputs make all three information densities equal to a single quantity, the
normalized entropy density of the noise, so that the $\varepsilon$-capacity region is governed by the entropy spectrum of the
noise alone. The section also shows the three tools at work in their simplest form.

\subsection{General additive noise}

Let $\cX=\cY=\cZ=\cV=\{0,1,\ldots,m-1\}$ and
\begin{equation}\label{eq:additive}
Z^n=X^n+Y^n+V^n\pmod m\quad\mbox{(componentwise)},
\end{equation}
where $\mathbf{V}=\{V^n\}$ is a general noise process independent of the inputs, so that $W^n(\bz\mid\bx,\by)=P_{V^n}(\bz-\bx-\by)$.
There is no cost constraint (formally $c_\alpha\equiv0$), and the $\varepsilon$-capacity region is denoted $C_{\rm ad}(\varepsilon\mid W)$. For $s\in\R$ let
\begin{equation}\label{eq:Jad}
J_{\rm ad}(s)\equiv\limsup_{n\to\infty}\Pr\Bigl\{\log m-\frac1n\log\frac{1}{P_{V^n}(V^n)}\le s\Bigr\},
\end{equation}
which is nondecreasing in $s$.

\begin{theorem}\label{thm:additive}
For every $0\le\varepsilon<1$,
\begin{equation}\label{eq:additive-region}
C_{\rm ad}(\varepsilon\mid W)=\Cl_+\bigl\{R\in\R^2:J_{\rm ad}(R_1+R_2)\le\varepsilon\bigr\}.
\end{equation}
The region depends on $(R_1,R_2)$ only through the sum rate.
\end{theorem}

Theorem~\ref{thm:additive} holds for an arbitrary noise process, which need not be stationary or ergodic. The additive MAC thus
behaves like a single-user additive channel with input alphabet of size $m$ whose rate is shared between the two
users: the sum rate is limited by $\log m$ minus the spectral entropy rate of the noise, and every split of the sum rate
is achievable. As in Remark~\ref{rem:clplus}, the closure $\Cl_+$ in \eqref{eq:additive-region} cannot be replaced by $\Cl$ applied to the nonnegative part.
For i.i.d.\ uniform noise, $J_{\rm ad}(s)=\bone\{s\ge0\}$, so the set in \eqref{eq:additive-region} is $\{R_1+R_2<0\}$; $\Cl_+$ correctly gives the region
$\{(0,0)\}$, whereas its nonnegative part is empty.

\begin{proof}
\emph{Direct part.} Let $X^n,Y^n$ be independent and uniform on $\cX^n,\cY^n$. For every $(\bx,\bz)$, $P_{Z^n|X^n}(\bz\mid\bx)=
m^{-n}\sum_{\by}P_{V^n}(\bz-\bx-\by)=m^{-n}$, and likewise $P_{Z^n|Y^n}$ and $P_{Z^n}$ are uniform. Hence all three densities in \eqref{eq:densities}
equal $\log m-\frac1n\log(1/P_{V^n}(V^n))$, where $V^n=Z^n-X^n-Y^n$, and
\begin{equation}\label{eq:Jad-max}
J_W(R\mid\bX,\bY)=J_{\rm ad}\bigl(\max\{R_1,R_2,R_1+R_2\}\bigr).
\end{equation}
Let $R^*\in\R^2_+$ belong to the right-hand side of \eqref{eq:additive-region}. Then there are $R^{(k)}\to R^*$ with $J_{\rm ad}(s_k)\le\varepsilon$, where
$s_k\equiv R_1^{(k)}+R_2^{(k)}\to s^*\equiv R_1^*+R_2^*$. Put $t_k\equiv\max\{s^*-s_k,0\}\to0$ and $R'^{(k)}\equiv R^*-t_k$. Since $R_1^*,R_2^*\ge0$, each
of $R_1'^{(k)}$, $R_2'^{(k)}$ and $R_1'^{(k)}+R_2'^{(k)}$ is at most $s^*-t_k\le s_k$. By \eqref{eq:Jad-max} and monotonicity, $J_W(R'^{(k)}\mid\bX,\bY)\le J_{\rm ad}(s_k)\le\varepsilon$.
Thus $R'^{(k)}\in A_\varepsilon(\bX,\bY)$ and $R'^{(k)}\to R^*$, so $R^*\in C_{\rm ad}(\varepsilon\mid W)$ by Theorem~\ref{thm:general}.

\emph{Converse part.} Let $(\bX,\bY)$ be any independent input and $\gamma>0$. Apply Lemma~\ref{lem:com} with $S$ trivial, $U=Z^n$ and $Q$
the uniform pmf on $\cZ^n$: with probability at least $1-e^{-n\gamma}$, $\frac1n\log(1/P_{Z^n}(Z^n))\le\log m+\gamma$. On this event,
\[
\iota_3^n=\frac1n\log\frac{1}{P_{Z^n}(Z^n)}-\frac1n\log\frac{1}{P_{V^n}(V^n)}\le\log m+\gamma-\frac1n\log\frac{1}{P_{V^n}(V^n)},
\]
where we used $W^n(Z^n\mid X^n,Y^n)=P_{V^n}(V^n)$. Since $\{\iota_3^n\le R_1+R_2\}\subseteq E_n(R_1,R_2,R_1+R_2\mid\bX,\bY)$,
\begin{equation}\label{eq:additive-conv}
J_W(R\mid\bX,\bY)\ge\limsup_{n\to\infty}\Bigl[\Pr\Bigl\{\log m-\frac1n\log\frac{1}{P_{V^n}(V^n)}\le R_1+R_2-\gamma\Bigr\}-e^{-n\gamma}\Bigr]=J_{\rm ad}(R_1+R_2-\gamma)
\end{equation}
for every $R\in\R^2$. Hence $R\in A_\varepsilon(\bX,\bY)$ implies $R-\gamma/2\in\{R':J_{\rm ad}(R_1'+R_2')\le\varepsilon\}$. Since $\gamma>0$ is arbitrary,
the union in \eqref{eq:general} is contained in the closure of this set, and Theorem~\ref{thm:general} gives ``$\subseteq$'' in \eqref{eq:additive-region}.
\end{proof}

\subsection{Mixed memoryless additive noise}

Now let the noise be a mixture of stationary memoryless sources:
\begin{equation}\label{eq:mixnoise}
P_{V^n}(\bv)=\int_\Theta P^n_\theta(\bv)\,dw(\theta),\qquad P^n_\theta(\bv)=\prod_{i=1}^nP_\theta(v_i),
\end{equation}
where $w$ is an arbitrary probability measure on a parameter space $\Theta$ and $\theta\mapsto P_\theta(v)$ is measurable. Let $H(V_\theta)$
denote the entropy of $P_\theta$. The following theorem extends \cite[Example 3.4.2]{Han03} to the MAC.

\begin{theorem}\label{thm:mixadd}
For every $0\le\varepsilon<1$, the $\varepsilon$-capacity region $C^{\rm m}_{\rm ad}(\varepsilon\mid W)$ of the mixed additive MAC is
\begin{equation}\label{eq:mixadd}
C^{\rm m}_{\rm ad}(\varepsilon\mid W)=\Cl\bigl\{R\in\R^2_+:w\{\theta:\log m-H(V_\theta)<R_1+R_2\}\le\varepsilon\bigr\}.
\end{equation}
Equivalently, $C^{\rm m}_{\rm ad}(\varepsilon\mid W)=\Cl_+\{R\in\R^2:w\{\theta:\log m-H(V_\theta)\le R_1+R_2\}\le\varepsilon\}$.
\end{theorem}

The theorem follows from Theorem~\ref{thm:additive} and the following lemma on the entropy spectrum of a mixed memoryless
source.

\begin{lemma}\label{lem:entspec}
Let $h_n(\bv)\equiv\frac1n\log(1/P_{V^n}(\bv))$. For every $r\in\R$,
\begin{equation}\label{eq:entspec}
w\{\theta:H(V_\theta)>r\}\le\liminf_{n\to\infty}\Pr\{h_n(V^n)\ge r\}\le\limsup_{n\to\infty}\Pr\{h_n(V^n)\ge r\}\le w\{\theta:H(V_\theta)\ge r\}.
\end{equation}
\end{lemma}

The proof, given in Appendix~\ref{app:B}, combines Lemma~\ref{lem:mixcomp} (for the upper bound) with Lemma~\ref{lem:com} (for the lower bound)
and the weak law of large numbers for each component.

\begin{proof}[Proof of Theorem~\ref{thm:mixadd}]
Let $\varphi(s)\equiv w\{\theta:\log m-H(V_\theta)\le s\}$ and $\psi(s)\equiv w\{\theta:\log m-H(V_\theta)<s\}$. Since
$J_{\rm ad}(s)=\limsup_n\Pr\{h_n(V^n)\ge\log m-s\}$, Lemma~\ref{lem:entspec} with $r=\log m-s$ gives $\psi(s)\le J_{\rm ad}(s)\le\varphi(s)$. Hence
\[
\{R:\varphi(R_1+R_2)\le\varepsilon\}\subseteq\{R:J_{\rm ad}(R_1+R_2)\le\varepsilon\}\subseteq\{R:\psi(R_1+R_2)\le\varepsilon\}.
\]
Moreover $\varphi(s)\le\psi(s+\eta)$ for every $\eta>0$, so $\psi(R_1+R_2)\le\varepsilon$ implies $\varphi((R_1-\eta/2)+(R_2-\eta/2))\le\varepsilon$. Thus the
outer two sets have the same closure, and Theorem~\ref{thm:additive} gives $C^{\rm m}_{\rm ad}(\varepsilon\mid W)=\Cl_+\{R\in\R^2:\psi(R_1+R_2)\le\varepsilon\}$,
which is also equal to the corresponding expression with $\varphi$; this is the second form.

It remains to show that $\Cl_+\{R\in\R^2:\psi(R_1+R_2)\le\varepsilon\}$ equals the right-hand side of \eqref{eq:mixadd}; only ``$\subseteq$'' needs
proof. Since $H(V_\theta)\le\log m$, we have $\psi(0)=0$, so the origin belongs to the set in \eqref{eq:mixadd}. Let $R\in\R^2_+$ be
the limit of $R^{(k)}$ with $\psi(s_k)\le\varepsilon$, $s_k\equiv R_1^{(k)}+R_2^{(k)}$. If $s^*\equiv R_1+R_2=0$, then $R=0$. Otherwise put $\lambda_k\equiv
\max\{0,\min\{1,s_k/s^*\}\}\to1$ and $b^{(k)}\equiv\lambda_kR\in\R^2_+$. If $s_k\ge0$, then $b_1^{(k)}+b_2^{(k)}\le s_k$ and $\psi(b_1^{(k)}+b_2^{(k)})\le\psi(s_k)\le\varepsilon$
by monotonicity; if $s_k<0$, then $b^{(k)}=0$ and $\psi(0)=0$. In both cases $b^{(k)}$ lies in the set in \eqref{eq:mixadd}, and $b^{(k)}\to R$.
\end{proof}

The strict inequality in \eqref{eq:mixadd} matters. With ``$\le$'' and $\Cl$ on $\R^2_+$, the origin would be lost for a single component
with uniform noise ($H(V_\theta)=\log m$), as in the example after Theorem~\ref{thm:additive}.

The region \eqref{eq:mixadd} has the outage form described in the Introduction. For component $\theta$ alone, the capacity region is
the triangle $R_1+R_2\le\log m-H(V_\theta)$ (see \eqref{eq:additive-comp} in Section~\ref{sec:examples}). Theorem~\ref{thm:mixadd} states that a rate pair is $\varepsilon$-achievable
exactly when the components whose triangle fails to contain it have total weight at most $\varepsilon$. The reason is that the uniform
input is optimal for every component simultaneously; Section~\ref{sec:outer} shows that this is a special case of a
general principle (Corollary~\ref{cor:verifiable}(a)), and that the region is the same in all four CSI scenarios.

%% ====================================================================
\section{Inner Bound for Mixed Memoryless MACs}\label{sec:inner}
\subsection{Mixed memoryless MAC and the result}

From now on, $\cX$, $\cY$ and $\cZ$ are finite. Let $(\Theta,w)$ be an arbitrary
probability space and, for each $\theta\in\Theta$, let $W_\theta:\cX\times\cY\to\cZ$ be a stochastic matrix such that $\theta\mapsto W_\theta(z\mid x,y)$ is
measurable. The mixed memoryless MAC $W=\{W^n\}$ with components $W_\theta$ is
\begin{equation}\label{eq:mixed}
W^n(\bz\mid\bx,\by)=\int_\Theta W^n_\theta(\bz\mid\bx,\by)\,dw(\theta),\qquad W^n_\theta(\bz\mid\bx,\by)=\prod_{i=1}^nW_\theta(z_i\mid x_i,y_i).
\end{equation}
The cost functions are additive, $c_1(\bx)=\sum_ic_1(x_i)$ and $c_2(\by)=\sum_ic_2(y_i)$, with $c_\alpha\ge0$ on letters. We make the
following standing assumption:
\begin{equation}\label{eq:standing}
\Gamma_1>c_{1,\min}\equiv\min_{x\in\cX}c_1(x),\qquad\Gamma_2>c_{2,\min}\equiv\min_{y\in\cY}c_2(y).
\end{equation}

\begin{remark}\label{rem:standing}
Assumption \eqref{eq:standing} entails no loss of generality. If $\Gamma_1<c_{1,\min}$, no codeword is admissible. If $\Gamma_1=c_{1,\min}$,
every admissible codeword and every admissible single-letter input uses only letters in $\cX_0\equiv\{x:c_1(x)=c_{1,\min}\}$,
and every sequence in $\cX_0^n$ meets the constraint. Replacing $\cX$ by $\cX_0$ thus removes the constraint on user 1, and the
proofs below go through without the truncation for user 1.
\end{remark}

Let $\cP_{\Gamma}$ denote the set of triples $(X,Y,T)$ of random variables such that $T$ takes values in a finite set $\cT$, $X\in\cX$
and $Y\in\cY$ are conditionally independent given $T$ (written $X-T-Y$), $\E c_1(X)\le\Gamma_1$, and $\E c_2(Y)\le\Gamma_2$. For
$(X,Y,T)\in\cP_{\Gamma}$ and $\theta\in\Theta$, let $Z_\theta$ be the output of $W_\theta$ due to $(X,Y)$, i.e., $P_{XYTZ_\theta}=P_TP_{X|T}P_{Y|T}W_\theta$, and write
\begin{equation}\label{eq:Itheta}
I^{(1)}_\theta\equiv I(X;Z_\theta\mid YT),\qquad I^{(2)}_\theta\equiv I(Y;Z_\theta\mid XT),\qquad I^{(3)}_\theta\equiv I(XY;Z_\theta\mid T).
\end{equation}
These are measurable functions of $\theta$ with values in $[0,\log|\cZ|]$.
Let
\begin{equation}\label{eq:pentagon}
\Lambda_\theta(X,Y,T)\equiv\bigl\{R\in\R^2:R_1\le I^{(1)}_\theta,\ R_2\le I^{(2)}_\theta,\ R_1+R_2\le I^{(3)}_\theta\bigr\}
\end{equation}
be the Ahlswede--Liao pentagon of component $\theta$ for the input $(X,Y,T)$, extended to $\R^2$, and define
\begin{equation}\label{eq:inner-region}
\cI_{\Gamma}(\varepsilon)\equiv\Cl\Bigl(\bigcup_{(X,Y,T)\in\cP_{\Gamma}}\bigl\{R\in\R^2_+:w\{\theta:R\notin\Lambda_\theta(X,Y,T)\}\le\varepsilon\bigr\}\Bigr).
\end{equation}

\begin{theorem}[Inner bound]\label{thm:inner}
For a finite-alphabet mixed memoryless MAC $W$ with additive cost constraint $\Gamma$ satisfying
\eqref{eq:standing}, and every $0\le\varepsilon<1$,
\begin{equation}\label{eq:inner}
C_{\Gamma}(\varepsilon\mid W)\supseteq\cI_{\Gamma}(\varepsilon).
\end{equation}
\end{theorem}

In words, a rate pair is achievable with error probability $\varepsilon$ if, for some single-letter time-sharing input, the $w$-probability
of the components in which the pair falls outside the Ahlswede--Liao pentagon is at most $\varepsilon$. The essential
restriction is that one input $(X,Y,T)$ is used for all components, as it must be when neither encoder knows the
realized component. A component may be ``in outage'' either because it is weak or because the chosen input is
poorly matched to it. Theorem~\ref{thm:inner} holds for an arbitrary mixing measure; in particular, $\Theta$ may be uncountable and
$w$ may have no atoms.

In the proof, the outage event $\{R\notin\Lambda_\theta(X,Y,T)\}$ is the union of the strict inequalities $I^{(1)}_\theta<R_1$, $I^{(2)}_\theta<R_2$ and
$I^{(3)}_\theta<R_1+R_2$; it is used directly.

\subsection{Outline of the proof and three lemmas}\label{sec:outline}

By Theorem~\ref{thm:general}, it suffices to exhibit, for each single-letter triple $(X,Y,T)$, a general input whose spectral outage
probability $J_W(R\mid\bX,\bY)$ is at most the outage probability $w\{\theta:R\notin\Lambda_\theta(X,Y,T)\}$, up to vanishing margins. The natural candidate is the time-sharing i.i.d.\ input built
from $(X,Y,T)$. The difficulty is that the information densities \eqref{eq:densities} are defined through the mixture $W^n$, whereas
the law of large numbers applies to the densities of the realized component $W^n_\theta$. Lemma~\ref{lem:reduction} below bridges the two:
for almost every component, the mixture densities are, up to a vanishing margin, no smaller than the component
densities, and the latter converge to the component mutual informations. Two technical points remain, and both are dealt with once and for all in Lemma~\ref{lem:single}: the i.i.d.\
input meets the cost constraint only with high probability, and Lemma~\ref{lem:trunc} shows that conditioning it on the constraint
costs nothing asymptotically; inputs that meet the cost constraint with equality are handled by a small perturbation
and the continuity of mutual information.

For a general input $(\bX,\bY)$ with values in $\cX^n\times\cY^n$, let $Z^n_\theta$ be the output of the component $W^n_\theta$. We evaluate
the densities $\iota^n_k$ of \eqref{eq:densities}, which are defined through the mixture $W^n$, at $(X^n,Y^n,Z^n_\theta)$ and write $E_n(a\mid\bX,\bY;\theta)$ for
the event \eqref{eq:En} so obtained. Since $Z^n$ is generated by first drawing $\theta\sim w$,
\begin{equation}\label{eq:mixrep}
\Pr\{E_n(a\mid\bX,\bY)\}=\int_\Theta\Pr\{E_n(a\mid\bX,\bY;\theta)\}\,dw(\theta).
\end{equation}
Let $E^\theta_n(a\mid\bX,\bY)$ denote the event \eqref{eq:En} formed with the component densities
\begin{equation}\label{eq:compdens}
\iota^n_{1,\theta}\equiv\frac1n\log\frac{W^n_\theta(Z^n_\theta\mid X^n,Y^n)}{P_{Z^n_\theta|Y^n}(Z^n_\theta\mid Y^n)},\qquad
\iota^n_{2,\theta}\equiv\frac1n\log\frac{W^n_\theta(Z^n_\theta\mid X^n,Y^n)}{P_{Z^n_\theta|X^n}(Z^n_\theta\mid X^n)},\qquad
\iota^n_{3,\theta}\equiv\frac1n\log\frac{W^n_\theta(Z^n_\theta\mid X^n,Y^n)}{P_{Z^n_\theta}(Z^n_\theta)}.
\end{equation}
Let $\Theta^*_n$ be as in Lemma~\ref{lem:mixcomp} with $\cA=\cX\times\cY$, $\cB=\cZ$ and $Q_\theta=W_\theta$, and let $\bone\equiv(1,1,1)$.

\begin{lemma}[Reduction to components]\label{lem:reduction}
For every $\theta\in\Theta^*_n$, every independent $(X^n,Y^n)$, every $a\in\R^3$ and every
$\gamma>0$,
\begin{equation}\label{eq:reduction}
\Pr\{E_n(a\mid\bX,\bY;\theta)\}\le\Pr\bigl\{E^\theta_n\bigl(a+(\gamma+n^{-1/2})\bone\mid\bX,\bY\bigr)\bigr\}+3e^{-n\gamma}.
\end{equation}
\end{lemma}

Lemma~\ref{lem:mixcomp} controls the numerators of the densities and Lemma~\ref{lem:com} their denominators; the details are in Appendix~\ref{app:B}.
Lemma~\ref{lem:reduction} is the instance $c_n=e^{-\sqrt n}$, $G_{\theta,n}=\cX^n\times\cY^n\times\cZ^n$ $(\theta\in\Theta^*_n)$ of the bridge lemma of Part I \cite[Lemma~2]{HY-I};
the margin $n^{-1/2}=\frac1n\log\frac1{c_n}$ is the rate lost by not knowing the component.

The next lemma handles the truncation needed to meet the cost constraint with probability one. It holds for an
arbitrary MAC.

\begin{lemma}[Truncation]\label{lem:trunc}
Let $(X^n,Y^n)$ be independent, let $\cC_1\subseteq\cX^n$ and $\cC_2\subseteq\cY^n$ with $\pi_1\equiv\Pr\{X^n\in\cC_1\}>0$ and
$\pi_2\equiv\Pr\{Y^n\in\cC_2\}>0$, and let $(\bar X^n,\bar Y^n)$ be independent with $\bar X^n$ distributed as $X^n$ conditioned on $\{X^n\in\cC_1\}$
and $\bar Y^n$ as $Y^n$ conditioned on $\{Y^n\in\cC_2\}$. Then, for every $a\in\R^3$,
\begin{equation}\label{eq:trunc}
\Pr\{E_n(a\mid\bar{\bX},\bar{\bY})\}\le\frac{1}{\pi_1\pi_2}\Pr\{E_n(a+\delta_n\mid\bX,\bY)\},\qquad\delta_n\equiv\frac1n\Bigl(\log\frac{1}{\pi_1},\ \log\frac{1}{\pi_2},\ \log\frac{1}{\pi_1\pi_2}\Bigr).
\end{equation}
The same holds if some thresholds $a_k$ equal $-\infty$, the corresponding events being empty.
\end{lemma}

The proof, in Appendix~\ref{app:B}, compares the conditional output distributions of the truncated and untruncated inputs.

The two lemmas above are combined with the law of large numbers in the following lemma, which contains all the exception
handling of the direct parts (cost constraints met with equality, truncation to meet the constraint with probability one)
and is stated for the component densities only, one component at a time. It is used in the proofs of Theorems~\ref{thm:inner},
\ref{thm:csit} and their $K$-user versions.

\begin{lemma}[Single-letter input]\label{lem:single}
Let $(X,Y,T)\in\cP_{\Gamma}$ and $\eta>0$. There is a general input $(\bar{\bX},\bar{\bY})\in\cG_{\Gamma}$, depending only on $(X,Y,T)$ and $\eta$, such that,
for every $\theta\in\Theta$ and $k=1,2,3$, the component densities $\bar\iota^n_{k,\theta}$ of \eqref{eq:compdens} for $(\bar{\bX},\bar{\bY})$ satisfy
\begin{equation}\label{eq:single}
\lim_{n\to\infty}\Pr\{\bar\iota^n_{k,\theta}\le b\}=0\qquad\mbox{whenever }b<I^{(k)}_\theta-\eta,
\end{equation}
where $I^{(k)}_\theta$ are the quantities \eqref{eq:Itheta} for $(X,Y,T)$. The same holds for $K$ users, with the $2^K-1$ component densities
and the quantities $I^{(\cS)}_\theta$ of \eqref{eq:K-pentagon}.
\end{lemma}

\begin{proof}
\emph{Perturbation to strict costs.} Fix $x_0\in\cX$ and $y_0\in\cY$ of minimum cost. For $0<\lambda\le1/4$, let $(X_\lambda,Y_\lambda,T)$ have the
same $P_T$ and
\[
P_{X_\lambda|T=t}\equiv(1-\lambda)P_{X|T=t}+\lambda\bone_{\{x_0\}},\qquad P_{Y_\lambda|T=t}\equiv(1-\lambda)P_{Y|T=t}+\lambda\bone_{\{y_0\}},
\]
with $X_\lambda-T-Y_\lambda$. By \eqref{eq:standing}, $\E c_1(X_\lambda)=(1-\lambda)\E c_1(X)+\lambda c_{1,\min}<\Gamma_1$, and likewise for user 2. For every $\theta$
and $t$, the conditional joint pmfs of $(X,Y,Z_\theta)$ and $(X_\lambda,Y_\lambda,Z_{\theta,\lambda})$ given $T=t$ share the channel $W_\theta$, so that their
variational distance is at most $2\lambda\le1/2$. Since $X-T-Y$, we have $I(X;Z_\theta\mid Y,T=t)=I(X;YZ_\theta\mid T=t)$ and $I(Y;Z_\theta\mid X,T=t)=
I(Y;XZ_\theta\mid T=t)$, and $I(XY;Z_\theta\mid T=t)$ is a mutual information between $(X,Y)$ and $Z_\theta$; in each case the product of the
two alphabet sizes is $N\equiv|\cX||\cY||\cZ|$. Lemma~\ref{lem:cont}, averaged over $t$, gives, for all $\theta$ and $k=1,2,3$,
\begin{equation}\label{eq:perturb}
\bigl|I^{(k)}_\theta-I^{(k)}_{\theta,\lambda}\bigr|\le\tau_N(2\lambda),
\end{equation}
where $I^{(k)}_{\theta,\lambda}$ are the quantities \eqref{eq:Itheta} for $(X_\lambda,Y_\lambda,T)$. Choose $\lambda$ with $\tau_N(2\lambda)\le\eta/2$; the bound does not depend on $\theta$.

\emph{Time-sharing input and law of large numbers.} We may assume $\cT=\{1,\ldots,L\}$ and $P_T(t)>0$. Partition $\{1,\ldots,n\}$ into
consecutive blocks $I_1,\ldots,I_L$ with $|I_t|=\lfloor n\sum_{s\le t}P_T(s)\rfloor-\lfloor n\sum_{s<t}P_T(s)\rfloor$, so that $|I_t|/n\to P_T(t)$, and let $X^n$
and $Y^n$ be independent with independent components $X_i\sim P_{X_\lambda|T=t}$ and $Y_i\sim P_{Y_\lambda|T=t}$ for $i\in I_t$. Fix $\theta$. Since $W^n_\theta$
is memoryless and the $X_i$ are independent and independent of $Y^n$, $P_{Z^n_\theta|Y^n}(\bz\mid\by)=\prod_iP_{Z_{\theta,i}|Y_i}(z_i\mid y_i)$, so that
$\iota^n_{1,\theta}$ is a weighted sum of $L$ block averages of bounded i.i.d.\ random variables with means $I(X_\lambda;Z_{\theta,\lambda}\mid Y_\lambda,T=t)$. By
the weak law of large numbers, $\iota^n_{k,\theta}\to I^{(k)}_{\theta,\lambda}$ in probability for $k=1,2,3$.

\emph{Truncation.} Let $\cC_1\equiv\{\bx:\frac1nc_1(\bx)\le\Gamma_1\}$ and $\cC_2\equiv\{\by:\frac1nc_2(\by)\le\Gamma_2\}$. Since the costs of $(X_\lambda,Y_\lambda,T)$ are strict,
the weak law of large numbers gives $\pi_1=\Pr\{X^n\in\cC_1\}\to1$ and $\pi_2\to1$. Let $(\bar X^n,\bar Y^n)$ be as in Lemma~\ref{lem:trunc} when $\pi_1\pi_2>0$, and
constant sequences of minimum-cost letters otherwise; then $(\bar{\bX},\bar{\bY})\in\cG_{\Gamma}$. Lemma~\ref{lem:trunc}, applied to the single MAC $W^n_\theta$ with
the thresholds for $k'\ne k$ equal to $-\infty$, gives $\Pr\{\bar\iota^n_{k,\theta}\le b\}\le\frac{1}{\pi_1\pi_2}\Pr\{\iota^n_{k,\theta}\le b+\delta_{n,k}\}$, which tends to zero whenever
$b<I^{(k)}_{\theta,\lambda}$, and in particular whenever $b<I^{(k)}_\theta-\eta$ by \eqref{eq:perturb}.

\emph{$K$ users.} Mixing each $P_{X_k|T=t}$ with a minimum-cost letter with weight $\lambda\le1/(4K)$ changes the conditional joint pmf by at most
$K\lambda\le1/4$; since $X_\cS$ and $X_{\cS^c}$ are conditionally independent given $T$, $I(X_\cS;Z_\theta\mid X_{\cS^c},T=t)=I(X_\cS;X_{\cS^c}Z_\theta\mid T=t)$, and
Lemma~\ref{lem:cont} gives \eqref{eq:perturb} with $\tau_{N_K}(K\lambda)$, where $N_K\equiv|\cX_1|\cdots|\cX_K||\cZ|$ (Sec.~\ref{sec:kuser}). The rest is unchanged.
\end{proof}

In engineering terms, Lemma~\ref{lem:single} says that a single i.i.d.\ (time-sharing) codebook distribution, slightly perturbed and conditioned
on the cost constraint, delivers to \emph{every} component the mutual informations of that component up to $\eta$; the
components differ only in which of the three thresholds they can support. No uniformity in $\theta$ is claimed or needed.

\subsection{Proof of Theorem~\ref{thm:inner}}

Since $C_{\Gamma}(\varepsilon\mid W)$ is closed, it suffices to show that every $R\in\R^2_+$ with $w\{\theta:R\notin\Lambda_\theta(X,Y,T)\}\le\varepsilon$ for some
$(X,Y,T)\in\cP_{\Gamma}$ belongs to $C_{\Gamma}(\varepsilon\mid W)$. Fix such $R$ and $(X,Y,T)$, and let $\gamma>0$. Let $(\bar{\bX},\bar{\bY})\in\cG_{\Gamma}$ be the input of
Lemma~\ref{lem:single} with $\eta=\gamma$.

By \eqref{eq:mixrep}, Lemma~\ref{lem:reduction} for $\theta\in\Theta^*_n$ and Lemma~\ref{lem:mixcomp}, for every $a\in\R^3$ and every $n$ with $n^{-1/2}\le\gamma$,
\[
\Pr\{E_n(a\mid\bar{\bX},\bar{\bY})\}\le w(\Theta\setminus\Theta^*_n)+3e^{-n\gamma}+\int_\Theta\Pr\{E^\theta_n(a+2\gamma\bone\mid\bar{\bX},\bar{\bY})\}\,dw(\theta).
\]
Take $a\equiv(R_1-4\gamma,R_2-4\gamma,R_1+R_2-8\gamma)$, so that $a+2\gamma\bone=(R_1-2\gamma,R_2-2\gamma,R_1+R_2-6\gamma)$. By \eqref{eq:single}, $\Pr\{E^\theta_n(a+2\gamma\bone\mid
\bar{\bX},\bar{\bY})\}\to0$ unless $I^{(1)}_\theta\le R_1-\gamma$, $I^{(2)}_\theta\le R_2-\gamma$ or $I^{(3)}_\theta\le R_1+R_2-5\gamma$, and each of these implies $R\notin\Lambda_\theta(X,Y,T)$.
The reverse Fatou lemma therefore gives
\[
J_W(R-4\gamma\mid\bar{\bX},\bar{\bY})\le w\{\theta:R\notin\Lambda_\theta(X,Y,T)\}\le\varepsilon,
\]
that is, $R-4\gamma\in A_\varepsilon(\bar{\bX},\bar{\bY})$; note that $R-4\gamma$ may have negative components, as admitted in \eqref{eq:general}. Letting $\gamma\downarrow0$,
$R\in\Cl_+\bigl(\bigcup_{(\bX,\bY)\in\cG_{\Gamma}}A_\varepsilon(\bX,\bY)\bigr)=C_{\Gamma}(\varepsilon\mid W)$ by Theorem~\ref{thm:general}.\hfill$\Box$

The proof uses the strict outage event $\{R\notin\Lambda_\theta(X,Y,T)\}$ directly: the margins $4\gamma$ and $8\gamma$ absorb the non-strict
thresholds of the information spectrum, so that no auxiliary region with non-strict outage conditions is needed.

%% ====================================================================
\section{Capacity Region of Mixed Memoryless MACs}\label{sec:cap}

For $(X,Y,T)\in\cP_{\Gamma}$, let
\begin{equation}\label{eq:Pi}
\Lambda^*(X,Y,T)\equiv\bigl\{R\in\R^2:R_1\le\essinf I^{(1)}_\theta,\ R_2\le\essinf I^{(2)}_\theta,\ R_1+R_2\le\essinf I^{(3)}_\theta\bigr\}.
\end{equation}

\begin{theorem}[Capacity region]\label{thm:cap}
For a finite-alphabet mixed memoryless MAC $W$ with additive cost constraint $\Gamma$
satisfying \eqref{eq:standing},
\begin{align}
C_{\Gamma}(0\mid W)&=\Cl\Bigl(\bigcup_{(X,Y,T)\in\cP_{\Gamma}}\bigl(\Lambda^*(X,Y,T)\cap\R^2_+\bigr)\Bigr)\label{eq:cap}\\
&=\Cl_+\Bigl(\bigcup_{(X,Y,T)\in\cP_{\Gamma}}\Lambda^*(X,Y,T)\Bigr).\label{eq:cap-clplus}
\end{align}
\end{theorem}

With $\varepsilon=0$ the code must be reliable on almost every component, so the mixed MAC behaves like a compound
MAC over the components, except that $w$-null sets of components are ignored. For single-user compound channels
with CSI at the receiver, the corresponding sup--inf formula, with the infimum over all states, is \cite[Theorem 12]{LC16}.
The capacity region is obtained by choosing one input and taking, for each of the three constraints, the worst
component up to $w$-null sets. In particular, components of zero $w$-measure do not matter, however weak they are.
The order ``first choose the input, then take the essential infimum'' cannot be reversed. The set of rate pairs that lie
in the capacity region of $w$-almost every component is in general strictly larger (Remark~\ref{rem:compare-cap}).

Written out, \eqref{eq:cap} is the closure of the union, over $(X,Y,T)\in\cP_{\Gamma}$, of the pentagons
\[
0\le R_1\le\essinf I(X;Z_\theta\mid YT),\qquad0\le R_2\le\essinf I(Y;Z_\theta\mid XT),\qquad R_1+R_2\le\essinf I(XY;Z_\theta\mid T);
\]
this is the form in which the result is described in \cite[Sec.~II-D]{HY-I}. The set $\Lambda^*(X,Y,T)$ differs from the pentagon $\Lambda_\theta(X,Y,T)$ of \eqref{eq:pentagon} in that the essential
infima over $\theta$ are taken.

The two forms \eqref{eq:cap} and \eqref{eq:cap-clplus}, and the converse below, rest on the following elementary lemma, stated for $K$ users.

\begin{lemma}[Positive part]\label{lem:proj}
Let $\{I_\cS\}_{\emptyset\ne\cS\subseteq[1:K]}$ be nonnegative and nondecreasing in $\cS$, and let $a\in\R^K$ satisfy $a_\cS\le I_\cS$ for all nonempty $\cS$, where
$a_\cS\equiv\sum_{k\in\cS}a_k$. Then $a^+\equiv(\max\{a_k,0\})_{k=1}^K$ satisfies $a^+_\cS\le I_\cS$ for all nonempty $\cS$, and $\max_k|a_k-a^+_k|=\max_k\max\{-a_k,0\}$.
\end{lemma}

\begin{proof}
Let $\mathcal{N}_a\equiv\{k:a_k\ge0\}$. If $\cS\cap\mathcal{N}_a=\emptyset$, then $a^+_\cS=0\le I_\cS$. Otherwise $a^+_\cS=a_{\cS\cap\mathcal{N}_a}\le I_{\cS\cap\mathcal{N}_a}\le I_\cS$. The second claim is obvious.
\end{proof}

For two users, with $I_{\{1\}}=I^*_1$, $I_{\{2\}}=I^*_2$ and $I_{\{1,2\}}=I^*_3$ ($I^*_k\equiv\essinf I^{(k)}_\theta$), the monotonicity holds because $I^{(3)}_\theta=I(Y;Z_\theta\mid T)+I^{(1)}_\theta\ge I^{(1)}_\theta$
and $I^{(3)}_\theta\ge I^{(2)}_\theta$ for every $\theta$ by the chain rule, so that $I^*_3\ge\max\{I^*_1,I^*_2\}$. Hence $a\in\Lambda^*(X,Y,T)$ implies $a^+\in\Lambda^*(X,Y,T)\cap\R^2_+$, and
$a^{(k)+}\to R$ whenever $a^{(k)}\to R\in\R^2_+$; this gives the equality of \eqref{eq:cap} and \eqref{eq:cap-clplus}.

\begin{remark}[Polymatroid structure of $\Lambda^*$]\label{rem:polymatroid}
Unlike the pentagons $\Lambda_\theta(X,Y,T)$ of \eqref{eq:pentagon}, the region $\Lambda^*(X,Y,T)$ defined
by \eqref{eq:Pi} is in general not of polymatroid form: its three right-hand sides need not constitute a polymatroid rank
function. Nevertheless, $\Lambda^*(X,Y,T)\cap\R^2_+$ is a polymatroid, with a truncated rank function. Write
\begin{equation}\label{eq:Istar}
I^*_k\equiv\essinf I^{(k)}_\theta,\qquad k=1,2,3,
\end{equation}
so that $\Lambda^*(X,Y,T)=\{R\in\R^2:R_1\le I^*_1,\ R_2\le I^*_2,\ R_1+R_2\le I^*_3\}$. Since $0\le I^{(k)}_\theta\le\log|\cZ|$ for every $\theta$, each
$I^*_k$ is finite and $0\le I^*_k\le\log|\cZ|$. Define $f:2^{\{1,2\}}\to\R_+$ by
\begin{equation}\label{eq:rankfn}
f(\emptyset)\equiv0,\qquad f(\{1\})\equiv I^*_1,\qquad f(\{2\})\equiv I^*_2,\qquad f(\{1,2\})\equiv\min\{I^*_3,I^*_1+I^*_2\}.
\end{equation}
Then $f$ is a polymatroid rank function, and
\begin{equation}\label{eq:polymatroid}
\Lambda^*(X,Y,T)\cap\R^2_+=\Pi_f\equiv\Bigl\{R\in\R^2_+:\sum_{\alpha\in S}R_\alpha\le f(S)\mbox{ for all }S\subseteq\{1,2\}\Bigr\}.
\end{equation}
Indeed, $f$ is normalized and nonnegative. It is monotone: for every $\theta$, the chain rule gives $I^{(3)}_\theta=I(Y;Z_\theta\mid
T)+I^{(1)}_\theta\ge I^{(1)}_\theta$, hence $I^*_1\le I^*_3$; together with $I^*_2\ge0$ this gives $f(\{1\})\le f(\{1,2\})$, and $f(\{2\})\le f(\{1,2\})$
follows in the same way. It is submodular, since $f(\{1,2\})+f(\emptyset)\le f(\{1\})+f(\{2\})$ by definition. Finally, \eqref{eq:polymatroid}
holds because the constraints $R_1\le I^*_1$ and $R_2\le I^*_2$ already imply $R_1+R_2\le I^*_1+I^*_2$, so replacing $I^*_3$ by
$\min\{I^*_3,I^*_1+I^*_2\}$ does not change the set. In particular, the dominant face of $\Lambda^*(X,Y,T)\cap\R^2_+$ is the segment with
end points $(I^*_1,f(\{1,2\})-I^*_1)$ and $(f(\{1,2\})-I^*_2,I^*_2)$, and the maximal sum rate is $f(\{1,2\})$.

The truncation in $f(\{1,2\})$ cannot be omitted. For each $\theta$, the Markov chain $X-T-Y$ gives $I(Y;Z_\theta\mid T)\le
I(Y;Z_\theta\mid XT)$, hence $I^{(3)}_\theta\le I^{(1)}_\theta+I^{(2)}_\theta$, so that each pentagon $\Lambda_\theta(X,Y,T)\cap\R^2_+$ is a polymatroid with rank
function $(I^{(1)}_\theta,I^{(2)}_\theta,I^{(3)}_\theta)$. The essential infimum, however, is only superadditive, $\essinf(I^{(1)}_\theta+I^{(2)}_\theta)\ge I^*_1+I^*_2$,
and $I^*_3>I^*_1+I^*_2$ may occur. For example, let $\cX=\cY=\cZ=\{0,1\}$, $\Theta=\{\theta_1,\theta_2\}$ with $w(\theta_1)=w(\theta_2)=1/2$,
$W_{\theta_1}$ the channel $Z=X$ and $W_{\theta_2}$ the channel $Z=Y$, let $T$ be constant, and let $X,Y$ be uniform. Then
$(I^{(1)}_\theta,I^{(2)}_\theta,I^{(3)}_\theta)$ equals $(\log2,0,\log2)$ for $\theta=\theta_1$ and $(0,\log2,\log2)$ for $\theta=\theta_2$, so $(I^*_1,I^*_2,I^*_3)=(0,0,\log2)$
and $\Lambda^*(X,Y,T)\cap\R^2_+=\{(0,0)\}$; the triple $(I^*_1,I^*_2,I^*_3)$ itself is not submodular. Equivalently, $\Lambda^*(X,Y,T)\cap\R^2_+$ is
the set of $R\in\R^2_+$ with $R\in\Lambda_\theta(X,Y,T)$ for $w$-almost every $\theta$, an essential intersection of polymatroids; for two
users such an intersection is again a polymatroid, with the rank function \eqref{eq:rankfn}.
\end{remark}

\begin{remark}[Special cases]\label{rem:special}
If $|\Theta|=1$ and there are no cost constraints, \eqref{eq:cap} is the capacity region of the discrete
memoryless MAC established by Ahlswede \cite{Ahl71} and Liao \cite{Liao72}. If $|\cY|=1$, the time-sharing variable can be dropped,
because $\sum_tP_T(t)I(P_{X|T=t},W_\theta)\le I(P_X,W_\theta)$ for every $\theta$ by the concavity of mutual information in the input
distribution. Then \eqref{eq:cap} reduces to the single-user formula
\begin{equation}\label{eq:singleuser-cap}
C_{\Gamma_1}(0\mid W)=\sup_{X:\ \E c_1(X)\le\Gamma_1}\essinf I(X;Z_\theta)
\end{equation}
of Ahlswede \cite[Theorem 1]{Ahl68} and Han \cite[Theorem 3.6.5]{Han03}.
\end{remark}

\begin{remark}[Cardinality of $\cT$]\label{rem:card}
When $\Theta$ is finite, the union in \eqref{eq:cap} may be restricted to $|\cT|\le3|\Theta|+2$. Indeed,
given $(X,Y,T)$, consider for each $t$ the vector in $\R^{3|\Theta|+2}$ formed by the $3|\Theta|$ conditional mutual informations
given $T=t$ and the two conditional expected costs. The $3|\Theta|$ quantities \eqref{eq:Itheta} and the two costs are the $P_T$-averages
of these vectors. The set of such vectors, as $(P_{X|T=t},P_{Y|T=t})$ ranges over a product of simplices, is connected.
By the Fenchel--Eggleston strengthening of Carath\'eodory's theorem, the same averages are therefore obtained with
at most $3|\Theta|+2$ values of $t$. The same reduction applies to Theorem~\ref{thm:inner}. We do not know a bound on $|\cT|$ that is
uniform in $\Theta$.
\end{remark}

\begin{remark}[The case $0<\varepsilon<1$]\label{rem:eps}
For $\varepsilon=0$ the essential-infimum characterization in Theorem~\ref{thm:cap} is exact. For
$0<\varepsilon<1$, a converse would have to control the tail of the distribution of the component mutual informations
under the mixture, not merely its essential infimum. No single-letter outer bound matching Theorem~\ref{thm:inner} is known in
general. Section~\ref{sec:outer} gives a single-letter outer bound (Theorem~\ref{thm:outer}), shows that it is the $\varepsilon$-capacity region with CSI
at the transmitters (Theorem~\ref{thm:csit}) and coincides with Theorem~\ref{thm:inner} for well-ordered mixed MACs (Theorem~\ref{thm:wellordered}), and
that it is not tight in general without CSI (Proposition~\ref{prop:gap}). Theorem~\ref{thm:mixadd} is an instance of Theorem~\ref{thm:wellordered}.
\end{remark}

\subsection{Proof of the direct part}

Let $(X,Y,T)\in\cP_{\Gamma}$ and recall $I^*_k\equiv\essinf I^{(k)}_\theta$ from \eqref{eq:Istar}. By the definition of the essential infimum,
$w\{\theta:I^{(k)}_\theta<I^*_k\}=0$. Hence, if $R\in\Lambda^*(X,Y,T)$, then $R_1\le I^{(1)}_\theta$, $R_2\le I^{(2)}_\theta$ and $R_1+R_2\le I^{(3)}_\theta$ for $w$-almost
every $\theta$, that is, $w\{\theta:R\notin\Lambda_\theta(X,Y,T)\}=0$. Thus $\Lambda^*(X,Y,T)\cap\R^2_+$ is contained in the set in \eqref{eq:inner-region} with $\varepsilon=0$,
and Theorem~\ref{thm:inner} with $\varepsilon=0$ gives ``$\supseteq$'' in \eqref{eq:cap}.\hfill$\Box$

\subsection{Proof of the converse part}

The converse applies Fano's inequality to the given code over each component $W_\theta$ separately. The resulting per-letter
bounds involve the input distributions of the code, which are the same for every $\theta$. The main technical step
is to replace the time index, whose alphabet grows with $n$, by a time-sharing variable with a fixed finite alphabet,
uniformly in $\theta$. This is done by quantizing the per-letter input distributions and using the uniform continuity of
mutual information. A compactness argument then yields a single limiting input, and Fatou's lemma transfers the
vanishing average error probability to almost every component.

Since the right-hand side of \eqref{eq:cap} is closed, it suffices to show that every $0$-achievable $R=(R_1,R_2)$ belongs to
it. Fix $\gamma>0$. There is a sequence of $(n,M_n^{(1)},M_n^{(2)},\varepsilon_n)$ codes satisfying the cost constraint with $\varepsilon_n\to0$ and
$\frac1n\log M_n^{(\alpha)}\ge R_\alpha-\gamma$ for all $n\ge n_0$.

\emph{Step 1: Per-component single-letter bounds.} Let the messages $J$ and $K$ be independent and uniform, and let
$X^n=\bu_J$, $Y^n=\bv_K$. Then $X^n$ and $Y^n$ are independent, and so are $X_i$ and $Y_i$ for each $i$. Let $Z^n_\theta$ be the output
of $W^n_\theta$, and let $\varepsilon_{\theta,n}$ be the error probability of the code over $W^n_\theta$. By \eqref{eq:mixed}, $\varepsilon_n=\int\varepsilon_{\theta,n}\,dw(\theta)$.

By Fano's inequality, $H(J\mid Z^n_\theta,K)\le\log2+\varepsilon_{\theta,n}\log M_n^{(1)}$. Since $Y^n$ is a function of $K$, and $Z^n_\theta$ depends on
$(J,K)$ only through $(X^n,Y^n)$,
\[
I(J;Z^n_\theta\mid K)\le I(X^n;Z^n_\theta\mid K)=H(Z^n_\theta\mid K)-H(Z^n_\theta\mid X^n,Y^n)\le I(X^n;Z^n_\theta\mid Y^n)\le\sum_{i=1}^nI(X_i;Z_{\theta,i}\mid Y_i),
\]
where the last step uses that $W^n_\theta$ is memoryless. Combining this with $\log M_n^{(1)}=H(J\mid K)=I(J;Z^n_\theta\mid
K)+H(J\mid Z^n_\theta,K)$ yields the first of the bounds below. The other two follow in the same way, using $H(J,K\mid
Z^n_\theta)\le\log2+\varepsilon_{\theta,n}\log(M_n^{(1)}M_n^{(2)})$ for the sum rate. For $n\ge n_0$ and all $\theta$,
\begin{align}
(1-\varepsilon_{\theta,n})(R_1-\gamma)&\le\frac1n\sum_{i=1}^nI(X_i;Z_{\theta,i}\mid Y_i)+\frac{\log2}{n},\label{eq:fano1}\\
(1-\varepsilon_{\theta,n})(R_2-\gamma)&\le\frac1n\sum_{i=1}^nI(Y_i;Z_{\theta,i}\mid X_i)+\frac{\log2}{n},\label{eq:fano2}\\
(1-\varepsilon_{\theta,n})(R_1+R_2-2\gamma)&\le\frac1n\sum_{i=1}^nI(X_iY_i;Z_{\theta,i})+\frac{\log2}{n}.\label{eq:fano3}
\end{align}
Since every codeword meets the constraint, $\frac1n\sum_i\E c_1(X_i)\le\Gamma_1$ and $\frac1n\sum_i\E c_2(Y_i)\le\Gamma_2$.

\emph{Step 2: A time-sharing variable with bounded alphabet.} For pmfs $p$ on $\cX$ and $q$ on $\cY$, let $f^{(1)}_\theta(p,q)\equiv I(X;Z_\theta\mid
Y)$, $f^{(2)}_\theta(p,q)\equiv I(Y;Z_\theta\mid X)$ and $f^{(3)}_\theta(p,q)\equiv I(XY;Z_\theta)$, where $X\sim p$ and $Y\sim q$ are independent and $Z_\theta$ is
the output of $W_\theta$. The right-hand sides of \eqref{eq:fano1}--\eqref{eq:fano3} are $\frac1n\sum_if^{(k)}_\theta(p_i,q_i)+\frac{\log2}{n}$ with $p_i\equiv P_{X_i}$ and $q_i\equiv P_{Y_i}$. The
time index plays the role of a time-sharing variable whose alphabet grows with $n$, and it must be replaced by one
with a fixed finite alphabet that does not depend on $\theta$.

Choose $\delta\in(0,1/4]$ with $\tau_N(2\delta)\le\gamma$, where $N=|\cX||\cY||\cZ|$. By compactness, the set of pairs $(p,q)$ can be
partitioned into finitely many nonempty Borel cells $\cK_1,\ldots,\cK_L$ such that $d(p,p')\le\delta$ and $d(q,q')\le\delta$ whenever
$(p,q)$, $(p',q')$ lie in the same cell; $L$ depends only on $\delta$, $|\cX|$ and $|\cY|$. For each $n$, let $S_\ell\equiv\{i:(p_i,q_i)\in\cK_\ell\}$.
Define $(\bar X^n,\bar Y^n,T_n)$ by $P_{T_n}(\ell)\equiv|S_\ell|/n$, and, given $T_n=\ell$, let $\bar X^n$ and $\bar Y^n$ be independent with pmfs $\bar p_\ell\equiv
|S_\ell|^{-1}\sum_{i\in S_\ell}p_i$ and $\bar q_\ell\equiv|S_\ell|^{-1}\sum_{i\in S_\ell}q_i$. For empty $S_\ell$, let $(\bar p_\ell,\bar q_\ell)$ be any point of $\cK_\ell$. Then $(\bar X^n,\bar Y^n,T_n)\in\cP_{\Gamma}$,
since $\E c_1(\bar X^n)=\frac1n\sum_i\E c_1(X_i)\le\Gamma_1$ and likewise for user 2. For $i\in S_\ell$, the convexity of $d$ gives $d(p_i,\bar p_\ell)\le\delta$
and $d(q_i,\bar q_\ell)\le\delta$. As in the proof of Lemma~\ref{lem:single}, Lemma~\ref{lem:cont} then gives $|f^{(k)}_\theta(p_i,q_i)-f^{(k)}_\theta(\bar p_\ell,\bar q_\ell)|\le
\tau_N(2\delta)\le\gamma$ for all $\theta$. Averaging over $i$,
\begin{equation}\label{eq:quantize}
\frac1n\sum_{i=1}^nf^{(k)}_\theta(p_i,q_i)\le I^{(k)}_\theta(\bar X^n,\bar Y^n,T_n)+\gamma\qquad(k=1,2,3,\ \mbox{all }\theta),
\end{equation}
where $I^{(k)}_\theta(\cdot)$ denotes the quantities \eqref{eq:Itheta} for the indicated triple.

\emph{Step 3: Passage to the limit.} The parameters $(P_{T_n},(\bar p_\ell,\bar q_\ell)_{\ell=1}^L)$ range over a compact subset of a Euclidean space
of fixed dimension. Hence a subsequence $n_1<n_2<\cdots$ exists along which they converge to some $(P_T,(\bar p_\ell,\bar q_\ell)_{\ell=1}^L)$,
which defines a triple $(X,Y,T)$ with $X-T-Y$. The costs are linear in these parameters, so $(X,Y,T)\in\cP_{\Gamma}$. For
each fixed $\theta$, $I^{(k)}_\theta(\cdot)$ is continuous in these parameters, so $I^{(k)}_\theta(\bar X^{n_j},\bar Y^{n_j},T_{n_j})\to I^{(k)}_\theta(X,Y,T)$. The subsequence
does not depend on $\theta$.

\emph{Step 4: Fatou's lemma.} By Fatou's lemma, $\int\liminf_j\varepsilon_{\theta,n_j}\,dw(\theta)\le\liminf_j\varepsilon_{n_j}=0$. Hence $\Theta_0\equiv\{\theta:
\liminf_j\varepsilon_{\theta,n_j}=0\}$ has $w(\Theta_0)=1$. For $\theta\in\Theta_0$, pass to a further, $\theta$-dependent subsequence along which $\varepsilon_{\theta,n_j}\to0$.
Along it, \eqref{eq:fano1}--\eqref{eq:quantize} and Step 3 give
\[
R_1-2\gamma\le I^{(1)}_\theta(X,Y,T),\qquad R_2-2\gamma\le I^{(2)}_\theta(X,Y,T),\qquad R_1+R_2-3\gamma\le I^{(3)}_\theta(X,Y,T).
\]
Since these hold for $w$-almost every $\theta$, the point $R-2\gamma$ satisfies the three inequalities in \eqref{eq:Pi} with the essential
infima for $(X,Y,T)$. That is, $R-2\gamma\in\Lambda^*(X,Y,T)$. By Lemma~\ref{lem:proj}, $(R-2\gamma)^+\in\Lambda^*(X,Y,T)\cap\R^2_+$, and since
$R\in\R^2_+$, each component of $(R-2\gamma)^+$ differs from that of $R$ by at most $2\gamma$. The triple $(X,Y,T)$ depends on $\gamma$,
but $\gamma>0$ is arbitrary, so $R$ belongs to the right-hand side of \eqref{eq:cap}.\hfill$\Box$

%% ====================================================================
\section{Outer Bound and CSI at the Transmitters}\label{sec:outer}

Theorem~\ref{thm:inner} gives a single-letter inner bound for every $0\le\varepsilon<1$, and Theorem~\ref{thm:cap} shows that it is tight for $\varepsilon=0$.
For $0<\varepsilon<1$ the essential infimum is no longer the relevant statistic: a code may give up a set of components of
total weight up to $\varepsilon$, and the question is which components it should give up and with which input. This section
compares the inner bound with the natural outage region and determines when they coincide.

Throughout this section, $\cX$, $\cY$ and $\cZ$ are finite and $W$ is the mixed memoryless MAC \eqref{eq:mixed}. We first introduce four
scenarios of channel state information (CSI) in Section~\ref{sec:csi-scen}. We then apply Ahlswede's per-blocklength strong converse \cite{Ahl82}
to each component and obtain a single-letter outer bound that holds with or without CSI at either terminal (Theorem~\ref{thm:outer}).
The outer bound is the $\varepsilon$-capacity region when the encoders know the component and $\Theta$ is countable (Theorem~\ref{thm:csit});
in view of Part I \cite[Theorem~9]{HY-I}, this single-letterizes the multi-letter formula \cite[Eq.~(94)]{HY-I} for these regions. The outer bound coincides with the inner
bound of Theorem~\ref{thm:inner} for well-ordered mixed MACs, in which case all four $\varepsilon$-capacity regions coincide (Theorem~\ref{thm:wellordered}).
Without CSI it is not tight in general (Proposition~\ref{prop:gap}).

The section is organized as follows. Section~\ref{sec:csi-scen} introduces the CSI scenarios. Section~\ref{sec:outage} defines the outage region and restates the inner bound in the same
terms. Section~\ref{sec:ahlswede} extends Ahlswede's strong converse to cost constraints, and Section~\ref{sec:outerbound} derives the outer bound from it.
Section~\ref{sec:csit} shows that the outer bound is achieved with CSI at the transmitters, Section~\ref{sec:wellordered} identifies classes of mixtures
for which it is achieved even without CSI, and Section~\ref{sec:examples} gives examples. Section~\ref{sec:discussion} discusses the remaining gap.

\subsection{CSI scenarios}\label{sec:csi-scen}

The component $\theta$ is drawn once according to $w$ and held fixed for the whole block. We denote by $\tth$ the realized
component, a $\Theta$-valued random variable with law $w$. Besides the setting of Section~\ref{sec:prelim} (\emph{no-CSI}), we consider three
scenarios: in the \textbf{CSIR} scenario the decoder observes $\theta$, so that the decoding regions, denoted $D_{\theta,jk}$, may depend on $\theta$;
in the \textbf{CSIT} scenario the encoders observe $\theta$, so that the codewords, denoted $\bu_{\theta,j}$ and $\bv_{\theta,k}$, may depend on $\theta$, each
satisfying the cost constraint, while the decoder does not observe $\theta$; and in the \textbf{CSIRT} scenario both the encoders and the
decoder observe $\theta$. In every scenario the codebook sizes $M_n^{(1)}$ and $M_n^{(2)}$ do not depend on $\theta$, and all $\theta$-dependent
maps are measurable. Let $\varepsilon_{\theta,n}$ be the average error probability of the code used for $\theta$ over $W^n_\theta$. The average error
probability is then
\begin{equation}\label{eq:csi-avgerr}
\varepsilon_n=\int_\Theta\varepsilon_{\theta,n}\,dw(\theta).
\end{equation}
$\varepsilon$-achievability is defined as in Definition~\ref{def:achievable}. The $\varepsilon$-capacity regions are denoted $C^s_{\Gamma}(\varepsilon\mid W)$ for $s\in\{{\rm no},{\rm rx},{\rm tx},{\rm rt}\}$,
with $C^{\rm no}_{\Gamma}(\varepsilon\mid W)=C_{\Gamma}(\varepsilon\mid W)$. By definition,
\begin{equation}\label{eq:csi-incl}
C^{\rm no}_{\Gamma}\subseteq C^{\rm rx}_{\Gamma}\subseteq C^{\rm rt}_{\Gamma},\qquad C^{\rm no}_{\Gamma}\subseteq C^{\rm tx}_{\Gamma}\subseteq C^{\rm rt}_{\Gamma}.
\end{equation}
These are the scenarios of Part I \cite[Secs.~V and VIII-A]{HY-I}, specialized to finite-alphabet mixed memoryless MACs. For a
countable $\Theta$ and completely general components, Part I \cite[Theorem~9]{HY-I} shows that $C^{\rm no}_{\Gamma}=C^{\rm rx}_{\Gamma}$ and that
$C^{\rm tx}_{\Gamma}=C^{\rm rt}_{\Gamma}$ is given by a multi-letter information spectrum formula; the results of this section are finite-alphabet
single-letter counterparts.

\subsection{Component capacity regions and the outage region}\label{sec:outage}

For $(X,Y,T)\in\cP_{\Gamma}$ and $\theta\in\Theta$, let $\Lambda_\theta(X,Y,T)$ be the pentagon \eqref{eq:pentagon}, and define
\begin{equation}\label{eq:compcap}
\widehat C_{\Gamma}(W_\theta)\equiv\bigcup_{(X,Y,T)\in\cP_{\Gamma}}\Lambda_\theta(X,Y,T),\qquad C_{\Gamma}(W_\theta)\equiv\widehat C_{\Gamma}(W_\theta)\cap\R^2_+.
\end{equation}
By Theorem~\ref{thm:cap} with $|\Theta|=1$ and Remark~\ref{rem:card}, $C_{\Gamma}(W_\theta)$ is the capacity region of the discrete memoryless MAC $W_\theta$
under the cost constraint $\Gamma$, and the union in \eqref{eq:compcap} may be restricted to $|\cT|\le5$. On the restricted parameter set,
which is compact, the mutual informations and the costs are continuous, so $\widehat C_{\Gamma}(W_\theta)$ is closed. It is convex by time
sharing and \emph{down-closed}: if $R\in\widehat C_{\Gamma}(W_\theta)$ and $R'\le R$ componentwise, then $R'\in\widehat C_{\Gamma}(W_\theta)$.

The \emph{outage function} and the \emph{outage region} are
\begin{equation}\label{eq:outage-region}
o_{\Gamma}(R)\equiv w\{\theta:R\notin C_{\Gamma}(W_\theta)\},\qquad\cO_{\Gamma}(\varepsilon)\equiv\{R\in\R^2_+:o_{\Gamma}(R)\le\varepsilon\}.
\end{equation}
These are well defined. For two stochastic matrices $V,V':\cX\times\cY\to\cZ$ with $\max_{x,y}d(V(\cdot\mid x,y),V'(\cdot\mid x,y))\le
\delta\le1/2$, Lemma~\ref{lem:cont} shows that each of the three mutual informations of every $(X,Y,T)$ changes by at most
$\tau_N(\delta)$, with $N=|\cX||\cY||\cZ|$. Hence $\widehat C_{\Gamma}(V)-\tau_N(\delta)\subseteq\widehat C_{\Gamma}(V')$, so the set $\{V:R\in\widehat C_{\Gamma}(V)\}$ is closed, and
$\theta\mapsto\bone\{R\in C_{\Gamma}(W_\theta)\}$ is measurable.

\begin{lemma}\label{lem:lsc}
$o_{\Gamma}$ is lower semicontinuous on $\R^2_+$ and nondecreasing in each coordinate. Hence $\cO_{\Gamma}(\varepsilon)$ is closed and
down-closed in $\R^2_+$.
\end{lemma}

\begin{proof}
Let $R^{(k)}\to R$ in $\R^2_+$. If $R\notin C_{\Gamma}(W_\theta)$, then $R^{(k)}\notin C_{\Gamma}(W_\theta)$ for all large $k$, since $C_{\Gamma}(W_\theta)$ is closed.
Hence $\bone\{R\notin C_{\Gamma}(W_\theta)\}\le\liminf_k\bone\{R^{(k)}\notin C_{\Gamma}(W_\theta)\}$, and Fatou's lemma gives $o_{\Gamma}(R)\le\liminf_k o_{\Gamma}(R^{(k)})$.
Monotonicity follows from down-closedness.
\end{proof}

Recall the inner bound $\cI_{\Gamma}(\varepsilon)$ of \eqref{eq:inner-region}. Comparing \eqref{eq:outage-region} with \eqref{eq:inner-region}: in
$\cO_{\Gamma}(\varepsilon)$ the union over input triples is taken inside the outage event, so each component may use its own triple. In
$\cI_{\Gamma}(\varepsilon)$ it is taken outside, so one triple must serve all components.

\subsection{Ahlswede's strong converse with cost constraints}\label{sec:ahlswede}

This subsection concerns a single discrete memoryless MAC $V:\cX\times\cY\to\cZ$, without mixture. Ahlswede
\cite{Ahl82} showed, without costs, that every $(n,M^{(1)},M^{(2)},\varepsilon_n)$ code with $\varepsilon_n\le\bar\lambda<1$ satisfies $(\log M^{(1)},\log M^{(2)})\in
(n+O(\sqrt n\log n))C(V)$, where $C(V)$ is the capacity region of $V$ without cost constraint. We need it in additive form and with cost constraints,
with $\widehat C_{\Gamma}(V)$ of \eqref{eq:compcap} ($W_\theta$ replaced by $V$) in place of $C(V)$.

\begin{lemma}\label{lem:ahlswede}
Let $V$ be a discrete memoryless MAC and $\bar\lambda\in[0,1)$. There are $n_0$ and $K'$ with the following property.
Let $n\ge n_0$, and consider any $(n,M^{(1)},M^{(2)},\varepsilon_n)$ code for $V$ that satisfies the cost constraint $\Gamma$ and has $\varepsilon_n\le\bar\lambda$.
Then there is $(X,Y,T)\in\cP_{\Gamma}$, with $T$ uniform on $\{1,\ldots,n\}$, such that, with $\kappa_n\equiv K'(\log n)/\sqrt n$ and $Z$ the output
of $V$,
\begin{equation}\label{eq:ahlswede}
\begin{gathered}
\tfrac1n\log M^{(1)}-\kappa_n\le I(X;Z\mid YT),\qquad\tfrac1n\log M^{(2)}-\kappa_n\le I(Y;Z\mid XT),\\
\tfrac1n\log\bigl(M^{(1)}M^{(2)}\bigr)-\kappa_n\le I(XY;Z\mid T).
\end{gathered}
\end{equation}
In particular, $(\frac1n\log M^{(1)},\frac1n\log M^{(2)})-\kappa_n\in\widehat C_{\Gamma}(V)$.
\end{lemma}

The triple in Lemma~\ref{lem:ahlswede} has $|\cT|=n$. This is consistent with \eqref{eq:compcap}, where $\widehat C_{\Gamma}(V)$ is the union over all finite $\cT$; the
restriction to $|\cT|\le5$ in Section~\ref{sec:outage} is used only to show that $\widehat C_{\Gamma}(V)$ is closed. Lemma~\ref{lem:ahlswede} is the case
of two users of the $K$-user bound of Lemma~\ref{lem:K-ahlswede} in Section~\ref{sec:kuser}, with the constant $K'$ of \eqref{eq:K-consts}
for two users; we prove it only once, for general $K$, in Appendix~\ref{app:K}. The proof follows \cite{Ahl82} step by step (expurgation, wringing,
single-letterization, removal of the residual dependence). Only Augustin's bound and the wringing lemma are taken from the literature; the passage
from the wrung subcode to the rates, the removal of the residual dependence and the bookkeeping of the cost constraint are written out. For two users,
Ahlswede's two-variable form of the wringing lemma \cite[Cor.~2]{Ahl82} could be used instead of the form of Fong and Tan used there; this
changes only the constant.

The constants $n_0$ and $K'$ depend only on $\bar\lambda$ and the alphabet sizes. We do not
need this: the proof of Theorem~\ref{thm:outer} uses only the following consequence, applied to each component separately, and
no uniformity over components is required (Remark~\ref{rem:needs}).

\begin{lemma}[Strong converse property of a component]\label{lem:scprop}
Let $R\in\R^2_+$ with $R\notin C_{\Gamma}(V)$. Every sequence of codes
for $V$ that satisfies the cost constraint $\Gamma$ and has $\liminf_n\frac1n\log M_n^{(\alpha)}\ge R_\alpha$ $(\alpha=1,2)$ has $\varepsilon_n\to1$.
\end{lemma}

\begin{proof}
Since $\widehat C_{\Gamma}(V)$ is closed and $R\notin\widehat C_{\Gamma}(V)$, there is $\rho>0$ with $R-2\rho\notin\widehat C_{\Gamma}(V)$. Suppose $\liminf_n\varepsilon_n<\bar\lambda<1$,
and pass to a subsequence with $\varepsilon_n\le\bar\lambda$. For large $n$ in it, $\frac1n(\log M_n^{(1)},\log M_n^{(2)})\ge R-\rho$ componentwise and
$\kappa_n\le\rho$. Lemma~\ref{lem:ahlswede} and down-closedness then give $R-2\rho\in\widehat C_{\Gamma}(V)$, a contradiction. Hence $\liminf_n\varepsilon_n\ge\bar\lambda$ for
every $\bar\lambda<1$.
\end{proof}

\medskip\noindent\textbf{Comment (A new operational meaning of the strong converse).}
Ahlswede's strong converse for the discrete memoryless MAC \cite{Ahl82} and Fong and Tan's strong converse for the Gaussian MAC \cite{FT16}
were both stated as $\varepsilon$-strong converses, i.e., as the statement that the $\varepsilon$-capacity region does not depend on $\varepsilon$. Their proofs,
however, give bounds that hold at every blocklength, and they establish the stronger strong converse property in the sense of
\cite[Definition~2]{HY-I}: every sequence of codes whose rate pair lies outside the capacity region has error probability tending to one.
For Ahlswede's bound this is made explicit by Lemmas~\ref{lem:ahlswede} and \ref{lem:scprop}; for the bound of Fong and Tan, by
\cite[Remark~5]{HY-I}, and an information spectrum proof for $K$ users is given in Part III \cite{HY-III}. The two results play the same key role in
the series. The strong converse of Fong and Tan gives the converse parts of the coding theorems for the quasi-static fading Gaussian MAC
\cite[Theorems~4, 6 and 7]{HY-I}, and Ahlswede's gives the outer bound for finite-alphabet mixed MACs (Theorem~\ref{thm:outer}). In both
cases the mechanism is the same: the error probability of a mixed MAC is the $w$-average of the error probabilities of its components
\cite[Eq.~(28)]{HY-I}; a component in outage has error probability tending to one, as a limit; and Fatou's lemma bounds the limit inferior of
the average from below by the outage probability. The $\varepsilon$-strong converse alone would give only $\limsup_n\varepsilon_{\theta,n}=1$ for such a
component, which does not bound the average (Remark~\ref{rem:needs}, \cite[Remarks~6 and 7]{HY-I}). In this sense the results of Parts I and II
reveal a new operational meaning of the strong converse property of \cite[Definition~2]{HY-I}: it is exactly the property of the components that
makes the outage probability a lower bound on the error probability of the mixed channel, and hence the outage region an outer bound on its
$\varepsilon$-capacity regions, which for the quasi-static fading Gaussian MAC is the $\varepsilon$-capacity region itself.

\subsection{Outer bound}\label{sec:outerbound}

\begin{theorem}[Outer bound]\label{thm:outer}
Let $(\Theta,w)$ be an arbitrary probability space and $\Gamma$ any cost constraint. For every
$0\le\varepsilon<1$ and every $s\in\{{\rm no},{\rm rx},{\rm tx},{\rm rt}\}$,
\begin{equation}\label{eq:outer}
C^s_{\Gamma}(\varepsilon\mid W)\subseteq\cO_{\Gamma}(\varepsilon)=\bigl\{R\in\R^2_+:w\{\theta:R\notin C_{\Gamma}(W_\theta)\}\le\varepsilon\bigr\}.
\end{equation}
\end{theorem}

The idea of the proof is that, once both terminals know $\theta$, the code used on component $\theta$ is an ordinary code
for the discrete memoryless MAC $W_\theta$. By the strong converse, it fails with probability tending to one whenever
the rate pair lies outside the capacity region of $W_\theta$. The total weight of such components is therefore at most $\varepsilon$.
Since CSI can only help, the same bound holds in the other three scenarios.

\begin{proof}
By \eqref{eq:csi-incl} it suffices to treat CSIRT. Let $R$ be $\varepsilon$-achievable under CSIRT. There are codes, one for each $\theta$, with
common sizes $M_n^{(1)}$ and $M_n^{(2)}$ such that $\liminf_n\frac1n\log M_n^{(\alpha)}\ge R_\alpha$ and $\limsup_n\varepsilon_n\le\varepsilon$. Fix $\theta$ with $R\notin C_{\Gamma}(W_\theta)$.
Since the decoder knows $\theta$, the code used for $\theta$ is an ordinary code for the discrete memoryless MAC $W_\theta$, with
error probability $\varepsilon_{\theta,n}$. Lemma~\ref{lem:scprop} gives $\varepsilon_{\theta,n}\to1$. By \eqref{eq:csi-avgerr} and Fatou's lemma,
\begin{equation}\label{eq:outer-fatou}
\varepsilon\ \ge\ \liminf_{n\to\infty}\int_\Theta\varepsilon_{\theta,n}\,dw(\theta)\ \ge\ \int_\Theta\liminf_{n\to\infty}\varepsilon_{\theta,n}\,dw(\theta)\ \ge\ w\{\theta:R\notin C_{\Gamma}(W_\theta)\}=o_{\Gamma}(R).
\end{equation}
Thus every $\varepsilon$-achievable pair lies in $\cO_{\Gamma}(\varepsilon)$. Since $\cO_{\Gamma}(\varepsilon)$ is closed (Lemma~\ref{lem:lsc}), it also contains $C^{\rm rt}_{\Gamma}(\varepsilon\mid W)$.
\end{proof}

In Theorem~\ref{thm:outer}, $C_{\Gamma}(W_\theta)$ denotes the set defined in \eqref{eq:compcap}; the proof uses only this definition. Under \eqref{eq:standing}, or after
the reduction of Remark~\ref{rem:standing}, it is the capacity region of $W_\theta$.

\begin{remark}[What the proof needs]\label{rem:needs}
The proof uses, for each fixed $\theta$, that $\varepsilon_{\theta,n}\to1$ whenever $R\notin C_{\Gamma}(W_\theta)$; no
uniformity in $\theta$ is required, because Fatou's lemma uses pointwise limits only. The weaker property that the $\varepsilon$-capacity
region of $W_\theta$ does not depend on $\varepsilon$ would give only $\limsup_n\varepsilon_{\theta,n}=1$, which does not bound the $\liminf$
of the integral in \eqref{eq:outer-fatou}. This is the finite-alphabet counterpart of \cite[Remark~6]{HY-I} for the Gaussian MAC.
\end{remark}

\begin{remark}[Comparison with Theorem~\ref{thm:cap} at $\varepsilon=0$]\label{rem:compare-cap}
Theorem~\ref{thm:outer} with $\varepsilon=0$ gives $C_{\Gamma}(0\mid W)\subseteq\cO_{\Gamma}(0)=\{R\in
\R^2_+:R\in C_{\Gamma}(W_\theta)\mbox{ for $w$-a.e.\ }\theta\}$. This is weaker than the converse of Theorem~\ref{thm:cap}, which requires one triple
for almost every $\theta$, and it can be strictly weaker (Section~\ref{sec:examples}). The converse of Theorem~\ref{thm:cap} uses the common code:
Fano's inequality is applied to the code's own input distribution, which does not depend on $\theta$, and the time-sharing
variable is fixed before $\theta$ enters, in Steps 1--3 of Section~\ref{sec:cap}. For $\varepsilon>0$ the Fano term $\varepsilon_{\theta,n}\log M_n^{(1)}$ does not
vanish, and replacing Fano's inequality by Lemma~\ref{lem:ahlswede} brings in $\theta$-dependent subcodes. Section~\ref{sec:discussion} discusses why
this obstruction is genuine.
\end{remark}

\subsection{CSI at the transmitters}\label{sec:csit}

\begin{theorem}[CSIT and CSIRT regions]\label{thm:csit}
Let $\Theta$ be countable and let $\Gamma$ satisfy \eqref{eq:standing}. For every $0\le\varepsilon<1$,
\begin{equation}\label{eq:csit-region}
C^{\rm tx}_{\Gamma}(\varepsilon\mid W)=C^{\rm rt}_{\Gamma}(\varepsilon\mid W)=\cO_{\Gamma}(\varepsilon).
\end{equation}
\end{theorem}

For a single user ($|\cY|=1$), Theorem~\ref{thm:csit} gives $C^{\rm tx}_{\Gamma}(\varepsilon\mid W)=C^{\rm rt}_{\Gamma}(\varepsilon\mid W)=\sup\{R:w\{\theta:c_\theta<R\}\le\varepsilon\}$ for
every finite-alphabet mixed memoryless channel with countable $\Theta$, where $c_\theta$ is the capacity--cost function of
$W_\theta$. Corollary~\ref{cor:womc} below shows that, for well-ordered mixed channels, this is also the $\varepsilon$-capacity without CSI.
For $\varepsilon=0$ it equals $\essinf c_\theta$, the mixed-channel counterpart of the worst-case capacity $\inf_sC(s)$ of a compound
channel \cite[(29)]{LC16}, which for discrete memoryless compound channels is the capacity with full CSI at the transmitter \cite{Wol78}.

Theorem~\ref{thm:csit} has a simple operational meaning. If the encoders know $\theta$, each of them can use an input matched
to the realized component, so a rate pair is lost only on components whose capacity region does not contain it.
The decoder does not need to know $\theta$: since $\Theta$ is countable, it can estimate $\theta$ jointly with the messages from a
list of candidate components that grows subexponentially with $n$, at no cost in rate. By Part I \cite[Theorem~9]{HY-I}, the
common region in \eqref{eq:csit-region} is also given by the multi-letter information spectrum formula \cite[Eq.~(94)]{HY-I};
Theorem~\ref{thm:csit} thus identifies that multi-letter region with the single-letter outage region. Remark~\ref{rem:where-countable} indicates
where countability is used.

The direct part rests on the CSIT coding lemma of Part I in component form \cite[Lemma~8]{HY-I}, which is derived there from the CSIT coding
lemma \cite[Lemma~7]{HY-I} and which we restate in the present notation. Let $\Theta$ be countable,
and for each $\theta$ let $X^n_\theta$ and $Y^n_\theta$ be independent inputs that satisfy the cost constraint $\Gamma$ with probability one; let $Z^n_\theta$ be
the output of $W^n_\theta$ due to $(X^n_\theta,Y^n_\theta)$, and let $\iota^n_{k,\theta}$ $(k=1,2,3)$ be the component densities \eqref{eq:compdens} for these inputs.
For a nonempty ${\cA}\subseteq\{1,2\}$ let, as in \cite[Sec.~VIII-D]{HY-I},
\[
\cL_{\cA}\equiv\bigl\{\cS:\emptyset\ne\cS\subseteq\cA\bigr\}\cup\bigl\{\{1,2\}\bigr\};
\]
for ${\cA}=\{1,2\}$ it consists of all three subsets, for ${\cA}=\{1\}$ of $\{1\}$ and $\{1,2\}$, and for ${\cA}=\{2\}$ of $\{2\}$ and $\{1,2\}$.
Identifying $\{1\}$, $\{2\}$ and $\{1,2\}$ with the indices $1$, $2$ and $3$ of \eqref{eq:densities}, let
\begin{equation}\label{eq:ES}
E^{(\theta,{\cA})}_n(a\mid\bX_\theta,\bY_\theta)\equiv\bigcup_{k\in\cL_{\cA}}\{\iota^n_{k,\theta}\le a_k\},
\end{equation}
with $k$ running over the indices of the sets in $\cL_{\cA}$. This is the event formed with the component densities, with the event for
$\{\alpha\}$ omitted when $\alpha\notin {\cA}$ and the event for $\{1,2\}$ always kept. For ${\cA}=\{1,2\}$ it is the event $E^\theta_n(a\mid\bX_\theta,\bY_\theta)$ of Section~\ref{sec:outline}.

The family $\cL_{\cA}$ is needed only to handle rate pairs on the axes. If $R_1,R_2>0$, then $\cA=\{1,2\}$, $\cL_{\cA}$ consists of all three sets, and
\eqref{eq:ES} is the usual event $E^\theta_n$. If, say, $R_2=0$, then $\cA=\{1\}$, and the event for $\{2\}$ is omitted. User~2 has a single message, so
no error corresponds to this event, and it must be omitted, because its threshold $R_2+4\gamma=4\gamma$ in Lemma~\ref{lem:compform} has no rate margin:
the event need not be small even in a component that supports $(R_1,0)$, for example one whose output does not depend on the input of user~2,
in which $\iota^n_{2,\theta}=0$. The event for $\{1,2\}$ is kept, with the positive threshold $R_1+4\gamma$: with CSI at the transmitters even the single
codeword of user~2 depends on $\theta$, so that a wrong component changes the codewords of both users \cite[Sec.~VIII-D]{HY-I}. Rate pairs on the
axes cannot simply be approximated by pairs with both rates positive: if components of the kind just described have total weight more than
$\varepsilon$, then $\cO_{\Gamma}(\varepsilon)$ may contain $(R_1,0)$ but no pair $(R_1,R_2)$ with $R_2>0$.

\begin{lemma}[CSIT coding lemma in component form; the case $K=2$ of {\cite[Lemma~8]{HY-I}}]\label{lem:compform}
In the setup above, let $R\in\R^2_+$, ${\cA}\equiv\{\alpha:R_\alpha>0\}\ne\emptyset$, and $\gamma>0$. Then
there are CSIT codes satisfying the cost constraint $\Gamma$, with $M_n^{(\alpha)}=\lceil e^{nR_\alpha}\rceil$ for $\alpha\in {\cA}$ and $M_n^{(\alpha)}=1$ for $\alpha\notin {\cA}$, such
that
\begin{equation}\label{eq:compform}
\limsup_{n\to\infty}\varepsilon_n\le\limsup_{n\to\infty}\sum_{\theta\in\Theta}w(\theta)\Pr\bigl\{E^{(\theta,{\cA})}_n\bigl(R_1+4\gamma,\ R_2+4\gamma,\ R_1+R_2+4\gamma\bigm|\bX_\theta,\bY_\theta\bigr)\bigr\}.
\end{equation}
\end{lemma}

The proof in Part I is a random-coding argument in which the decoder, which does not know $\theta$, searches jointly over the
message pair and over a finite list of candidate components.

\begin{proof}[Proof of Theorem~\ref{thm:csit}, converse part]
By \eqref{eq:csi-incl}, $C^{\rm tx}_{\Gamma}(\varepsilon\mid W)\subseteq C^{\rm rt}_{\Gamma}(\varepsilon\mid W)$, and by Theorem~\ref{thm:outer}, whose proof
treats the CSIRT scenario directly, $C^{\rm rt}_{\Gamma}(\varepsilon\mid W)\subseteq\cO_{\Gamma}(\varepsilon)$. Neither step uses the countability of $\Theta$.
\end{proof}

\begin{proof}[Proof of Theorem~\ref{thm:csit}, direct part]
Let $R\in\cO_{\Gamma}(\varepsilon)$ with $R\ne0$; the origin is trivially achievable. Fix $\delta\in(0,1)$ and put $R'\equiv(1-\delta)R$. Since
$C^{\rm tx}_{\Gamma}(\varepsilon\mid W)$ is closed and $\delta$ is arbitrary, it suffices to show that $R'$ is $\varepsilon$-achievable under CSIT. Let $\cA\equiv\{\alpha:R_\alpha>0\}$
be the set of active users and $r\equiv\min_{\alpha\in\cA}R_\alpha>0$, so that $R_1+R_2\ge r$.

\emph{Step 1 (Inputs).} Let $\Theta_{\rm good}\equiv\{\theta:R\in C_{\Gamma}(W_\theta)\}$, so that $w(\Theta\setminus\Theta_{\rm good})=o_{\Gamma}(R)\le\varepsilon$. For $\theta\in\Theta_{\rm good}$,
Section~\ref{sec:outage} gives $(X_\theta,Y_\theta,T_\theta)\in\cP_{\Gamma}$ with $R\in\Lambda_\theta(X_\theta,Y_\theta,T_\theta)$; let $(\bar{\bX}_\theta,\bar{\bY}_\theta)$ be the input of Lemma~\ref{lem:single}
for this triple with $\eta\equiv\delta r/4$. For $\theta\notin\Theta_{\rm good}$, let $\bar X^n_\theta$ and $\bar Y^n_\theta$ be constant sequences of minimum-cost letters.

\emph{Step 2 (Thresholds).} Take $\gamma\equiv\delta r/16$. For $\theta\in\Theta_{\rm good}$ and $\alpha\in\cA$, $R_\alpha\le I^{(\alpha)}_\theta$ and $R_\alpha\ge r$ give
\[
R'_\alpha+4\gamma=R_\alpha-\delta R_\alpha+\tfrac14\delta r\le I^{(\alpha)}_\theta-\tfrac34\delta r<I^{(\alpha)}_\theta-\eta,
\]
and in the same way $R'_1+R'_2+4\gamma<I^{(3)}_\theta-\eta$. By \eqref{eq:single}, each of the events in \eqref{eq:ES} for $k\in\cL_{\cA}$, with the thresholds
of \eqref{eq:compform} for $R'$, has probability tending to zero; hence
\begin{equation}\label{eq:csit-good}
\lim_{n\to\infty}\Pr\bigl\{E^{(\theta,\cA)}_n\bigl(R'_1+4\gamma,\ R'_2+4\gamma,\ R'_1+R'_2+4\gamma\bigm|\bar{\bX}_\theta,\bar{\bY}_\theta\bigr)\bigr\}=0\qquad(\theta\in\Theta_{\rm good}).
\end{equation}
Only the events for the active users and for the sum are controlled; for $\alpha\notin\cA$ the threshold $4\gamma$ need not lie below
$I^{(\alpha)}_\theta$, and the event for $k=\alpha$ is not needed, since user $\alpha$ sends a single message.

\emph{Step 3 (Conclusion).} Since $\Theta$ is countable and $\sum_\theta w(\theta)=1$, dominated convergence and \eqref{eq:csit-good} bound the right-hand
side of \eqref{eq:compform}, for $R'$ and the inputs of Step 1, by $w(\Theta\setminus\Theta_{\rm good})\le\varepsilon$. Lemma~\ref{lem:compform} yields CSIT codes at rates $R'$
with $\limsup_n\varepsilon_n\le\varepsilon$.
\end{proof}

\begin{remark}[Where countability enters]\label{rem:where-countable}
Countability of $\Theta$ is used only in Lemma~\ref{lem:compform}: in its proof \cite[Appendix~B]{HY-I}, the decoder searches
over finite subsets $\cD_n\subseteq\Theta$ with $w(\Theta\setminus\cD_n)\to0$, and the passage to component densities uses $W^n\ge w(\theta)W^n_\theta$.
Theorem~\ref{thm:outer} holds for an arbitrary $(\Theta,w)$, and so does Theorem~\ref{thm:wellordered} below. Whether Theorem~\ref{thm:csit} extends to uncountable
$\Theta$ is open. By \cite[Example~2]{HY-I}, the first equality $C^{\rm tx}_{\Gamma}(\varepsilon)=C^{\rm rt}_{\Gamma}(\varepsilon)$ in \cite[Eq.~(94)]{HY-I} can fail for uncountable
$\Theta$, even under mild measurability conditions, but the components there are nonstationary \cite[Remark~20]{HY-I}. For CSI at both terminals, on the
other hand, the second equality in \cite[Eq.~(94)]{HY-I}, i.e., the multi-letter information spectrum formula for $C^{\rm rt}_{\Gamma}(\varepsilon)$, remains valid
for uncountable $\Theta$ under these conditions \cite[Proposition~1]{HY-I}.
\end{remark}

\subsection{Well-ordered mixed MACs: tightness of Theorem~\ref{thm:inner}}\label{sec:wellordered}

The bounds \eqref{eq:inner-region} and \eqref{eq:outer} coincide when, for every rate pair that matters, one triple serves all the components
that can support that pair. For $R\in\R^2_+$ let $S_R\equiv\{\theta:R\in C_{\Gamma}(W_\theta)\}$, and for $0\le\varepsilon<1$ consider:
\begin{itemize}
\item[$(C_\varepsilon)$] For every $R\in\cO_{\Gamma}(\varepsilon)$ there is $(X,Y,T)\in\cP_{\Gamma}$ such that $w\{\theta\in S_R:R\notin\Lambda_\theta(X,Y,T)\}=0$.
\end{itemize}

\begin{definition}\label{def:wellordered}
The mixed MAC $W$ is \emph{well-ordered} (under $\Gamma$) if $(C_\varepsilon)$ holds for every $0\le\varepsilon<1$.
\end{definition}

Definition~\ref{def:wellordered} is related to, but not the same as, the single-user well-ordered condition \eqref{eq:wo-single} below, which is a
simplified form of the condition of \cite[Definition 3]{YHN16}. By Proposition~\ref{prop:singleuser} below, \eqref{eq:wo-single} implies Definition~\ref{def:wellordered} when,
in addition, the minimum of the capacity--cost function over every nonempty $S_R$ is attained, and the converse
implication holds under further assumptions: a finite parameter space, a mixing measure of full support and unique
inputs achieving the capacity--cost functions.

\begin{theorem}\label{thm:wellordered}
Let $(\Theta,w)$ be arbitrary, let $\Gamma$ satisfy \eqref{eq:standing}, and let $0\le\varepsilon<1$. If $(C_\varepsilon)$ holds, then
\begin{equation}\label{eq:wellordered}
\cI_{\Gamma}(\varepsilon)=C^{\rm no}_{\Gamma}(\varepsilon\mid W)=C^{\rm rx}_{\Gamma}(\varepsilon\mid W)=C^{\rm tx}_{\Gamma}(\varepsilon\mid W)=C^{\rm rt}_{\Gamma}(\varepsilon\mid W)=\cO_{\Gamma}(\varepsilon).
\end{equation}
In particular, for a well-ordered mixed MAC the inner bound of Theorem~\ref{thm:inner} is the $\varepsilon$-capacity region for every $\varepsilon$,
and it does not depend on the availability of CSI.
\end{theorem}

\begin{proof}
Let $R\in\cO_{\Gamma}(\varepsilon)$ and take $(X,Y,T)$ from $(C_\varepsilon)$. Up to a $w$-null set, $\{\theta:R\notin\Lambda_\theta(X,Y,T)\}\subseteq\Theta\setminus S_R$, whose
measure is $o_{\Gamma}(R)\le\varepsilon$. By \eqref{eq:inner-region}, $R\in\cI_{\Gamma}(\varepsilon)$. Hence $\cO_{\Gamma}(\varepsilon)\subseteq\cI_{\Gamma}(\varepsilon)\subseteq C^{\rm no}_{\Gamma}(\varepsilon\mid W)$ by Theorem~\ref{thm:inner}. Theorem~\ref{thm:outer} gives
$C^{\rm rt}_{\Gamma}(\varepsilon\mid W)\subseteq\cO_{\Gamma}(\varepsilon)$, and \eqref{eq:csi-incl} closes the chain.
\end{proof}

Theorem~\ref{thm:wellordered} does not need countability, because the direct part is Theorem~\ref{thm:inner}, which holds for an arbitrary mixing
measure.

\begin{remark}[The triple may depend on $R$]\label{rem:triple}
$(C_\varepsilon)$ does not require the components to share an optimal input. For
each $R$ it asks for one triple that covers $R$ on almost all of $S_R$, and the triple may change with $R$. Typically it
achieves $R$ on the weakest component that supports $R$; for stronger components it need not be optimal, which
only loses slack that a fixed-rate code cannot use. A common optimal input is the special case in which the triple
depends neither on $R$ nor on $\theta$.
\end{remark}

$(C_\varepsilon)$ is restrictive: it fails as soon as optimal inputs disagree among components that support the same $R$.
Section~\ref{sec:examples} gives an example. It is also sufficient rather than necessary for Theorem~\ref{thm:wellordered}, as that example shows
for $\varepsilon\ge\min_\theta w(\theta)$. Two classes are well-ordered. The first consists of mixtures with a \emph{common optimal input}: one
$(X^\star,Y^\star,T^\star)\in\cP_{\Gamma}$ with $\Lambda_\theta(X^\star,Y^\star,T^\star)\cap\R^2_+=C_{\Gamma}(W_\theta)$ for $w$-a.e.\ $\theta$, as for the mixed additive MACs of
Section~\ref{sec:additive} (Section~\ref{sec:examples}). The second consists of degradedly ordered mixtures.

The family $\{W_\theta\}$ is \emph{degradedly ordered} if $\Theta$ carries a total order $\preceq$ such that $\theta\preceq\theta'$ implies $W_\theta=U\circ W_{\theta'}$ for
some stochastic matrix $U:\cZ\to\cZ$, that is, $W_\theta(z\mid x,y)=\sum_{z'}U(z\mid z')W_{\theta'}(z'\mid x,y)$.

\begin{proposition}\label{prop:degraded}
Let $\{W_\theta\}$ be degradedly ordered, and assume that every nonempty $S_R$ has a $\preceq$-least element. Then
$W$ is well-ordered.
\end{proposition}

\begin{proof}
Let $\theta\preceq\theta'$ and $(X,Y,T)\in\cP_{\Gamma}$. Generate $Z_\theta$ from $Z_{\theta'}$ through $U$, independently of everything else. Then
$(X,Y,T)-Z_{\theta'}-Z_\theta$, and conditionally on $(Y,T)$ also $X-Z_{\theta'}-Z_\theta$. The data-processing inequality gives $I^{(k)}_\theta\le I^{(k)}_{\theta'}$
for $k=1,2,3$. Hence $\Lambda_\theta(X,Y,T)\subseteq\Lambda_{\theta'}(X,Y,T)$ for every triple, and $S_R$ is an up-set. If $S_R\ne\emptyset$, let $\theta_0$ be
its least element and choose $(X,Y,T)\in\cP_{\Gamma}$ with $R\in\Lambda_{\theta_0}(X,Y,T)$. For every $\theta\in S_R$ we have $\theta_0\preceq\theta$, so
$R\in\Lambda_\theta(X,Y,T)$.
\end{proof}

The least-element assumption holds when $\Theta$ is finite. It also holds when $\Theta$ is a compact interval of $\R$ with its
usual order and $\theta\mapsto W_\theta$ is continuous. In that case $S_R$ is a closed up-set, by the continuity shown in Section~\ref{sec:outage},
and therefore contains its infimum.

The two classes give verifiable hypotheses for Theorem~\ref{thm:wellordered}.

\begin{corollary}\label{cor:verifiable}
Let $(\Theta,w)$ be arbitrary and let $\Gamma$ satisfy \eqref{eq:standing}. The conclusion \eqref{eq:wellordered} holds for every $0\le\varepsilon<1$ in each
of the following cases:
\begin{itemize}
\item[(a)] common optimal input: there is $(X^\star,Y^\star,T^\star)\in\cP_{\Gamma}$ with $\Lambda_\theta(X^\star,Y^\star,T^\star)\cap\R^2_+=C_{\Gamma}(W_\theta)$ for $w$-a.e.\ $\theta$;
\item[(b)] degraded ordering: $\{W_\theta\}$ is degradedly ordered and every nonempty $S_R$ has a $\preceq$-least element.
\end{itemize}
\end{corollary}

\begin{proof}
In case (a), the triple $(X^\star,Y^\star,T^\star)$ satisfies $(C_\varepsilon)$ for every $R$: for $w$-a.e.\ $\theta\in S_R$ we have $R\in C_{\Gamma}(W_\theta)\subseteq
\Lambda_\theta(X^\star,Y^\star,T^\star)$. Case (b) is Proposition~\ref{prop:degraded}. Theorem~\ref{thm:wellordered} completes the proof.
\end{proof}

Corollary~\ref{cor:verifiable} parallels \cite[Propositions 16 and 18]{LC16}. For a compound channel with CSI at the receiver, the worst-case
capacity equals the compound capacity when a saddle-point property holds and the channel is uniform, and
also when the compound channel is degraded; in either case full CSI at the transmitter brings no advantage. Case
(a) plays the role of the saddle-point property and case (b) that of degradedness. Here, however, the error probability
is averaged over $w$ and $\varepsilon>0$ is allowed.

\subsection{Relation to the well-ordered channels of \cite{YHN16}}\label{sec:yhn}
For a single user ($|\cY|=1$), let $c_\theta\equiv\max\{I(P,W_\theta):
\E c_1(X_P)\le\Gamma_1\}$ be the capacity--cost function of $W_\theta$, attained by an input $P_\theta$. Yagi, Han and Nomura \cite[Definition 3]{YHN16}
call the mixture well-ordered if the family $\{W_\theta\}$ is closed and, for every $\theta$ and every input $P$
achieving $c_\theta$, $c_\theta=I(P,W_{\theta'})$ whenever $c_\theta=c_{\theta'}$ and $c_\theta<I(P,W_{\theta'})$ whenever $c_\theta<c_{\theta'}$. We use the following
simplified, non-strict form of this condition:
\begin{equation}\label{eq:wo-single}
c_\theta\le I(P_\theta,W_{\theta'})\qquad\mbox{for all }\theta'\in\Theta\mbox{ with }c_\theta\le c_{\theta'}.
\end{equation}
Every mixture that is well-ordered in the sense of \cite[Definition 3]{YHN16} satisfies \eqref{eq:wo-single}. This class typically includes mixtures
of binary symmetric channels with crossover probability $\theta$ as components \cite{Kie07}, \cite{PPV11}, and mixtures of binary erasure
channels with erasure probability $\theta$ as components; outside the finite-alphabet setting, it also includes single-user
quasi-static fading Gaussian channels with fading coefficient $\theta$. For one user, $S_R=\{\theta:c_\theta\ge R\}$, and $(C_\varepsilon)$ asks
for one input $P$ with $I(P,W_\theta)\ge R$ for $w$-a.e.\ $\theta\in S_R$. In \eqref{eq:wo-single} the input $P_\theta$ also varies with $\theta$.

\begin{proposition}\label{prop:singleuser}
Let $|\cY|=1$.
\begin{enumerate}
\item If one input achieves $c_\theta$ for every $\theta$, then \eqref{eq:wo-single} holds.
\item If \eqref{eq:wo-single} holds and, for every $R$ with $S_R\ne\emptyset$, the minimum of $c_\theta$ over $S_R$ is attained, then $W$ is well-ordered
in the sense of Definition~\ref{def:wellordered}.
\item Suppose $W$ is well-ordered in the sense of Definition~\ref{def:wellordered} and, for every $\theta$, the input $P_\theta$ achieving $c_\theta$ is unique.
Then, for every atom $\theta$ of $w$, $c_\theta\le I(P_\theta,W_{\theta'})$ for $w$-a.e.\ $\theta'$ with $c_\theta\le c_{\theta'}$. In particular, if $\Theta$ is finite, $w$ has
full support and every $P_\theta$ is unique, then Definition~\ref{def:wellordered} is equivalent to \eqref{eq:wo-single}.
\item If $\{W_\theta\}$ is degradedly ordered and $c_\theta$ is strictly increasing along $\preceq$, then \eqref{eq:wo-single} holds.
\item The converses of items 1) and 4) fail in general.
\end{enumerate}
\end{proposition}

The proof is given in Appendix~\ref{app:E}.

Thus, for one user, Definition~\ref{def:wellordered} agrees with \eqref{eq:wo-single} up to $w$-null sets and the choice of capacity-achieving inputs,
under the assumptions of items 2) and 3). Together with Theorem~\ref{thm:wellordered}, item 2) gives the coincidence of the four $\varepsilon$-capacities,
the finite-alphabet counterpart of \cite[Theorem~6]{HY-I}:

\begin{corollary}[Well-ordered mixed channels]\label{cor:womc}
Let $|\cY|=1$, let the cost function be additive with $\Gamma_1>\min_{x\in\cX}c_1(x)$, and let $W$ satisfy \eqref{eq:wo-single} and be such that, for every $R$ with
$S_R=\{\theta:c_\theta\ge R\}\ne\emptyset$, the minimum of $c_\theta$ over $S_R$ is attained. Then, for every $0\le\varepsilon<1$,
\begin{equation}\label{eq:womc}
C^{\rm no}_{\Gamma_1}(\varepsilon\mid W)=C^{\rm rx}_{\Gamma_1}(\varepsilon\mid W)=C^{\rm tx}_{\Gamma_1}(\varepsilon\mid W)=C^{\rm rt}_{\Gamma_1}(\varepsilon\mid W)=\sup\bigl\{R:\Pr\{c_{\tth}\le R\}\le\varepsilon\bigr\}.
\end{equation}
\end{corollary}

\begin{proof}
By Proposition~\ref{prop:singleuser}, item 2), $W$, viewed as a MAC with $|\cY|=1$, is well-ordered in the sense of Definition~\ref{def:wellordered}. Hence
Theorem~\ref{thm:wellordered} applies: the four $\varepsilon$-capacities coincide with the supremum of the single-user outage region $\cO_{\Gamma_1}(\varepsilon)=\{R\ge0:w\{\theta:c_\theta<R\}\le\varepsilon\}$ (\eqref{eq:outage-region} with $|\cY|=1$). This
supremum equals the right-hand side of \eqref{eq:womc}: since $\{c_\theta<R\}\subseteq\{c_\theta\le R\}$, it is not smaller, and if $w\{c_\theta<R\}\le\varepsilon$, then
$w\{c_\theta\le R'\}\le\varepsilon$ for every $R'<R$.
\end{proof}

As in \cite[Theorem~6]{HY-I}, the point of Corollary~\ref{cor:womc} is the inequality $C^{\rm no}_{\Gamma}\ge C^{\rm rt}_{\Gamma}$: the direct part uses one input
that achieves $R$ on the weakest component supporting $R$, which exists by the attainment condition, and the converse
part uses the strong converse property of each component (Lemma~\ref{lem:scprop}). For mixed channels that are well-ordered in the
original sense of \cite[Definition 3]{YHN16}, Yagi et al.~\cite[Theorem 3]{YHN16} showed that the $\varepsilon$-capacity without CSI reduces to
\begin{equation}\label{eq:yhn}
C^{\rm no}_{\Gamma_1}(\varepsilon\mid W)=\sup\Bigl\{R\Bigm|\int_{\{\theta\,|\,c_\theta\le R\}}dw(\theta)\le\varepsilon\Bigr\}=\sup\bigl\{R:\Pr\{c_{\tth}\le R\}\le\varepsilon\bigr\};
\end{equation}
Corollary~\ref{cor:womc} extends this to the condition \eqref{eq:wo-single} under the attainment condition, and to all four CSI scenarios. The
right-hand side of \eqref{eq:womc} is the $\varepsilon$-outage capacity for finite alphabets; it corresponds to \cite[Corollary~2]{HY-I} for Gaussian mixed
channels. Without the attainment condition, \eqref{eq:wo-single} alone is not known to imply $(C_\varepsilon)$. For the MAC, \eqref{eq:wo-single} has no
direct analogue, because a component has a family of optimal triples, one for each boundary point of $C_{\Gamma}(W_\theta)$,
instead of a single optimal input. Definition~\ref{def:wellordered} replaces ``the optimal input of the weaker component'' by ``a triple
achieving $R$ on the weakest component that supports $R$''.

\subsection{Examples}\label{sec:examples}

\paragraph{Mixed additive MACs}
Consider the mixed additive MAC of Section~\ref{sec:additive}. For every $\theta$ and every independent
input, $H(Z_\theta\mid Y)$, $H(Z_\theta\mid X)$ and $H(Z_\theta)$ are at most $\log m$, while $H(Z_\theta\mid X,Y)=H(V_\theta)$. Uniform inputs attain
all three bounds at once, so
\begin{equation}\label{eq:additive-comp}
C_{\Gamma}(W_\theta)=\bigl\{R\in\R^2_+:R_1+R_2\le\log m-H(V_\theta)\bigr\},
\end{equation}
and the uniform pair is a common optimal input. Formally, there is no cost constraint, i.e., $c_\alpha\equiv0$ with any $\Gamma_\alpha>0$, so that
\eqref{eq:standing} holds. By Corollary~\ref{cor:verifiable}(a), the region \eqref{eq:mixadd} of Theorem~\ref{thm:mixadd} is the $\varepsilon$-capacity region in all four CSI scenarios,
for an arbitrary mixing measure.

\paragraph{Degradedly ordered families and the $\varepsilon$-quantile}
Let $W_0:\cX\times\cY\to\{0,1\}$ be a fixed MAC and $W_\theta\equiv
{\rm BSC}(\theta)\circ W_0$ for $\theta\in\Theta=[0,1/2]$, with an arbitrary $w$ and a cost constraint satisfying \eqref{eq:standing}. For $\theta\le\theta'$, ${\rm BSC}(\theta')$
is a degraded version of ${\rm BSC}(\theta)$, so the family is degradedly ordered by the reversed order. Since $\theta\mapsto W_\theta$ is
continuous and $\Theta$ is compact, Corollary~\ref{cor:verifiable}(b) applies. The regions are nested, $C_{\Gamma}(W_{\theta'})\subseteq C_{\Gamma}(W_\theta)$ for $\theta\le\theta'$.
Hence $S_R=[0,\theta^*(R)]$ for some $\theta^*(R)$ and $o_{\Gamma}(R)=w((\theta^*(R),1/2])$. With the upper $\varepsilon$-quantile
\[
q_\varepsilon\equiv\inf\bigl\{t\in[0,\tfrac12]:w\bigl((t,\tfrac12]\bigr)\le\varepsilon\bigr\},
\]
right-continuity of $t\mapsto w((t,1/2])$ shows that $o_{\Gamma}(R)\le\varepsilon$ if and only if $\theta^*(R)\ge q_\varepsilon$, that is, $R\in C_{\Gamma}(W_{q_\varepsilon})$.
Therefore
\begin{equation}\label{eq:quantile}
C^s_{\Gamma}(\varepsilon\mid W)=C\bigl(W_{q_\varepsilon}\bigr)\qquad(s\in\{{\rm no},{\rm rx},{\rm tx},{\rm rt}\}).
\end{equation}
The $\varepsilon$-capacity region is the capacity region of a single component, the one at the $\varepsilon$-quantile of the noise level.
This is the MAC counterpart of the $\varepsilon$-outage capacity of well-ordered single-user mixed channels \eqref{eq:womc}, \eqref{eq:yhn}, \cite{YHN16}.
The same holds with ${\rm BSC}(\theta)$ replaced by any continuous chain of degrading channels, such as ${\rm BEC}(\theta)$.

\paragraph{A gap without CSI}
The following shows that Theorem~\ref{thm:outer} is not tight in general without CSI.

\begin{figure}[t]
\centering
\includegraphics[width=0.36\textwidth]{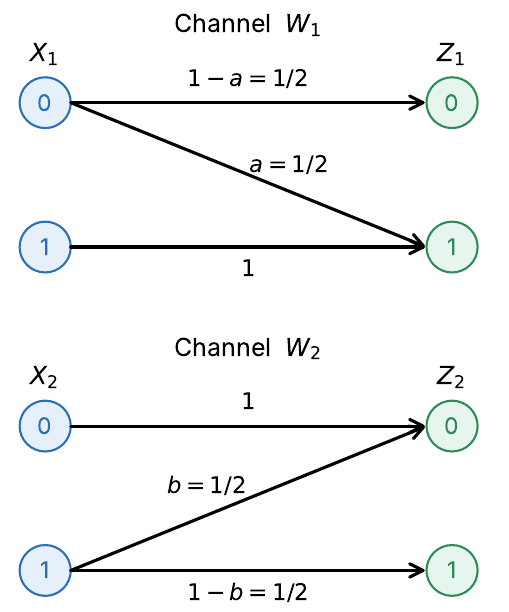}
\caption{Bipartite transition diagrams of the two component Z-channels $W_1$ and $W_2$ of Proposition~\ref{prop:gap} (the example of \cite[Example~1]{HY-I}).
The channel $W_2$ is obtained from $W_1$ by interchanging the labels $0$ and $1$ both at the input and at the output.}
\label{fig:zchannels}
\end{figure}

\begin{figure}[t]
\centering
\includegraphics[width=0.62\textwidth]{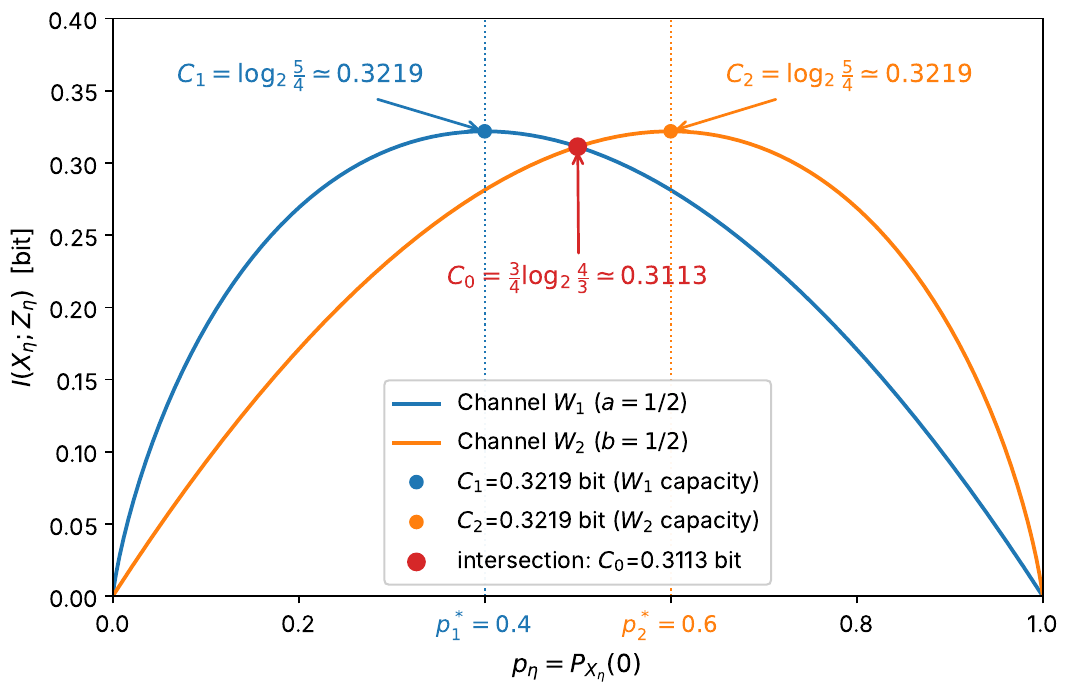}
\caption{Mutual information $I(P,W_\theta)$ (in bits) as a function of $p=P(0)$ for the components $W_1$ (blue, $\theta=1$) and $W_2$ (orange, $\theta=2$)
of Fig.~\ref{fig:zchannels}. The capacities are $c_1=c_2=C=\log_2\frac54\simeq0.3219$, attained at $p=2/5$ and $p=3/5$, respectively. The two curves are
mirror images of each other about $p=1/2$, where they intersect at the height $C_0=\frac34\log_2\frac43\simeq0.3113$.}
\label{fig:mi}
\end{figure}

\begin{proposition}\label{prop:gap}
There is a finite-alphabet mixed memoryless MAC with $|\Theta|=2$ and an interval of $\varepsilon$ on which
\[
C^{\rm no}_{\Gamma}(\varepsilon\mid W)\subsetneq\cO_{\Gamma}(\varepsilon)=C^{\rm tx}_{\Gamma}(\varepsilon\mid W)=C^{\rm rt}_{\Gamma}(\varepsilon\mid W).
\]
\end{proposition}

\begin{proof}
Take the single-user example of \cite[Example~1]{HY-I}, viewed as a MAC with $|\cY|=1$ and no costs. Here $\Theta=\{1,2\}$,
$W_1$ is the Z-channel with $W_1(1\mid0)=\frac12$, $W_2$ is its mirror image, $W_2(z\mid x)=W_1(1-z\mid1-x)$, so that $W_2(0\mid1)=\frac12$, and the
weights $w(1),w(2)>0$ are arbitrary (Fig.~\ref{fig:zchannels}). With $p=P(0)$, $I(P,W_1)=h(\frac p2)-p\log2$ and $I(P,W_2)$ is obtained by $p\to1-p$.
The capacities are $c_1=c_2=C\equiv\log\frac54\approx0.2231$ nat ($0.3219$ bit), attained at the different inputs $p=\frac25$ and
$p=\frac35$, respectively. Since $I(P,W_1)$ is decreasing on $[\frac25,1]$, $I(P,W_2)$ is increasing on $[0,\frac35]$ and the two coincide at $p=\frac12$,
$\max_P\min_\theta I(P,W_\theta)=C_0\equiv\frac34\log\frac43\approx0.2158$ nat ($0.3113$ bit), attained at $p=\frac12$ (Fig.~\ref{fig:mi}). By the first-order $\varepsilon$-capacity formula for mixed
memoryless channels \cite[Theorem 1]{YHN16}, $C^{\rm no}_{\Gamma}(\varepsilon)=\sup_P\sup\{R:w\{\theta:I(P,W_\theta)\le R\}\le\varepsilon\}$ (see also \cite[Eqs.~(111)--(112)]{HY-I}).
For $\varepsilon<\min_\theta w(\theta)$, neither component may be in outage, so $C^{\rm no}_{\Gamma}(\varepsilon)=\max_P\min_\theta I(P,W_\theta)=C_0$.
The outage function is $o_{\Gamma}(R)=w\{\theta:c_\theta<R\}$, which vanishes for $R\le C$ and equals one for $R>C$, so $\cO_{\Gamma}(\varepsilon)=[0,C]$ for every $0\le\varepsilon<1$.
Theorem~\ref{thm:csit} identifies $\cO_{\Gamma}(\varepsilon)$ with $C^{\rm tx}_{\Gamma}$ and $C^{\rm rt}_{\Gamma}$.
\end{proof}

\begin{table}[t]
\centering
\caption{The $\varepsilon$-capacities of the Z-channel example in Proposition~\ref{prop:gap}.}
\label{tab:zchannel}
\small
\begin{tabular}{lll}
\toprule
$\varepsilon$ & $C^{\rm no}_{\Gamma}(\varepsilon)$ & $\cO_{\Gamma}(\varepsilon)=C^{\rm tx}_{\Gamma}(\varepsilon)=C^{\rm rt}_{\Gamma}(\varepsilon)$\\
\midrule
$0\le\varepsilon<\min_\theta w(\theta)$ & $C_0=\frac34\log\frac43\approx0.2158$ nat ($0.3113$ bit) & $C=\log\frac54\approx0.2231$ nat ($0.3219$ bit)\\
$\min_\theta w(\theta)\le\varepsilon<1$ & $C=\log\frac54\approx0.2231$ nat ($0.3219$ bit) & $C=\log\frac54\approx0.2231$ nat ($0.3219$ bit)\\
\bottomrule
\end{tabular}
\end{table}

In this example $(C_\varepsilon)$ fails for every $\varepsilon$: every $R\in(C_0,C]$ lies in both component regions, which are in fact equal, but
$\max_P\min_\theta I(P,W_\theta)=C_0<R$. Since $c_1=c_2$, the outage region $\cO_{\Gamma}(\varepsilon)$ involves no outage at all and does not depend on $\varepsilon$; the gap is
thus caused solely by the mismatch of the optimal inputs, not by a weaker component. For $\varepsilon\ge\min_\theta w(\theta)$, nevertheless $C^{\rm no}_{\Gamma}=\cO_{\Gamma}$, since a
code without CSI may give up the component whose optimal input it does not use; this is the sense in which $(C_\varepsilon)$ is sufficient but not
necessary. The equality $C^{\rm tx}_{\Gamma}=C^{\rm rt}_{\Gamma}=\cO_{\Gamma}$ in this example is also obtained directly in \cite[Example~1]{HY-I}.

\subsection{Discussion: why a per-component strong converse cannot close the gap}\label{sec:discussion}

In the proof of Theorem~\ref{thm:outer}, Lemma~\ref{lem:ahlswede} is applied to each $W_\theta$ separately, and inside its proof (Appendix~\ref{app:K} with
two users) every reduction depends on $\theta$: the expurgated set $\cE$ of Step 1, the wringing coordinates of Step 2, and hence the triple of Step 5.
The argument thus certifies $R\in C_{\Gamma}(W_\theta)$ with a $\theta$-dependent input, which is exactly $\cO_{\Gamma}(\varepsilon)$. Proposition~\ref{prop:gap} shows
that information about the common code is lost at this point.

Part of the reduction can be made independent of $\theta$. Expurgating with the mixture error $e(j,k)\equiv\int W^n_\theta(D^c_{jk}\mid
\bu_j,\bv_k)\,dw(\theta)$ gives a set $\cE$ that does not depend on $\theta$, and the wringing step (Step 2 of Appendix~\ref{app:K}) uses only $|\cE|$. The obstacle
is the next step. Augustin's bound \cite[Cor.~1]{Ahl82}, \cite{Aug66} needs, for each $\theta$, a subcode whose maximal error over $W_\theta$
is below one. Extracting it is a second, $\theta$-dependent expurgation, and removing a constant fraction of pairs can
change the Fano distribution substantially.

The following heuristic suggests that for one user the obstacle disappears; we do not make it rigorous here.
Every code contains a constant-composition subcode, of polynomially small relative size, whose codewords each
have mixture error at most $\varepsilon+\eta$ for a given $\eta>0$. Every $\theta$-dependent subset of it has the same type, so the
strong converse for constant-composition codes is applied to each $W_\theta$ with a common input; the conclusion would
be consistent with \cite{YHN16}. For the MAC, the joint type of $(\bu_j,\bv_k)$ varies over pairs, and removing its correlation
is precisely what wringing does \cite{Dueck81}, \cite{Ahl82}. A converse for $0<\varepsilon<1$ without CSI would need a $\theta$-independent
decomposition into joint-type classes that is compatible with wringing. We leave this open.

%% ====================================================================
%% ====================================================================
\section{Extension to $K$ Users}\label{sec:kuser}

The two-user restriction of Sections~\ref{sec:additive}--\ref{sec:outer} is essential in only one place: the outer bound rests on Ahlswede's
strong converse (Lemmas~\ref{lem:ahlswede} and \ref{lem:scprop}), which was proved for two users. In this section we extend that bound to $K$
users with cost constraints (Lemma~\ref{lem:K-ahlswede}), and then indicate how the results of the preceding sections extend. Two points do
not carry over literally (Remark~\ref{rem:K-polymatroid}), and the result for CSI at the transmitters uses a $K$-user
version of a lemma of Part I (Lemma~\ref{lem:K-compform}), which is proved there for $K$ users \cite[Lemma~8 and Appendix~B]{HY-I}.

The picture is that of two users. Each nonempty set $\cS$ of users whose messages may be decoded wrongly contributes one constraint
$R_\cS\le I^{(\cS)}_\theta$ to the region of a component, so that the pentagon becomes the polymatroid of \eqref{eq:K-pentagon} below, and a rate tuple
is in outage when the realized component violates the constraint of some $\cS$. In the proofs, the three events of two users become $2^K-1$
events, one for each $\cS$, and they are handled one at a time. The only place where the sets must be treated together is Ahlswede's bound
(Lemma~\ref{lem:K-ahlswede}), in which all $2^K-1$ inequalities must hold for a common input.

\subsection{Setting}

Let $K\ge2$, let $\cX_1,\ldots,\cX_K$ and $\cZ$ be finite, and write $\cX_{[1:K]}\equiv\cX_1\times\cdots\times\cX_K$, $L\equiv|\cX_{[1:K]}|$ and $N_K\equiv L|\cZ|$. For a subset
$\cS\subseteq[1:K]\equiv\{1,\ldots,K\}$ let $\cS^c\equiv[1:K]\setminus\cS$, $x_\cS\equiv(x_k)_{k\in\cS}$, $R_\cS\equiv\sum_{k\in\cS}R_k$ and $M_\cS\equiv\prod_{k\in\cS}M_k$. The
mixed memoryless $K$-user MAC is defined as in \eqref{eq:mixed}, with components $W_\theta:\cX_{[1:K]}\to\cZ$, and the codes, the average error probability,
$\varepsilon$-achievability and the four CSI scenarios are defined as in Sections~\ref{sec:prelim} and \ref{sec:csi-scen}, with $K$ codebooks and decoding regions
indexed by message tuples. The cost functions are additive, $c_k(\bx)=\sum_tc_k(x_t)$, the cost constraint is $\bGamma=(\Gamma_1,\ldots,\Gamma_K)$, and we
assume, as in \eqref{eq:standing}, that $\Gamma_k>\min_{x\in\cX_k}c_k(x)$ for every $k$. The $\varepsilon$-capacity regions are denoted $C^s_{\bGamma}(\varepsilon\mid W)$.

Let $\cP_{\bGamma}$ be the set of tuples $(X_1,\ldots,X_K,T)$ with $T$ finite-valued, $X_1,\ldots,X_K$ conditionally independent given $T$, and
$\E c_k(X_k)\le\Gamma_k$; for $K=2$ it is the set $\cP_{\Gamma}$ of Section~\ref{sec:inner}. For such a tuple and $\theta\in\Theta$, let $Z_\theta$ be the output of $W_\theta$, $I^{(\cS)}_\theta\equiv I(X_\cS;Z_\theta\mid X_{\cS^c}T)$ for nonempty $\cS$, and
\begin{equation}\label{eq:K-pentagon}
\Lambda_\theta(X_{[1:K]},T)\equiv\bigl\{R\in\R^K:R_\cS\le I^{(\cS)}_\theta\mbox{ for all nonempty }\cS\subseteq[1:K]\bigr\},
\end{equation}
the $K$-user counterpart of \eqref{eq:pentagon}. The inner bound and the outage region are, as in \eqref{eq:inner-region} and \eqref{eq:outage-region},
\begin{gather}
\cI_{\bGamma}(\varepsilon)\equiv\Cl\Bigl(\bigcup_{(X_{[1:K]},T)\in\cP_{\bGamma}}\bigl\{R\in\R^K_+:w\{\theta:R\notin\Lambda_\theta(X_{[1:K]},T)\}\le\varepsilon\bigr\}\Bigr),\label{eq:K-inner}\\
\cO_{\bGamma}(\varepsilon)\equiv\bigl\{R\in\R^K_+:w\{\theta:R\notin C_{\bGamma}(W_\theta)\}\le\varepsilon\bigr\},\qquad
C_{\bGamma}(W_\theta)\equiv\Bigl(\bigcup_{(X_{[1:K]},T)\in\cP_{\bGamma}}\Lambda_\theta(X_{[1:K]},T)\Bigr)\cap\R^K_+.\label{eq:K-outage}
\end{gather}
The auxiliary quantities $\widehat C_{\bGamma}(W_\theta)$ and $o_{\bGamma}(R)$ are defined by \eqref{eq:compcap} and \eqref{eq:outage-region} with $\cP_{\Gamma}$ replaced by $\cP_{\bGamma}$
and $\Lambda_\theta$ by \eqref{eq:K-pentagon}, so that $\cO_{\bGamma}(\varepsilon)=\{R\in\R^K_+:o_{\bGamma}(R)\le\varepsilon\}$. The chain rule gives, for $\cS\subseteq\cS'$,
\begin{equation}\label{eq:K-monotone}
I^{(\cS')}_\theta=I(X_{\cS'\setminus\cS};Z_\theta\mid X_{\cS'^c}T)+I^{(\cS)}_\theta\ge I^{(\cS)}_\theta,
\end{equation}
so that $\cS\mapsto I^{(\cS)}_\theta$ is nondecreasing; since $X_1,\ldots,X_K$ are conditionally independent given $T$, it is also submodular, and
$\Lambda_\theta(X_{[1:K]},T)\cap\R^K_+$ is a polymatroid.

\subsection{Ahlswede's strong converse for $K$ users}

\begin{lemma}[$K$-user Ahlswede-type bound with cost constraints]\label{lem:K-ahlswede}
Let $V:\cX_{[1:K]}\to\cZ$ be a discrete memoryless $K$-user MAC and $\bar\lambda\in[0,1)$. Put $\beta\equiv2\bar\lambda/(1-\bar\lambda)$,
\begin{equation}\label{eq:K-consts}
n_0\equiv\max\bigl\{3,\ 16(K+1)^2,\ L^2,\ \lfloor\beta^{-2}\rfloor+1\bigr\},\qquad
K'\equiv\frac{6L}{1-\bar\lambda}+\frac{2\bar\lambda}{1-\bar\lambda}+\log\frac{1+\bar\lambda}{1-\bar\lambda}+8(K+1)\bigl(\log N_K+2\bigr).
\end{equation}
Let $n\ge n_0$, and consider any $(n,M_1,\ldots,M_K,\varepsilon_n)$ code for $V$ that satisfies the cost constraint $\bGamma$ and has $\varepsilon_n\le\bar\lambda$. Then there is
$(X_1,\ldots,X_K,T)\in\cP_{\bGamma}$, with $T$ uniform on $\{1,\ldots,n\}$, such that, with $\kappa_n\equiv K'(\log n)/\sqrt n$ and $Z$ the output of $V$,
\begin{equation}\label{eq:K-ahlswede}
\frac1n\log M_\cS-\kappa_n\le I(X_\cS;Z\mid X_{\cS^c}T)\qquad\mbox{for every nonempty }\cS\subseteq[1:K].
\end{equation}
In particular, $(\frac1n\log M_1,\ldots,\frac1n\log M_K)-\kappa_n\in\widehat C_{\bGamma}(V)$.
\end{lemma}

(For $\bar\lambda=0$, the term $\lfloor\beta^{-2}\rfloor+1$ is read with $\bar\lambda$ replaced by any positive value; see Step~0 of Appendix~\ref{app:K}.) The proof,
given in Appendix~\ref{app:K}, follows Ahlswede \cite{Ahl82} step by step, except for the form of the wringing lemma: instead of Ahlswede's
two-variable corollary \cite[Cor.~2]{Ahl82}, we follow Fong and Tan \cite[Lemma~3]{FT16}, who apply the wringing lemma to the product of the
users' (quantized) alphabets with the product of the users' codebook laws as reference measure. Here this is done without quantization and once
for all subsets of users simultaneously, because all $2^K-1$ inequalities must hold for a common input; the residual dependence is then
controlled by a total-variation argument and the continuity of entropy. For $K=2$, Lemma~\ref{lem:K-ahlswede} is
Lemma~\ref{lem:ahlswede}.

As in Section~\ref{sec:outage}, the union defining $\widehat C_{\bGamma}(V)$ may be restricted to $|\cT|\le2^K-1+K$ by the Fenchel--Eggleston theorem (the
$2^K-1$ conditional mutual informations and the $K$ conditional costs given $T=t$ range over a connected set), so that $\widehat C_{\bGamma}(V)$ is closed;
it is also down-closed. Exactly as Lemma~\ref{lem:scprop} follows from Lemma~\ref{lem:ahlswede}, we obtain:

\begin{corollary}[Strong converse property, $K$ users]\label{cor:K-scprop}
Let $R\in\R^K_+$ with $R\notin C_{\bGamma}(V)$. Every sequence of codes for $V$ that satisfies the cost constraint $\bGamma$ and has
$\liminf_n\frac1n\log M_{k,n}\ge R_k$ $(k=1,\ldots,K)$ has $\varepsilon_n\to1$.
\end{corollary}

\subsection{Positive part and polymatroids}

Lemma~\ref{lem:proj} is stated for $K$ users; by \eqref{eq:K-monotone}, both $I^{(\cS)}_\theta$ and its essential infimum $I^*_\cS\equiv\essinf I^{(\cS)}_\theta$
are nondecreasing in $\cS$, so that it applies verbatim in the equivalence of \eqref{eq:cap} and \eqref{eq:cap-clplus} and in Step~4 of the converse part
of Theorem~\ref{thm:cap}.

\begin{remark}[Remark~\ref{rem:polymatroid} does not extend]\label{rem:K-polymatroid}
For two users, $\Lambda^*(X,Y,T)\cap\R^2_+$ is a polymatroid (Remark~\ref{rem:polymatroid}). For $K\ge3$, the region
$\{R\in\R^K_+:R_\cS\le I^*_\cS\ \forall\cS\}$, an essential intersection of polymatroids, need not be a polymatroid. For example, let $K=3$, $\cX_k=\{0,1\}$,
$\Theta=\{\theta_1,\theta_2\}$ with $w(\theta_1),w(\theta_2)>0$, let $W_{\theta_1}$ be the deterministic channel $Z=(X_1\oplus X_2,X_3)$ and $W_{\theta_2}$ the deterministic
channel $Z=(X_1,X_2\oplus X_3)$, let $T$ be constant and the inputs uniform. In units of $\log2$, $(I^*_{\{1\}},I^*_{\{2\}},I^*_{\{3\}})=(1,1,1)$,
$I^*_{\{1,2\}}=I^*_{\{2,3\}}=1$ and $I^*_{\{1,3\}}=I^*_{\{1,2,3\}}=2$. The region then has the maximal points $(1,0,1)$ and $(0,1,0)$, whose component sums
differ, so it is not a polymatroid. This does not affect any of the results below, which do not use a polymatroid structure of the essential
intersection.
\end{remark}

\subsection{The extended results}

\begin{theorem}[$K$-user versions]\label{thm:K}
Let $W$ be a finite-alphabet mixed memoryless $K$-user MAC with an arbitrary probability space $(\Theta,w)$ and an additive cost constraint $\bGamma$
as above.
\begin{itemize}
\item[(a)] (Additive MACs; Theorems~\ref{thm:additive} and \ref{thm:mixadd}.) For $Z^n=X^n_1+\cdots+X^n_K+V^n\pmod m$ with a general noise process, the
$\varepsilon$-capacity region is $\Cl_+\{R\in\R^K:J_{\rm ad}(R_{[1:K]})\le\varepsilon\}$; for mixed memoryless noise it is
$\Cl\{R\in\R^K_+:w\{\theta:\log m-H(V_\theta)<R_{[1:K]}\}\le\varepsilon\}$.
\item[(b)] (Inner bound; Theorem~\ref{thm:inner}.) $\cI_{\bGamma}(\varepsilon)\subseteq C^{\rm no}_{\bGamma}(\varepsilon\mid W)$ for every $0\le\varepsilon<1$.
\item[(c)] (Capacity region; Theorem~\ref{thm:cap}.) $C^{\rm no}_{\bGamma}(0\mid W)=\Cl\bigl(\bigcup_{(X_{[1:K]},T)\in\cP_{\bGamma}}\bigl(\Lambda^*(X_{[1:K]},T)\cap\R^K_+\bigr)\bigr)$, where
$\Lambda^*(X_{[1:K]},T)\equiv\{R\in\R^K:R_\cS\le I^*_\cS\ \forall\cS\}$.
\item[(d)] (Outer bound; Theorem~\ref{thm:outer}.) $C^s_{\bGamma}(\varepsilon\mid W)\subseteq\cO_{\bGamma}(\varepsilon)$ for every $0\le\varepsilon<1$ and $s\in\{{\rm no},{\rm rx},{\rm tx},{\rm rt}\}$.
\item[(e)] (Well-ordered mixtures; Theorem~\ref{thm:wellordered}, Proposition~\ref{prop:degraded} and Corollary~\ref{cor:verifiable}.) If condition $(C_\varepsilon)$ of
Section~\ref{sec:wellordered} holds with $\cP_{\bGamma}$ and \eqref{eq:K-pentagon}, then
$\cI_{\bGamma}(\varepsilon)=C^{\rm no}_{\bGamma}(\varepsilon\mid W)=C^{\rm rx}_{\bGamma}(\varepsilon\mid W)=C^{\rm tx}_{\bGamma}(\varepsilon\mid W)=C^{\rm rt}_{\bGamma}(\varepsilon\mid W)=\cO_{\bGamma}(\varepsilon)$. This is the
case for every $\varepsilon$ if there is a common optimal input, or if $\{W_\theta\}$ is degradedly ordered and every nonempty $S_R$ has a least element.
\end{itemize}
\end{theorem}

\begin{proof}[Proof (outline)]
In each part the proof of the cited two-user result applies with the following changes.

(a) For independent uniform inputs, $P_{Z^n|X^n_{\cS^c}}$ is uniform on $\cZ^n$ for every nonempty $\cS$, because at least one uniform input enters the
sum. Hence all $2^K-1$ densities equal $\log m-\frac1n\log(1/P_{V^n}(V^n))$, and $J_W(R\mid\bX)=J_{\rm ad}(\max_{\cS}R_\cS)$. In the direct part, with $R^*\in\R^K_+$
and $R'^{(k)}\equiv R^*-t_k$, every $R'^{(k)}_\cS=R^*_\cS-|\cS|t_k\le s^*-t_k\le s_k$, as in the proof of Theorem~\ref{thm:additive}. The converse part uses only the
density for $\cS=[1:K]$. The mixed case follows from Lemma~\ref{lem:entspec} as before, with $R_1+R_2$ replaced by $R_{[1:K]}$ and $\lambda_kR$ defined as before.

(b) Lemmas~\ref{lem:mixcomp}, \ref{lem:reduction}, \ref{lem:trunc} and \ref{lem:single} hold with $\cA=\cX_{[1:K]}$, with the term $3e^{-n\gamma}$ in \eqref{eq:reduction} replaced by
$(2^K-1)e^{-n\gamma}$, and with $\delta_n$ in \eqref{eq:trunc} replaced by $(\frac1n\log\frac1{\pi_\cS})_\cS$, $\pi_\cS\equiv\prod_{k\in\cS}\pi_k$. The proof of Theorem~\ref{thm:inner} then applies with the $2^K-1$ thresholds $R_\cS-4|\cS|\gamma$.

(c) The direct part follows from (b). In the converse part, Fano's inequality applied to the messages of the users in $\cS$, given the others, gives
$(1-\varepsilon_{\theta,n})(R_\cS-|\cS|\gamma)\le\frac1n\sum_iI(X_{\cS,i};Z_{\theta,i}\mid X_{\cS^c,i})+\frac{\log2}n$ for every nonempty $\cS$. The quantization of Step~2 is done on
$K$-tuples of pmfs with cells of diameter $\delta$ in each coordinate, and $\tau_{N_K}(K\delta)\le\gamma$; Steps~3 and 4 are unchanged. For finite $\Theta$ the union may be restricted to $|\cT|\le(2^K-1)|\Theta|+K$, as in Remark~\ref{rem:card}.

(d) The proof of Theorem~\ref{thm:outer} uses only the strong converse property of each component, the closedness of $\widehat C_{\bGamma}(W_\theta)$ and the lower
semicontinuity of $o_{\bGamma}$ (Lemma~\ref{lem:lsc}), whose proofs extend verbatim given the continuity of the $2^K-1$ mutual informations in the channel
(Lemma~\ref{lem:cont} with $N=N_K$). Corollary~\ref{cor:K-scprop} replaces Lemma~\ref{lem:scprop}.

(e) The proof of Theorem~\ref{thm:wellordered} combines (b) and (d). In Proposition~\ref{prop:degraded}, the data-processing inequality gives
$I^{(\cS)}_\theta\le I^{(\cS)}_{\theta'}$ for every nonempty $\cS$, since conditionally on $(X_{\cS^c},T)$ the Markov chain $X_\cS-Z_{\theta'}-Z_\theta$ holds; the rest is
unchanged, and so is the proof of Corollary~\ref{cor:verifiable}.
\end{proof}

\subsection{CSI at the transmitters}

The direct part of Theorem~\ref{thm:csit} rests on the CSIT coding lemma of Part I in component form, restated here for two users as Lemma~\ref{lem:compform}. Part I
states and proves it for $K$ users \cite[Lemma~8 and Appendix~B]{HY-I}; we restate the $K$-user form in the present notation. For a countable $\Theta$, let $X^n_\theta=(X^n_{1,\theta},\ldots,X^n_{K,\theta})$ be, for each $\theta$,
independent inputs that satisfy the cost constraint $\bGamma$ with probability one, let $Z^n_\theta$ be the output of $W^n_\theta$, and let $\iota^n_{\cS,\theta}$ be the
component densities, defined as in \eqref{eq:compdens} with
$\iota^n_{\cS,\theta}\equiv\frac1n\log\bigl(W^n_\theta(Z^n_\theta\mid X^n_\theta)/P_{Z^n_\theta|X^n_{\cS^c,\theta}}(Z^n_\theta\mid X^n_{\cS^c,\theta})\bigr)$. For a nonempty $\cA\subseteq[1:K]$ let
\begin{equation}\label{eq:K-LS}
\cL_\cA\equiv\bigl\{\cS:\emptyset\ne\cS\subseteq\cA\bigr\}\cup\bigl\{[1:K]\bigr\},\qquad
E^{(\theta,\cA)}_n(a\mid\bX_\theta)\equiv\bigcup_{\cS\in\cL_\cA}\bigl\{\iota^n_{\cS,\theta}\le a_\cS\bigr\}.
\end{equation}
For $K=2$, $\cL_\cA$ is the family of Section~\ref{sec:csit}, and \eqref{eq:K-LS} is \eqref{eq:ES}.

\begin{lemma}[$K$-user CSIT coding lemma in component form]\label{lem:K-compform}
In the setup above, let $R\in\R^K_+$, $\cA\equiv\{k:R_k>0\}\ne\emptyset$, and $\gamma>0$. Then there are CSIT codes satisfying the cost constraint $\bGamma$, with
$M_{k,n}=\lceil e^{nR_k}\rceil$ for $k\in\cA$ and $M_{k,n}=1$ for $k\notin\cA$, such that
\begin{equation}\label{eq:K-compform}
\limsup_{n\to\infty}\varepsilon_n\le\limsup_{n\to\infty}\sum_{\theta\in\Theta}w(\theta)\Pr\bigl\{E^{(\theta,\cA)}_n\bigl((R_\cS+4\gamma)_{\cS\in\cL_\cA}\bigm|\bX_\theta\bigr)\bigr\}.
\end{equation}
\end{lemma}

Lemma~\ref{lem:K-compform} is \cite[Lemma~8]{HY-I}, with the same family $\cL_\cA$ as in \cite[Sec.~VIII-D]{HY-I}; for $K=2$ it is Lemma~\ref{lem:compform}. In the random-coding
argument of \cite[Appendix~B]{HY-I}, the decoder searches jointly over the message tuple and a finite list $\cD_n$ of candidate components. A wrong message tuple
for the true component differs from the transmitted one on a nonempty set $\cS\subseteq\cA$, since the users outside $\cA$ have a single message, and is
handled by the test for $\cS$; a wrong component changes the codewords of all users and is handled by the test for $[1:K]$. Each of the
$|\cL_\cA|\le2^K$ tests contributes $e^{-n\gamma}$, and the list contributes $|\cD_n|e^{-n\gamma}$.

\begin{theorem}[$K$-user CSIT and CSIRT regions]\label{thm:K-csit}
Let $\Theta$ be countable and let $\bGamma$ satisfy $\Gamma_k>\min_{x\in\cX_k}c_k(x)$ for every $k$. For every $0\le\varepsilon<1$,
\begin{equation}\label{eq:K-csit}
C^{\rm tx}_{\bGamma}(\varepsilon\mid W)=C^{\rm rt}_{\bGamma}(\varepsilon\mid W)=\cO_{\bGamma}(\varepsilon).
\end{equation}
\end{theorem}

\begin{proof}
\emph{Converse part.} $C^{\rm tx}_{\bGamma}(\varepsilon\mid W)\subseteq C^{\rm rt}_{\bGamma}(\varepsilon\mid W)\subseteq\cO_{\bGamma}(\varepsilon)$ by the definitions and part (d) of Theorem~\ref{thm:K}.

\emph{Direct part.} We follow the direct part of the proof of Theorem~\ref{thm:csit}, with Lemma~\ref{lem:K-compform} in place of Lemma~\ref{lem:compform}. Let
$R\in\cO_{\bGamma}(\varepsilon)$, $R\ne0$, fix $\delta\in(0,1)$, and put $R'\equiv(1-\delta)R$, $\cA\equiv\{k:R_k>0\}$ and $r\equiv\min_{k\in\cA}R_k>0$. Every $\cS\in\cL_\cA$ meets
$\cA$, so that
\begin{equation}\label{eq:K-RJ}
R_\cS\ge r\qquad(\cS\in\cL_\cA).
\end{equation}

\emph{Steps 1--3.} Let $\Theta_{\rm good}\equiv\{\theta:R\in C_{\bGamma}(W_\theta)\}$, so that $w(\Theta\setminus\Theta_{\rm good})=o_{\bGamma}(R)\le\varepsilon$. For $\theta\in\Theta_{\rm good}$ choose
$(X_{[1:K],\theta},T_\theta)\in\cP_{\bGamma}$ with $R\in\Lambda_\theta(X_{[1:K],\theta},T_\theta)$, and let $\bar{\bX}_\theta$ be the input of Lemma~\ref{lem:single} ($K$ users) for it with
$\eta\equiv\delta r/4$; for $\theta\notin\Theta_{\rm good}$ use constant sequences of minimum-cost letters. With $\gamma\equiv\delta r/16$, $R_\cS\le I^{(\cS)}_\theta$ and \eqref{eq:K-RJ}
give $R'_\cS+4\gamma\le I^{(\cS)}_\theta-\frac34\delta r<I^{(\cS)}_\theta-\eta$ for $\cS\in\cL_\cA$, so that, by \eqref{eq:single},
\[
\lim_{n\to\infty}\Pr\bigl\{E^{(\theta,\cA)}_n\bigl((R'_\cS+4\gamma)_{\cS\in\cL_\cA}\bigm|\bar{\bX}_\theta\bigr)\bigr\}=0\qquad(\theta\in\Theta_{\rm good}).
\]
As in the two-user case, the events for $\cS\notin\cL_\cA$ are not controlled and are not needed.

\emph{Step 4.} Since $\Theta$ is countable, dominated convergence bounds the right-hand side of \eqref{eq:K-compform} by $w(\Theta\setminus\Theta_{\rm good})\le\varepsilon$,
and Lemma~\ref{lem:K-compform} yields CSIT codes at rates $R'$ with $\limsup_n\varepsilon_n\le\varepsilon$. Since $\delta$ is arbitrary and $C^{\rm tx}_{\bGamma}(\varepsilon\mid W)$ is
closed, $R\in C^{\rm tx}_{\bGamma}(\varepsilon\mid W)$.
\end{proof}

As for two users (Remark~\ref{rem:where-countable}), countability of $\Theta$ is used only in Lemma~\ref{lem:K-compform}. Proposition~\ref{prop:gap} is a
single-user example and applies to $K$ users as it stands, so that, without CSI, $C^{\rm no}_{\bGamma}(\varepsilon\mid W)\subsetneq C^{\rm tx}_{\bGamma}(\varepsilon\mid W)$ is possible.

%% ====================================================================
\section{Concluding Remarks}\label{sec:conc}

Starting from the single information spectrum formula of Theorem~\ref{thm:general}, we obtained single-letter characterizations
and bounds for the $\varepsilon$-capacity regions of finite-alphabet mixed MACs with an arbitrary mixing measure. The
specialization required only a change-of-measure bound, a type-counting comparison between a mixture and its
components, and the continuity of mutual information, together with Ahlswede's strong converse for the outer
bound.

For mixed MACs with general components and a countable parameter space, Part I \cite[Theorem~9]{HY-I}
shows that CSI at the receiver alone never enlarges the $\varepsilon$-capacity region and adds nothing to CSI at the transmitters,
and gives the region with CSI at the transmitters as a multi-letter formula; Theorem~\ref{thm:csit} single-letterizes this formula
for finite-alphabet mixed memoryless MACs.

The results can be summarized in terms of the two regions of the Introduction. The outage region $\cO_{\Gamma}(\varepsilon)$, in which
each component may use its own input, is an outer bound in every CSI scenario and is achieved when the encoders
know the component. The region $\cI_{\Gamma}(\varepsilon)$, in which one input serves all components, is achieved without any CSI.
For $\varepsilon=0$ the no-CSI capacity region equals $\cI_{\Gamma}(0)$, and the two regions coincide for well-ordered mixtures, which
include degradedly ordered families and mixtures with a common optimal input. In these cases the familiar outage
formula is the exact $\varepsilon$-capacity region, and informing the transmitters or the receiver of the realized component
does not enlarge it. In general, however, the value of CSI at the transmitters can be strictly positive (Proposition~\ref{prop:gap}).

The main open problem is a single-letter converse for $0<\varepsilon<1$ without CSI; Section~\ref{sec:discussion} explains why a
per-component strong converse cannot provide it. All of these results extend to $K$ users with cost constraints (Theorems~\ref{thm:K} and \ref{thm:K-csit}), the key step being the $K$-user Ahlswede-type bound of
Lemma~\ref{lem:K-ahlswede}; the polymatroid structure of Remark~\ref{rem:polymatroid} is specific to two users (Remark~\ref{rem:K-polymatroid}).
A second open problem is whether Theorem~\ref{thm:csit} extends to uncountable
parameter spaces for stationary memoryless components (Remark~\ref{rem:where-countable}). The Gaussian and quasi-static fading cases, and mixed MACs with
general components, are treated in Part I \cite{HY-I}. The strong converse for the $K$-user Gaussian MAC in the sense of \cite[Definition~2]{HY-I}, credited to Fong and Tan \cite{FT16}
(cf.\ \cite[Remark~5]{HY-I}), is given an elementary information spectrum proof in Part III \cite{HY-III}.

%% ====================================================================
\appendices

\section{Proofs of the Three Tools}\label{app:tools}

\begin{proof}[Proof of Lemma~\ref{lem:com}]
For each $s$, let $B_s\equiv\{u:P_{U|S}(u\mid s)<e^{-a}Q(u\mid s)\}$. Then $\Pr\{U\in B_s\mid S=s\}=
\sum_{u\in B_s}P_{U|S}(u\mid s)\le e^{-a}\sum_uQ(u\mid s)\le e^{-a}$; average over $s$. For densities, the sums are replaced by integrals with
respect to the reference measure.
\end{proof}

\begin{proof}[Proof of Lemma~\ref{lem:mixcomp}]
$Q^n_\theta(\bb\mid\ba)=\prod_{(a,b)}Q_\theta(b\mid a)^{nP(a,b)}$ depends on $(\ba,\bb)$ only through its joint type $P$; write
it $f_P(\theta)$. Then $Q^n(\bb\mid\ba)=\int f_P\,dw$, and by Markov's inequality $w\{\theta:f_P(\theta)>e^{\sqrt n}\int f_P\,dw\}\le e^{-\sqrt n}$; if
$\int f_P\,dw=0$ the set on the left is $w$-null. Since there are at most $(n+1)^{|\cA||\cB|}$ joint types, the union bound gives
the claim.
\end{proof}

\begin{proof}[Proof of Lemma~\ref{lem:cont}]
For pmfs on a finite set $\cA$ with $|\cA|\ge2$ and $d\le1-1/|\cA|$, Zhang \cite{Zha07} showed $|H(P)-
H(\tilde P)|\le d\log(|\cA|-1)+h(d)$. The marginals of $P$ and $\tilde P$ are also within distance $\delta$, and $\delta\log N+h(\delta)$ is
nondecreasing on $[0,1/2]$. Hence each of $H(U)$, $H(V)$ and $H(U,V)$ changes by at most $\delta\log N+h(\delta)$ with $N=
|\cU||\cV|$ (trivially so if an alphabet has one element). The claim follows from $I(U;V)=H(U)+H(V)-H(U,V)$.
\end{proof}

\section{Proofs of Lemmas~\ref{lem:entspec}, \ref{lem:reduction} and \ref{lem:trunc}}\label{app:B}

\begin{proof}[Proof of Lemma~\ref{lem:entspec}]
Let $V^n_\theta\sim P^n_\theta$; then $\Pr\{h_n(V^n)\ge r\}=\int\Pr\{h_n(V^n_\theta)\ge r\}\,dw(\theta)$. Let $g^\theta_n\equiv\frac1n\log(1/P^n_\theta(V^n_\theta))$,
which converges to $H(V_\theta)$ in probability by the weak law of large numbers.

\emph{Upper bound.} Let $\Theta^*_n$ be as in Lemma~\ref{lem:mixcomp} with $|\cA|=1$ and $Q_\theta=P_\theta$. For $\theta\in\Theta^*_n$ we have $P^n_\theta\le e^{\sqrt n}P_{V^n}$, hence
$h_n(V^n_\theta)\le g^\theta_n+n^{-1/2}$. Therefore
\[
\Pr\{h_n(V^n)\ge r\}\le w(\Theta\setminus\Theta^*_n)+\int_\Theta\Pr\{g^\theta_n\ge r-n^{-1/2}\}\,dw(\theta),
\]
and the last probability tends to 0 whenever $H(V_\theta)<r$. Since $w(\Theta\setminus\Theta^*_n)\to0$ by Lemma~\ref{lem:mixcomp}, the reverse Fatou lemma
gives $\limsup_n\Pr\{h_n(V^n)\ge r\}\le w\{\theta:H(V_\theta)\ge r\}$.

\emph{Lower bound.} Fix $\gamma>0$. By Lemma~\ref{lem:com}, $P_{V^n}(V^n_\theta)\le e^{n\gamma}P^n_\theta(V^n_\theta)$, i.e., $h_n(V^n_\theta)\ge g^\theta_n-\gamma$, with probability at
least $1-e^{-n\gamma}$. Hence $\Pr\{h_n(V^n_\theta)\ge r\}\ge\Pr\{g^\theta_n\ge r+\gamma\}-e^{-n\gamma}$, which tends to 1 whenever $H(V_\theta)>r+\gamma$.
By Fatou's lemma, $\liminf_n\Pr\{h_n(V^n)\ge r\}\ge w\{\theta:H(V_\theta)>r+\gamma\}$. Letting $\gamma\downarrow0$ gives the first inequality in
\eqref{eq:entspec}.
\end{proof}

\begin{proof}[Proof of Lemma~\ref{lem:reduction}]
For $\theta\in\Theta^*_n$, Lemma~\ref{lem:mixcomp} gives $\frac1n\log W^n(Z^n_\theta\mid X^n,Y^n)\ge\frac1n\log W^n_\theta(Z^n_\theta\mid X^n,Y^n)-n^{-1/2}$. By
Lemma~\ref{lem:com} with $S=Y^n$, $U=Z^n_\theta$ and $Q=P_{Z^n|Y^n}$, the event $G_1\equiv\{P_{Z^n|Y^n}(Z^n_\theta\mid Y^n)\le e^{n\gamma}P_{Z^n_\theta|Y^n}(Z^n_\theta\mid Y^n)\}$
has probability at least $1-e^{-n\gamma}$. On $G_1$ we get $\iota^n_1(X^n,Y^n,Z^n_\theta)\ge\iota^n_{1,\theta}-\gamma-n^{-1/2}$, so
\[
\{\iota^n_1\le a_1\}\subseteq\{\iota^n_{1,\theta}\le a_1+\gamma+n^{-1/2}\}\cup G_1^c.
\]
The same argument with $S=X^n$, and with $S$ trivial, handles $k=2,3$. The union bound completes the proof.
\end{proof}

\begin{proof}[Proof of Lemma~\ref{lem:trunc}]
Let $\bar Z^n$ be the output due to $(\bar X^n,\bar Y^n)$. For $\by\in\cC_2$, $P_{Z^n|Y^n}(\bz\mid\by)\ge\sum_{\bx\in\cC_1}P_{X^n}(\bx)W^n(\bz\mid
\bx,\by)=\pi_1P_{\bar Z^n|\bar Y^n}(\bz\mid\by)$. Similarly, $P_{Z^n|X^n}\ge\pi_2P_{\bar Z^n|\bar X^n}$ on $\cC_1$ and $P_{Z^n}\ge\pi_1\pi_2P_{\bar Z^n}$. Hence, on $\cC_1\times\cC_2\times\cZ^n$,
each density for $(X^n,Y^n)$ is at most the corresponding density for $(\bar X^n,\bar Y^n)$ plus the corresponding component
of $\delta_n$; that is, $\iota^n_k\le\bar\iota^n_k+\delta_{n,k}$ there. So the event on the left of \eqref{eq:trunc} is contained in the set $G$ of $(\bx,\by,\bz)$ in the event on the right.
Finally, $P_{\bar X^n\bar Y^n\bar Z^n}\le(\pi_1\pi_2)^{-1}P_{X^nY^nZ^n}$ pointwise, which gives \eqref{eq:trunc}. Thresholds $a_k=-\infty$ give empty events on
both sides.
\end{proof}

\section{Proof of Proposition~\ref{prop:singleuser}}\label{app:E}

\begin{proof}[Proof of Proposition~\ref{prop:singleuser}]
Item 1): if $P^\star$ achieves every $c_\theta$, then $I(P^\star,W_{\theta'})=c_{\theta'}\ge c_\theta$.

Item 2): given $R$, let $\theta_0$ minimize $c_\theta$ over $S_R$. For $\theta\in S_R$, $c_{\theta_0}\le c_\theta$, so \eqref{eq:wo-single} gives $I(P_{\theta_0},W_\theta)\ge c_{\theta_0}\ge R$.

Item 3): let $\theta$ be an atom and $R=c_\theta$. Then $o_{\Gamma}(R)\le1-w(\{\theta\})<1$, so $R\in\cO_{\Gamma}(\varepsilon)$ for some $\varepsilon<1$. Let $(X,T)$ be the
input from $(C_\varepsilon)$ (with $|\cY|=1$, the triple reduces to $(X,T)$), and let $\bar P\equiv\sum_tP_T(t)P_{X|T=t}$ be its average input
distribution. Since $\theta$ is an atom, $\theta\in S_R$ is not $w$-null, so $R\in\Lambda_\theta(X,T)$, i.e., $c_\theta=R\le I(X;Z_\theta\mid T)$. By the concavity of mutual
information in the input distribution, $I(X;Z_\theta\mid T)=\sum_tP_T(t)I(P_{X|T=t},W_\theta)\le I(\bar P,W_\theta)$, and $\E c_1(X_{\bar P})=\E c_1(X)\le\Gamma_1$.
Hence $\bar P$ achieves $c_\theta$, and $\bar P=P_\theta$ by uniqueness. For $w$-a.e.\ $\theta'\in S_R$, i.e., with $c_{\theta'}\ge c_\theta$, $(C_\varepsilon)$ and concavity give
$c_\theta=R\le I(X;Z_{\theta'}\mid T)\le I(\bar P,W_{\theta'})=I(P_\theta,W_{\theta'})$. If $\Theta$ is finite and $w$ has full support, every $\theta$ is an atom and
``$w$-a.e.'' means ``every'', so Definition~\ref{def:wellordered} implies \eqref{eq:wo-single}; the converse is item 2), the minimum being attained since $\Theta$ is
finite.

Item 4): if $c_\theta\le c_{\theta'}$, strict monotonicity excludes $\theta'\prec\theta$, so $\theta\preceq\theta'$, and data processing gives $I(P_\theta,W_{\theta'})\ge
I(P_\theta,W_\theta)=c_\theta$.

Item 5): the Z-channels $Z(a)$ with $Z(a)(1\mid0)=a$ are degradedly ordered, because $Z(a')=Z(b)\circ Z(a)$ with $a'=a+(1-a)b$,
and their capacities decrease strictly in $a$. By item 4) they satisfy \eqref{eq:wo-single}, although their optimal inputs differ.
A mixture of ${\rm BSC}(p)$ and ${\rm BEC}(h(p))$ has the uniform input optimal for both, so it satisfies \eqref{eq:wo-single} by item 1). For $p=0.11$,
so that $h(p)\approx0.5$ bit, neither channel is a degraded version of the other: ${\rm BSC}(p)$ is a degraded version of ${\rm BEC}(e)$ if and only if
$p\ge e/2$, and ${\rm BEC}(e)$ with $e<1$ is never a degraded version of ${\rm BSC}(p)$ with $p>0$; see \cite{Nair10}.
\end{proof}

\section{Proof of Lemmas~\ref{lem:ahlswede} and \ref{lem:K-ahlswede}}\label{app:K}

This appendix also proves Lemma~\ref{lem:ahlswede}, which is the case $K=2$. In two-user terms the proof runs as follows. Keep only the
message pairs that are decoded with small error (Step~1); this subcode has maximal error below one, but its codewords $\bu_j$ and $\bv_k$ are no
longer independent. Fix a few coordinates of the codewords (Step~2, wringing), after which the two users' letters are nearly independent at every
remaining time. Then each of the three bounds is a single-user bound (Step~3): fixing the message of user~2 leaves a code for user~1 over the
channel with input $\bu_j$, which gives the bound for $\{1\}$; symmetrically for $\{2\}$; and the whole subcode gives the bound for $\{1,2\}$. Finally,
the near independence is turned into exact independence at a small cost in rate (Steps~4--6). For $K$ users nothing changes but the number of
bounds: Step~3 gives one bound for each nonempty set $\cS$, by fixing the messages of the users outside $\cS$, and the wringing of Step~2 is done once,
on the product of all users' alphabets, so that all $2^K-1$ bounds hold for a common input. We use two results from \cite{Ahl82}. The first is Augustin's bound
\cite[Cor.~1]{Ahl82}, \cite{Aug66}:

\emph{Augustin's bound.} Let $0<\lambda<1$, and consider a nonstationary memoryless channel $\bx\mapsto\prod_{t=1}^nV_t(\cdot\mid x_t)$ with a finite input
alphabet $\cU$. Every code for it with $M$ codewords and maximal error probability at most $\lambda$ satisfies
$\log M\le\sum_{t=1}^nI(U_t;Z_t)+\frac{3}{1-\lambda}|\cU|\sqrt n$, where $U^n$ is uniform on the codewords and $Z^n$ is the corresponding output.

The second is Ahlswede's wringing lemma, in the form stated by Fong and Tan \cite[Lemma~2]{FT16}:

\emph{Wringing lemma.} Let $\cU$ be a finite set, let $p$ and $u$ be pmfs on $\cU^n$ with $p\le(1+c)u$ pointwise for some $c>0$, and let
$0<\nu<1$ and $0<\gamma<c$. Then there are $\ell\le c/\gamma$ coordinates $t_1,\ldots,t_\ell$ and letters $\bar x_{t_1},\ldots,\bar x_{t_\ell}\in\cU$ such that, with
$G\equiv\{x_{t_i}=\bar x_{t_i},\ i=1,\ldots,\ell\}$, (I) $p(G)\ge\nu^\ell$, and (II) for every $t\notin\{t_1,\ldots,t_\ell\}$ and every $x\in\cU$,
$p_{X_t|G}(x)\le\max\{(1+\gamma)u_{X_t|G}(x),\ \nu\}$.

\emph{Step 0 (Parameters).} A code with $\varepsilon_n\le0$ has $\varepsilon_n\le\bar\lambda$ for every $\bar\lambda>0$, so we may assume $0<\bar\lambda<1$. Write
\[
\lambda\equiv\frac{1+\bar\lambda}{2},\qquad\lambda^*\equiv\frac{2\bar\lambda}{1+\bar\lambda},\qquad\beta=\frac{\lambda^*}{1-\lambda^*}=\frac{2\bar\lambda}{1-\bar\lambda},
\]
so that $1-\lambda=(1-\bar\lambda)/2>0$, $1-\lambda^*=(1-\bar\lambda)/(1+\bar\lambda)>0$ and $1/(1-\lambda^*)=1+\beta$.

\emph{Step 1 (Expurgation).} Let $\cE\equiv\{\bj\in\mathcal{M}:V^n(D_{\bj}\,|\,\bu(\bj))\ge1-\lambda\}$. The tuples outside $\cE$ have error probability greater than $\lambda$,
and the average error probability is at most $\bar\lambda$. By Markov's inequality they form a fraction of at most $\bar\lambda/\lambda=\lambda^*$ of
$\mathcal{M}$, so that $|\cE|\ge(1-\lambda^*)M_{[1:K]}$.

\emph{Step 2 (Wringing on the product alphabet).} This step follows Fong and Tan \cite[Lemma~3]{FT16}, without quantization. Let $p$ be the law of the codeword tuple $\bu(\bj)\in\cX_{[1:K]}^n$ when $\bj$ is uniform on $\cE$, and let
$u\equiv\prod_{k=1}^Ku_k$ be its law when $\bj$ is uniform on $\mathcal{M}$, i.e., when $j_1,\ldots,j_K$ are independent and uniform; here $u_k$ is the law of
$\bu_k(j_k)$ for $j_k$ uniform on $[1:M_k]$, and letters of $\cX_{[1:K]}^n$ are identified with $K$-tuples of words. Then, for every $\bx=(\bx_1,\ldots,\bx_K)$,
\[
p(\bx)=\frac{|\{\bj\in\cE:\bu(\bj)=\bx\}|}{|\cE|}\le\frac{M_{[1:K]}}{|\cE|}\cdot\frac{|\{\bj\in\mathcal{M}:\bu(\bj)=\bx\}|}{M_{[1:K]}}\le(1+\beta)\,u(\bx).
\]
Apply the wringing lemma with the alphabet $\cX_{[1:K]}$, $c\equiv\beta$, $\gamma\equiv\gamma_n\equiv n^{-1/2}$ and $\nu\equiv\nu_n\equiv n^{-1}$; the requirement
$\gamma_n<\beta$ holds for $n>\beta^{-2}$. We obtain $\ell\le\beta\sqrt n$ coordinates and letters $\bar x_{t_i}=(\bar x_{1,t_i},\ldots,\bar x_{K,t_i})$. Let
$\bar\cE\subseteq\cE$ be the set of $\bj\in\cE$ with $\bu(\bj)\in G$. By (I),
\begin{equation}\label{eq:K-wrung}
|\bar\cE|=p(G)\,|\cE|\ge n^{-\beta\sqrt n}(1-\lambda^*)M_{[1:K]}.
\end{equation}
Let $\bj=(J_1,\ldots,J_K)$ be uniform on $\bar\cE$, $\bar X^n_k\equiv\bu_k(J_k)$, $\bar X^n\equiv(\bar X^n_1,\ldots,\bar X^n_K)$, and let $\bar Z^n$ be the output of $V^n$ due to
$\bar X^n$. Then $\bar X^n$ has the law $p_{\,\cdot\,|G}$. Write $\bar P_t$ for the joint pmf of $\bar X_t=(\bar X_{1,t},\ldots,\bar X_{K,t})$ on $\cX_{[1:K]}$, and
$\bar P_{k,t}$ for its marginals.

The event $G$ is a product event, $G=\bigcap_k\{\bx_k:x_{k,t_i}=\bar x_{k,t_i},\ i=1,\ldots,\ell\}$, and $u$ is a product measure. Hence $u_{\,\cdot\,|G}$ is again
a product measure, and its one-letter marginal at $t$ is $Q_t\equiv\prod_{k=1}^Kq_{k,t}$ for some pmfs $q_{k,t}$ on $\cX_k$. By (II),
\begin{equation}\label{eq:K-wring-ineq}
\bar P_t(x)\le\max\{(1+\gamma_n)Q_t(x),\ \nu_n\}\le(1+\gamma_n)Q_t(x)+\nu_n\qquad\mbox{for all }x\in\cX_{[1:K]}
\end{equation}
for $t\notin\{t_1,\ldots,t_\ell\}$. For $t=t_i$, both $\bar P_t$ and $Q_t$ are the point mass at $\bar x_{t_i}$, so \eqref{eq:K-wring-ineq} holds for every $t$.

\emph{Step 3 (Single-letter bounds).} Every $\bj\in\bar\cE\subseteq\cE$ has error probability at most $\lambda$, and the decoding regions are disjoint.
Fix a nonempty $\cS\subseteq[1:K]$. For each tuple $\bm w$ of messages of the users in $\cS^c$, let $\cB(\bm w)\equiv\{\bj_\cS:(\bj_\cS,\bm w)\in\bar\cE\}$ and
$p_{\bm w}\equiv|\cB(\bm w)|/|\bar\cE|=\Pr\{\bj_{\cS^c}=\bm w\}$. (If $\cS=[1:K]$, there is only the empty tuple $\bm w$ and $\cB(\bm w)=\bar\cE$.) If
$\cB(\bm w)\ne\emptyset$, then $\{(\bu_\cS(\bj_\cS),D_{(\bj_\cS,\bm w)}):\bj_\cS\in\cB(\bm w)\}$ is a code with $|\cB(\bm w)|$ messages and maximal error at most $\lambda$
for the nonstationary memoryless channel $\bx_\cS\mapsto V^n(\cdot\,|\,\bx_\cS,\bu_{\cS^c}(\bm w))$, whose input alphabet $\prod_{k\in\cS}\cX_k$ has at most $L$
letters. Given $\bj_{\cS^c}=\bm w$, $\bj_\cS$ is uniform on $\cB(\bm w)$, and Augustin's bound gives
\[
\log|\cB(\bm w)|\le\sum_{t=1}^nI(\bar X_{\cS,t};\bar Z_t\,|\,\bj_{\cS^c}=\bm w)+\frac{3}{1-\lambda}L\sqrt n.
\]
Average over $\bm w$ with the weights $p_{\bm w}$. On the left, since $\bj_{\cS^c}$ takes at most $M_{\cS^c}$ values,
\[
\sum_{\bm w}p_{\bm w}\log|\cB(\bm w)|=\sum_{\bm w}p_{\bm w}\log\bigl(p_{\bm w}|\bar\cE|\bigr)=\log|\bar\cE|-H(\bj_{\cS^c})\ge\log|\bar\cE|-\log M_{\cS^c}.
\]
On the right, $\bar X_{\cS^c,t}$ is a function of $\bj_{\cS^c}$, and, given $\bj$, the output $\bar Z_t$ has the law $V(\cdot\,|\,\bar X_t)$; hence
$H(\bar Z_t\,|\,\bar X_{\cS,t},\bj_{\cS^c})=H(\bar Z_t\,|\,\bar X_t)$ and
\[
I(\bar X_{\cS,t};\bar Z_t\,|\,\bj_{\cS^c})=H(\bar Z_t\,|\,\bj_{\cS^c})-H(\bar Z_t\,|\,\bar X_t)\le H(\bar Z_t\,|\,\bar X_{\cS^c,t})-H(\bar Z_t\,|\,\bar X_t)=I(\bar X_{\cS,t};\bar Z_t\,|\,\bar X_{\cS^c,t}).
\]
Combining these with \eqref{eq:K-wrung} and $M_{[1:K]}/M_{\cS^c}=M_\cS$ gives, for every nonempty $\cS$,
\begin{equation}\label{eq:K-singleK}
\log M_\cS\le\sum_{t=1}^nI(\bar X_{\cS,t};\bar Z_t\,|\,\bar X_{\cS^c,t})+\frac{3}{1-\lambda}L\sqrt n+\beta\sqrt n\log n+\log\frac{1}{1-\lambda^*}.
\end{equation}

\emph{Step 4 (Independence).} Let $\tilde P_t\equiv\prod_{k=1}^K\bar P_{k,t}$ be the product of the marginals of $\bar P_t$. We bound the variational distance
$d(P,Q)\equiv\frac12\sum_x|P(x)-Q(x)|=\sum_x(P(x)-Q(x))^+$. By \eqref{eq:K-wring-ineq},
\[
d(\bar P_t,Q_t)=\sum_{x\in\cX_{[1:K]}}\bigl(\bar P_t(x)-Q_t(x)\bigr)^+\le\sum_{x\in\cX_{[1:K]}}\bigl(\gamma_nQ_t(x)+\nu_n\bigr)=\gamma_n+L\nu_n\le2n^{-1/2},
\]
the last inequality for $n\ge L^2$. Marginalization does not increase the variational distance, so $d(\bar P_{k,t},q_{k,t})\le2n^{-1/2}$ for every $k$,
and, by the elementary bound for product measures, $d(\tilde P_t,Q_t)\le\sum_kd(\bar P_{k,t},q_{k,t})\le2Kn^{-1/2}$. By the triangle inequality,
\begin{equation}\label{eq:K-dK}
d(\bar P_t,\tilde P_t)\le\delta_n\equiv2(K+1)n^{-1/2}.
\end{equation}
Let $\tilde X_t=(\tilde X_{1,t},\ldots,\tilde X_{K,t})$ have the law $\tilde P_t$, so that $\tilde X_{1,t},\ldots,\tilde X_{K,t}$ are independent with the marginals of $\bar X_t$,
and let $\tilde Z_t$ be the output of $V$ due to $\tilde X_t$. The joint pmfs of $(\bar X_t,\bar Z_t)$ and $(\tilde X_t,\tilde Z_t)$ share the channel $V$, so they are
also within distance $\delta_n$, and so are all their marginals; for $n\ge16(K+1)^2$, $\delta_n\le1/2$.

Since $\bar X_{1,t},\ldots,\bar X_{K,t}$ need not be independent, we write each conditional mutual information as a signed sum of at most four
entropies of marginals of $(X_t,Z_t)$, with $X_t=(X_{1,t},\ldots,X_{K,t})$:
\[
I(X_{\cS,t};Z_t\,|\,X_{\cS^c,t})=H(X_{\cS^c,t}Z_t)-H(X_{\cS^c,t})-H(X_tZ_t)+H(X_t)
\]
(for $\cS=[1:K]$ the first two terms are replaced by $H(Z_t)$). All the alphabets involved have at most $N_K=L|\cZ|$ letters. By Zhang's bound
\cite{Zha07}, $|H(P)-H(P')|\le d\log(|\cA|-1)+h(d)$ for pmfs $P,P'$ on a finite set $\cA$ with $d=d(P,P')\le1-1/|\cA|$, and since $\delta\log N_K+h(\delta)$ is
nondecreasing on $[0,1/2]$, each of the entropies changes by at most $\delta_n\log N_K+h(\delta_n)$. Since $h(\delta)\le\delta\log(e/\delta)$ and
$\log(1/\delta_n)\le\frac12\log n$ (because $\delta_n\ge n^{-1/2}$), we have $h(\delta_n)\le\delta_n(1+\frac12\log n)$. For $n\ge3$, so that $\log n\ge1$,
each conditional mutual information therefore changes by at most
\begin{equation}\label{eq:K-indepK}
4\delta_n\Bigl(\log N_K+1+\frac12\log n\Bigr)\le4\delta_n(\log N_K+2)\log n=8(K+1)(\log N_K+2)\,n^{-1/2}\log n.
\end{equation}

\emph{Step 5 (Costs).} Each $\bar X^n_k$ is supported on codewords $\bu_k(j)$ with $\frac1nc_k(\bu_k(j))\le\Gamma_k$. Since the cost is additive and
$\tilde X_{k,t}$ has the law of $\bar X_{k,t}$,
\begin{equation}\label{eq:K-costsK}
\frac1n\sum_{t=1}^n\E c_k(\tilde X_{k,t})=\E\Bigl[\frac1nc_k(\bar X^n_k)\Bigr]\le\Gamma_k\qquad(k\in[1:K]).
\end{equation}
Let $T$ be uniform on $[1:n]$ and independent of everything else, and $X_k\equiv\tilde X_{k,T}$. Then $X_1,\ldots,X_K$ are conditionally independent
given $T$, $\E c_k(X_k)\le\Gamma_k$, so that $(X_1,\ldots,X_K,T)\in\cP_{\bGamma}$ (for $K=2$, $\cP_{\bGamma}=\cP_{\Gamma}$), and $\frac1n\sum_tI(\tilde X_{\cS,t};\tilde Z_t\,|\,\tilde X_{\cS^c,t})=I(X_\cS;Z\,|\,X_{\cS^c} T)$ for
every $\cS$.

\emph{Step 6 (Constants).} For $n\ge n_0$, all requirements used above hold: $\log n\ge1$, $\delta_n\le1/2$, $\gamma_n+L\nu_n\le2n^{-1/2}$ and $\gamma_n<\beta$.
Combine \eqref{eq:K-singleK}, \eqref{eq:K-indepK} summed over the $n$ coordinates, and Step~5, and divide by $n$. Since $\sqrt n\ge1$, $\log n\ge1$,
$3/(1-\lambda)=6/(1-\bar\lambda)$ and $\log(1/(1-\lambda^*))=\log((1+\bar\lambda)/(1-\bar\lambda))$, each remainder term is at most its coefficient in
\eqref{eq:K-consts} times $(\log n)/\sqrt n$, and \eqref{eq:K-ahlswede} follows. The constants $n_0$ and $K'$ depend only on $\bar\lambda$, $K$, $L$ and $|\cZ|$.
\hfill$\Box$

\section*{Acknowledgements}
\textit{Generative-AI use disclosure.} The authors used Anthropic's Claude (Opus 5.5) during the preparation of this manuscript.
It assisted in revising the exposition, including the introduction, the arrangement of proofs into appendices, and the redistribution of material
between Parts I and II (by which the theory of mixed MACs with general components was moved to Part I). It also assisted in drafting
Section~\ref{sec:kuser} and Appendix~\ref{app:K}, which extend the results to $K$ users. All mathematical results and proofs have been checked by
the authors, who take full responsibility for the content.

\end{document}